\RequirePackage{tikz}
\documentclass[pdflatex,sn-mathphys]{sn-jnl}
\usetikzlibrary{positioning,calc}
\usepackage{microtype}
\usepackage{mathtools}
\usepackage{stmaryrd}
\usepackage{booktabs}
\usepackage{multirow}
\usepackage{amsfonts}
\usepackage{amsbsy}
\usepackage{physics}
\usepackage{caption}
\usepackage{subcaption}
\usepackage{xcolor}
\usepackage{url}
\usepackage{natbib}

\newcommand{\Real}[0]{\mathbb{R}}
\newcommand{\Hilbert}[0]{\mathcal{H}}

\title{Resource Saving via Ensemble Techniques for Quantum Neural Networks}

\author*[1]{\fnm{Massimiliano} \sur{Incudini}}\email{massimiliano.incudini@univr.it}
\author[2]{\fnm{Michele} \sur{Grossi}}
\author[3]{\fnm{Andrea} \sur{Ceschini}}
\author[4]{\fnm{Antonio} \sur{Mandarino}}
\author[3]{\fnm{Massimo} \sur{Panella}}
\author[2]{\fnm{Sofia} \sur{Vallecorsa}}
\author[5]{\fnm{David} \sur{Windridge}}

\affil[1]{\orgdiv{Dipartimento di Informatica}, \orgname{Universit\`a di Verona}, \orgaddress{\street{Strada Le Grazie, 15}, \city{Verona}, \postcode{37134}, \country{Italy}}}

\affil[2]{\orgname{European Organization for Nuclear Research (CERN)}, \orgaddress{\street{Espl. des Particules, 1}, \city{Meyrin}, \postcode{1211}, \country{Switzerland}}}

\affil[3]{\orgdiv{Dipartimento di Ingegneria dell'Informazione, Elettronica e Telecomunicazioni (DIET)}, \orgname{Universit\`a di Roma ``La Sapienza''}, \orgaddress{\street{Via Eudossiana, 18}, \city{Roma}, \postcode{00184}, \country{Italy}}}

\affil[4]{\orgdiv{International Centre for Theory of Quantum Technologies (ICTQT)}, \orgname{University of Gdansk}, \orgaddress{\street{Jana Bażyńskiego 1A}, \city{Gdańsk}, \postcode{80-309}, \country{ Poland}}}

\affil[5]{\orgdiv{Department of Computer Science}, \orgname{Middlesex University}, \orgaddress{\street{The Burroughs}, \city{London}, \postcode{NW4 4BT}, \country{United Kingdom}}}

\begin{document}

\abstract{
Quantum neural networks hold significant promise for numerous applications, particularly as they can be executed on the current generation of quantum hardware. However, due to limited qubits or hardware noise, conducting large-scale experiments often requires significant resources. Moreover, the output of the model is susceptible to corruption by quantum hardware noise. To address this issue, we propose the use of ensemble techniques, which involve constructing a single machine learning model based on multiple instances of quantum neural networks. In particular, we implement bagging and AdaBoost techniques, with different data loading configurations, and evaluate their performance on both synthetic and real-world classification and regression tasks. To assess the potential performance improvement under different environments, we conduct experiments on both simulated, noiseless software and IBM superconducting-based QPUs, suggesting these techniques can mitigate the quantum hardware noise. Additionally, we quantify the amount of resources saved using these ensemble techniques. Our findings indicate that these methods enable the construction of large, powerful models even on relatively small quantum devices.
}

\keywords{Ensemble technique, Bagging, Boosting, Quantum Neural Network, Quantum Machine Learning}

\maketitle

\section{Introduction}

The emerging field of quantum machine learning \cite{biamonte2017quantum}  holds  promise for enhancing the accuracy and speed of machine learning  algorithms by utilizing quantum computing techniques. Although the potential of quantum machine learning is expected to be advantageous  for certain classes of problems in chemistry, physics, material science, and pharmacology \cite{cerezo2022challenges}, its applicability to more conventional use cases remains uncertain \cite{schuld2022quantum}. Notably, utilizable quantum machine learning algorithms generally need to be adapted to run on `NISQ' devices \cite{preskill2018quantum}, that are current noisy quantum computer, no error corrected and with modest number of qubits and circuit depth capabilities. In the quantum machine learning scenario, the quantum counterparts of classical neural networks, quantum neural networks \cite{abbas2021power}, have emerged as the de facto standard model for solving supervised and unsupervised learning tasks in the quantum domain.

While quantum neural networks have generated much interest, they presently have some issues. The first is {\em barren plateau} \cite{mcclean2018barren} characterised by the exponentially-fast decay of the loss gradient's variance with increasing system size. This problem  may be exacerbated by various factors, such as having overly-expressive quantum circuits \cite{holmes2022connecting}.  To address this issue, quantum neural networks need to be carefully designed \cite{larocca2022diagnosing} and to incorporate expressibility control techniques such as projection \cite{huang2021power} and bandwidth control \cite{canatar2022bandwidth}. The second problem, 
which is the one addressed in this work, concerns the amount of resources  required to run quantum neural networks (the limited number of total qubits -currently up to over a hundred-  and the low fidelity of operations on current quantum devices severely restrict the size of the quantum neural network in terms of input dimension and layers).

In order to address the latter issue, we propose employing of NISQ-appropriate implementation of ensemble learning \cite{zhang2012ensemble}, a widely used technique in classical machine learning for tuning the bias and variance of a specific machine learning mechanism via the construction of a stronger classifier using multiple weak components, such that the ensemble, as a whole, outperforms the best individual classifier. The effectiveness of ensemble systems has been extensively demonstrated empirically  and theoretically \cite{de2014essai}, although there does not currently exist any overarching theoretical framework capable of e.g. covering the requirements of ensemble components diversity to guarantee its out-performance. We here seek to provide and quantify a motivation for employing classical ensemble techniques in relation to NISQ-based quantum neural networks, which we address via the following three arguments.


The first argument concerns the potential for the superior performance  of an ensemble system composed of small quantum neural networks compared to a single larger quantum neural network. This notion is based on the rationale that while quantum neural networks are inherently powerful machine learning models, they exhibit intrinsic variance due to the nature of highly non-convex loss landscape, implying that different predictors will result from randomly-initialised stochastic gradient descent training, in common with classical neural networks. Modern deep learning practice often deliberately overparametrises the network in  order to render the loss more convex \cite{oymak2020toward}, with the asymptotic case of infinitely wide neural networks exhibiting a fully convex loss landscape, making it effectively a linear model \cite{jacot2018neural}. Although overparameterization in quantum neural networks has been studied theoretically  \cite{larocca2021theory, liu2022representation, incudini2023quantum} and has been shown to be beneficial to generalization performances within certain settings, the increase in resource requirements makes this approach almost completely impractical on NISQ devices. In the classical literature, however, it has been demonstrated that ensemble techniques can perform comparably to the largest (generally overparameterized) models with significantly fewer resources, especially in relation to overall model parameterization, c.f. for example \cite[Figure 2]{geiger2020scaling}. 

The second argument pertains to the resource savings achievable by ensemble systems, particularly in terms of the number of qubits, gates, and training samples required. For example, the boosting ensemble technique involves progressive dividing of the training dataset into multiple, partially overlapping subsets on the basis of their respective impact on the performance of the cumulative ensemble classifier created by summing of the partial weak classifiers trained on previously-selected data subsets. This enables the ensemble quantum neural network to be constructed in parallel with individual quantum neural networks operating on datasets of reduced size. The random subspace technique, by contrast, trains each base predictor on a random subset of features, but also provides an advantage in terms of the overall number of qubits and gates required. Employing the random subspace technique in a quantum machine learning setting would parallel the various quantum circuit splitting techniques (c.f. for example \cite{lowe2022fast}), and divide-and-conquer approaches, that have been utilized in the field of quantum chemistry \cite{yoshikawa2022quantum} and quantum optimization \cite{asproni2020accuracy}.
 

Our third argument, which is specific to quantum computing, examines the potential of ensembles' noise-canceling ability. Previous works have demonstrated that ensembles can enhance the performance of several noisy machine-learning tasks (see \cite{zhang2011robust}). Our investigation aims to determine whether and to what extent these techniques can reduce the impact of noise during the execution on a NISQ device \emph{at the applicative level}. This approach differs from most current approaches, which aim to reduce noise at a lower level, as described in \cite{larose2022mitiq}.


We here examine the impact of ensemble techniques based on bagging (bootstrap aggregation) and boosting ensembles in a quantum neural network setting across  seven variant data loading schemes. Bagging techniques are selected for their applicability in high-variance settings, i.e. those exhibiting significant fluctuations in relation to  differ initialisations and differ sample subselections; contrarily, boosting techniques are effective in relation to high-bias models, i.e. those which are relatively insensitive to data subsampling.

Our first objective is to quantify the amount of resources (in particular, the number of qubits, gates, parameters, and training samples) saved by the respective approaches. Secondly, we evaluate the performance using quantum neural networks as base predictors to solve a number of representative synthetic and real-world regression and classification tasks. Critically, the accuracy and loss performance of these approaches are assessed with respect to the number of layers of the quantum neural networks in a simulated environment. We thus obtain a layer-wise quantification of performance that addresses one of the fundamental questions in architecting deep neural systems, namely, how many layers of abstraction to incorporate? Note that this question is fundamentally different in a quantum setting compared to classical neural systems; in the latter, the possibility of multi-level feature learning exists, and thus the potential for indefinite performance improvement with neural layer depth \cite{incudini2023quantum}. This contrasts with the quantum neural networks, in which an increase in the number of layers affects the expressibility of the ansatz and thus might introduce a barren plateau \cite{holmes2022connecting}.

Finally, the noise-canceling capabilities of ensembles will be investigated by testing a synthetic linear regression task both on a simulated noisy environment mimicking IBM's real quantum device Lima and on IBM's superconductor-based quantum processing unit (QPU) Lagos.

\paragraph{Contributions}
Our contributions are the following:

\begin{itemize}
\item We evaluate various ensemble schemes that incorporate bagging and boosting techniques into quantum neural networks, and quantify the benefits in terms of resource savings, including the number of qubits, gates, and training samples required for these approaches.
\item We apply our approach both to a simulated noisy environment and to the IBM Lagos superconductor-based quantum processing unit to investigate the potential advantages of bagging techniques in mitigating the effects of noise during the execution of quantum circuits on NISQ devices.
\item We conduct a layer-wise analysis of quantum neural network performance in the ensemble setting with a view to determining the implicit trade-off between ensemble advantage and layer-wise depth.
\end{itemize}

\section{Related Works}

The quest for quantum algorithms able to be executed on noisy small-scale quantum systems led to the concept of Variational Quantum Circuits (VQCs), i.e. quantum circuits based on a hybrid quantum-classical optimization framework \cite{cerezo2021variational,mitarai_2018}. VQCs are currently believed to be promising candidates to harness the potential of QC and achieve a quantum advantage \cite{tilly2022variational,di2022quask,liu2021rigorous}. VQCs rely on a hybrid quantum-classical scheme, where a parameterized quantum circuit is iteratively optimized with the help of a classical co-processor. This way, low-depth quantum circuits can be efficiently designed and implemented on the available NISQ devices; the noisy components of the quantum process are mitigated by the low number of quantum gates present in the VQCs. The basic structure of a VQC include a data encoding stage, where classical data are embedded into a complex Hilbert space as quantum states, a processing of such quantum states via an ansatz made of parameterized rotation gates and entangling gates, and finally a measurement of the circuit to retrieve the expected outcome. Many different circuit architectures and ansatzes have been proposed for VQCs \cite{benedetti2021hardware,choquette2021quantum,farhi2014quantum,patil2022variational}, depending on the structure of the problem or on the underlying quantum hardware. VQCs demonstrated remarkable performances and a good resilience to noise in several optimization tasks and real-world applications. For example, researchers in \cite{schuld2020circuit} introduced a circuit-centric quantum classifier based on VQC that could effectively be implemented on a near-term quantum device. It correctly classified quantum encoded data and demonstrated to be robust against noise. Authors in \cite{mitarai_2018} proposed a VQC that successfully approximated high-dimensional regression and classification functions with a limited number of qubits.

VQCs are incredibly well-suited for the realization of quantum neural networks with a constraint on the number of qubits \cite{massoliALeap2022}. A quantum neural network is usually composed of a layered architecture able to encode input data into quantum states and perform heavy manipulations in a high-dimensional feature space. The encoding strategy and the choice of the circuit ansatz are critical for the achievement of superior performances over classical NNs: more complex data encoding with hard-to-simulate feature maps could lead to a concrete quantum advantage \cite{havlivcek2019supervised}, but too expressive quantum circuits may exhibit flatter cost landscapes and result in untrainable models \cite{holmes2022connecting}. An example of quantum neural network was given in \cite{macaluso2020variational}, where a shallow NN was employed to perform classification and regression tasks using both simulators and real quantum devices. In \cite{zhao2021qdnn}, authors proposed a multi-layer Quantum Deep Neural Network (QDNN) with three variational layers for an image classification task. They managed to prove that QDNNs have more representation capacity with respect to classical deep NN. A hybrid Quantum-classical Recurrent Neural Network (QRNN) was presented in \cite{ceschiniHybrid2022} to solve a time series prediction problem. The QRNN, composed of a quantum layer as well as two classical recurrent layers, demonstrated superior performances over the classical counterpart in terms of prediction error.

However, quantum neural networks suffer from some non-negligible problems, which deeply affect their performances and limit their impact in the quantum ecosystem. Firstly, they are still subject to quantum noise, and it gets worse as the number of layers (i.e., the depth of the quantum circuit) increases \cite{wang2022quantumnat,liang2021can}. Secondly, barren plateaus phenomena may occur depending on the ansatz and the number of qubits chosen, reducing the trainability of such models \cite{holmes2022connecting,cerezo2021cost,mcclean2018barren}. Finally, data encoding on NISQ devices continues to represent an obstacle when the number of features is considerable \cite{massoliALeap2022}, making them hard to implement and train \cite{ceschiniHybrid2022}.

In classical ML, ensemble learning has been investigated for years to improve generalization and robustness over a single estimator \cite{seni2010ensemble,zhang2012ensemble}. Ensembling is based on the so-called ``wisdom of the crowd'' principle, namely it combines the predictions of several base estimators with the same learning algorithm to build a single stronger model. Despite there are many different ensemble methods, the latter can be easily grouped into two different categories: bagging methods, which build and train several estimators independently and then compute an average of their predictions \cite{altman2017ensemble}, and boosting methods, which in turn train the estimators sequentially so that the each one corrects the predictions of the prior models and output a weighted average of such predictions \cite{buhlmann2012bagging}. Ensemble methods for NNs have also been extensively studied, yielding remarkable performances in both classification and regression tasks \cite{osman2020effective,sagi2018ensemble,berkhahn2019ensemble,dietterich2000ensemble,zhou2012ensemble,mas2014application,alzubi2019boosted,kumar2016ensemble,firdaus2018intent}. Authors in \cite{kim2021universal} have shown that overparameterization renders an optimization problem easier to train. 

In the quantum setting, the adoption of an ensemble strategy has received little consideration in the past few years, with very few approaches focusing on near-term quantum devices and VQC ensembles. In \cite{schuld2018quantum, abbas2020quantum}, the authors exploit the superposition principle to obtain an exponentially large ensemble wherein each instance is weighted according to its accuracy on the training dataset. However, they make use of a fault-tolerant approach rather than considering limited quantum resources. A similar approach is explored in \cite{leal2021training}, where authors create an ensemble of Quantum Binary Neural Networks (QBNNs) with reduced computational training cost without taking into consideration the amount of quantum resources necessary to build the circuit. An efficient strategy for bagging with quantum circuits is proposed in \cite{macaluso2020quantum} instead. Very recently, \cite{stein2022eqc} has proposed a distributed framework for ensemble learning on a variety of NISQ quantum devices, although it requires many NISQ devices to be actually implemented. A quantum Error-correcting output codes multiclass ensemble approach was proposed in \cite{Windridge2018QuantumEO}. In \cite{qin2022improving}, the authors investigated the performance enhancement of a majority-voting-based ensemble system in the quantum regime. Authors in \cite{krisnanda2023wisdom} studied the role of ensemble techniques in the context of quantum reservoir computing. Finally, an analysis of robustness to hardware error as applied to quantum reinforcement learning, and presenting compatible results, is given in \cite{skolik2023robustness}.

In this paper, we propose a classical ensemble learning approach to the outputs of several quantum neural networks in order to reduce the quantum resources for a given quantum model and provide superior performances in terms of error rate over single quantum neural network instances. To the best of our knowledge, no one has ever proposed such an ensemble framework for VQCs. We also compare both bagging and boosting strategy to provide an analysis on the most appropriate ensemble methods for quantum neural networks in a noiseless setting. An error analysis with respect to the number of layers of the quantum neural networks reveals that bagging models greatly outperform the baseline model with low number of layers, with remarkable performances as the number of layers increase; in fact, sufficiently complex bagging models allow to select better points on the bias-variance trade-off curve, such that one can maximise generalisability in a way not always possible with a single learner (especially ones with intrinsic constraints such as the QNN) \cite{rayana2016sequential,merentitis2014ensemble}. Finally, we apply our approach both to a simulated IBM Quantum Lima noisy backend and to the IBM Lagos superconductor-based QPU to investigate the potential advantages of bagging techniques in mitigating the effects of noise during the execution of quantum circuits on NISQ devices.

\section{Background and Notation}

We provide a brief introduction to the notation and concepts used in this work. The sets $\mathcal{X}$ and $\mathcal{Y}$ represent the set of features and targets, respectively. Typically, $\mathcal{X}$ is equal to $\Real^d$, with $d$ equal to the dimensionality in input, whereas $\mathcal{Y}$ is equal to $\Real$ for regression tasks and $\mathcal{Y}$ is equal to $\{ c_1, ..., c_k \}$ for $k$-ary classification tasks. Sequences of elements are indexed in the apex with $x^{(j)}$, where the $i$-th component is denoted as $x_i$. The notation $\epsilon \sim \mathcal{N}(\mu, \sigma^2)$ indicates that the value of $\epsilon$ is randomly sampled from a univariate normal distribution with mean $\mu$ and variance $\sigma^2$. We use the function $\llbracket P \rrbracket$ to denote one when the predicate $P$ is true and zero otherwise. 

\subsection{Models in quantum machine learning}

We define the state of a quantum system as the density matrix $\rho$ having unitary trace and belonging to the Hilbert space $\Hilbert \equiv \mathbb{C}^{2^n \times 2^n}$ where $n$ is the number of qubits. The system starts in the state $\rho_0 = \ketbra{0}{0}$. The evolution in a closed quantum system is described by a unitary transformation $U = \exp(-it H)$, $t \in \Real$, $H$ Hermitian operator, and acts like $\rho \mapsto U^\dagger \rho U$. The measurement of the system in its computational basis $\{ \Pi_i = \ketbra{i}{i} \}_{i=0}^{2^n-1}$ applied to the system in the state $\rho$ will give outcome $i \in 0, 1, ..., 2^n-1$ with probability $\Trace[\Pi_i \rho \Pi_i]$ after which the state collapses to $\rho' = \Pi_i \rho \Pi_i / \Trace[\Pi_i \rho \Pi_i]$. 
The expectation value of a Hermitian operator associated to a physical observable, $O = \sum_i \lambda_i \Pi_i$, acting on the system on the state $\rho$, is given by the Born rule $\expval{O} = \Trace[\rho O]$.

Quantum computation can be described using a quantum circuit, a sequence of gates (i.e. elementary operations) acting on one or more qubits of the system terminating with the measurement operation over some or all of its qubits. The output of the measurement can be post-processed using a classical function. ``The set of gates available shall be \emph{universal}'', i.e. the composition of such elementary operation allows the expression of any unitary transformation with arbitrary precision. An exemplar universal gate set is composed of parametric operators 
$R_x^{(i)}(\theta) = \mathrm{exp}(-i\frac{\theta}{2} \sigma_x^{(i)})$, 
$R_y^{(i)}(\theta) = \mathrm{exp}(-i\frac{\theta}{2} \sigma_y^{(i)})$, 
$R_z^{(i)}(\theta) = \mathrm{exp}(-i\frac{\theta}{2} \sigma_z^{(i)})$, and the operator 
$\mathrm{CNOT}^{(i,j)} = \mathrm{exp}(-i\frac{\pi}{4} \sigma_x^{(i)}\sigma_x^{(j)})$.
The gate $I$ is the identity.
The matrices $\sigma_x = \smqty(0 & 1 \\ 1 & 0), \sigma_y = \smqty(0 & -i \\ i & 0), \sigma_z = \smqty(1 & 0 \\ 0 & -1)$ are the Pauli matrices. The apex denotes explicitly the qubits in which the transformation acts. 

Quantum machine learning forms a broad family of algorithms, some of which require fault-tolerant quantum computation while others are ready to execute on current generation `NISQ' (noisy) quantum devices. The family of NISQ-ready techniques of interest in this document is denoted \emph{variational quantum algorithms} \cite{cerezo2021variational}. 
These  algorithms are based on the tuning of a cost function $C(\theta)$ dependent  on a set of parameters $\theta \in [0, 2\pi]^P$ and optimized classically (possibly via gradient descent-based techniques) to obtain the value $\theta^* = \arg\min_\theta C(\theta)$. Optimization through gradient-descent  thus involves computation of the gradient of $C$. This can be done using finite difference methods or else the parameter-shift rule \cite{schuld2019evaluating}. The parameter-shift rule is particularly well-suited for NISQ devices as it can utilise a large step size relative to finite difference methods, making it less sensitive to noise in calculations.

In general, $C(\theta)$ is a function corresponding to  a parametric quantum transformation $U(\theta)$ of a length polynomial in the number of qubits, the set of input states $\{ \rho_i \}$, and the set of observables $\{ O_k \}$. 
Specifically, a \emph{quantum neural network} is a function in the form
\begin{equation}
    f(x; \theta) = \Trace[U^\dagger(\theta)V^\dagger(x) \rho_0 V(x) U(\theta) O]
\end{equation}
where $\rho_0$ is the initial state of the system, $V(x)$ is a parametric quantum circuit depending on the input parameters $x \in \mathcal{X}$, $U(\theta)$ is a parametric quantum circuit named an \emph{ansatz} that depends on the trainable parameters $\theta \in [0, 2\pi)^P$, and $O$ is an observable. 
Given the training dataset $\{ (x^{(i)}, y^{(i)}) \}_{i=1}^M \in (\mathcal{X} \times \mathcal{Y})^M$, the cost function of a quantum neural network, being a supervised learning problem, is the empirical risk
\begin{equation}
    C(\theta) = \sum_{i=1}^M \ell(f(x^{(i)}; \theta), y^{(i)})
\end{equation}
where $\ell: \mathcal{Y} \times \mathcal{Y} \to \Real$ is any convex loss function, e.g. the Mean Squared Error (MSE). 

The quantum neural network constitutes a linear model in the Hilbert space of the quantum system as a consequence of the linearity of quantum dynamics. It behaves, in particular,  as a {\em kernel machine} that employs the unitary $V(x)$ as the feature map $\rho \mapsto \rho_x = V(x)\rho V^\dagger(x)$, while the variational ansatz $\rho \mapsto \rho_\theta = U(\theta)\rho U^\dagger(\theta)$ adjusts the model weights. Note that although the model is linear in the Hilbert space of the quantum system, the measurement projection makes it nonlinear in the parameter space, enabling a set of rich dynamics; nevertheless, this is not the only way of introducing nonlinearity into the quantum model \cite{inajetovic2023enabling}. Quantum neural networks can have a layer-wise structure, i.e., $U(\theta) = \prod_{i=1}^\ell U_i(\theta_i)$, which provides it with further degrees of freedom for optimization (however,  due to the lack of nonlinearity between the layers, the model does not possess the hierarchical feature learning capabilities of classical neural networks).

The selection of the ansatz is thus a crucial aspect in defining the quantum neural network, and it is required to adhere to certain classifier-friendly principles. Expressibility is one such, being the model's ability to approximate any quantum state in the Hilbert space. Although there are various ways to formalize expressibility, one of the most widely used definitions is based on the generation of the ensemble of states
$\{ \rho_\theta = U(\theta)\rho_0 U^\dagger(\theta) \mid \theta \in \Theta \}$ 
and the standard ensemble of random states induced by the Haar measure over the corresponding unitary group.
Expressible unitaries are those that make small the deviation between the former and the latter ensembles.
However, expressible circuits are susceptible to the barren plateau problem, where the variance of the gradient decreases exponentially with the number of qubits, making parameter training infeasible. The varieties of ansatz and their expressibilities are presented in \cite{sim2019expressibility}. Expressibility is tightly connected to the concept of controllability in quantum optimal control, and authors in \cite{larocca2022diagnosing} show that the asymptotic limit of the number of layers $\ell \to \infty$ in the expressible circuits are the controllable ones, i.e. those whose ansatz is underlied by a Lie algebra matching the space of skew-Hermitian matrices $\mathfrak{u}(2^n)$.

\subsection{Ensemble techniques}

The purpose of using ensemble systems is to improve the generalization performance through reducing the bias or variance of a decision system. Such a result is obtained by training several models and combining the outcomes according to a combination rule. A large body of literature on ensemble techniques exists; the reader is referred to \cite{zhang2012ensemble} for a general overview.

\begin{figure}[tb]
    \centering
    \scalebox{0.8}{
    \begin{tabular}{cccc}
    \hline
        \toprule
        \multicolumn{4}{c}{\bf Data selection strategy} \\ \midrule
        ~ & ~ & \multicolumn{2}{c}{\it Subset of features} \\
        ~ & ~ & No & Yes \\\cline{3-4}
        {\it Subset of} & No & / & Random subspace \\
        {\it samples} & Yes & Bootstrapping, pasting & Random patch \\ \bottomrule \\ \toprule
        \multicolumn{4}{c}{\bf Composition + training of single model instances} \\ \midrule
        ~ & ~ & \multicolumn{2}{c}{\it Model instances} \\
        ~ & ~ & Heterogeneous & Homogeneous \\\cline{3-4}
        {\it Type of} & Sequential & / & Boosting \\
        {\it processing} & Parallel & Stacking & Bagging \\ \bottomrule  \\ \toprule
        \multicolumn{4}{c}{\bf Combination rule of the outputs} \\ \midrule
        ~ & ~ & {\it Discrete} & {\it Continuous} \\\cline{3-4}
        ~ & ~ & Majority voting & Average \\
        ~ & ~ & Weighted majority voting & Weighted average \\
        ~ & ~ & Borda counts & Min Max \\ \bottomrule
    \end{tabular}}
    \caption{Taxonomy of the three aspects characterizing an ensemble system.}
    \label{fig:ensemble_taxonomy}
\end{figure}

The idea behind the ensemble system may be motivated by Condorcet's jury theorem \cite{de2014essai}: a jury of $m$ peers, each having probability $p = \frac{1}{2} + \epsilon, 0 < \epsilon \ll 1,$ of giving the correct answer, implies that the probability of the verdict given by majority voting to be correct is
\begin{equation}
    p_\text{jury} = \sum_{k = \lceil m/2 \rceil + 1}^m \binom{m}{k} p^k (1-p)^{m-k}
\end{equation}
and quickly approaches $1$ as $m\to \infty$. The theorem, broadly interpreted, suggests that a combination of small, individually ineffective machine learning models $h_1, ..., h_m$ (\emph{weak learners}) can be combined to constitute a more powerful one, with arbitrarily good performance depending on the nature of data manifold and the base classifiers $h_\text{ens}$ (\emph{strong learner}). 
According to \cite{zhang2012ensemble}, three aspects characterize an ensemble system: a data selection strategy, the composition plus training strategies of the single model instances, and the combination rule of its output. Some of the possible choices are summarized in Figure \ref{fig:ensemble_taxonomy}. 

The data selection strategy determines how the data should be distributed to the individual instances. If all instances are trained on the same dataset, their predictions will be highly correlated, resulting in similar output. The \emph{bootstrapping} technique creates smaller, overlapping subsets by sampling with replacement from the dataset, which are then assigned to different instances. Alternatively, the \emph{pasting} technique can be used for processing larger datasets by subsampling without replacement. Another approach is to divide the dataset by randomly assigning different sets of features with replacement, known as the random subspace technique (when the bootstrapping and random subspace techniques are combined, the result is the {\em random patch} technique).

\begin{figure}
    \centering
    \scalebox{0.60}{\begin{tikzpicture}
    \draw[black, fill=black!20!white] (0.0, 1.00) circle (3pt);
    \draw[black, fill=red!20!white] (0.0, 2.00) circle (3pt);
    \draw[black, fill=green!20!white] (0.0, 3.00) circle (3pt);
    \draw[black, fill=blue!20!white] (0.5, 0.50) circle (3pt);
    \draw[black, fill=yellow!20!white] (0.5, 1.50) circle (3pt);
    \draw[black, fill=black!20!white] (0.5, 2.50) circle (3pt);
    \draw[black, fill=purple!20!white] (0.5, 3.50) circle (3pt);
    \draw[black, fill=orange!20!white] (1.0, 0.99) circle (3pt);
    \draw[black, fill=green!20!white] (1.0, 1.99) circle (3pt);
    \draw[black, fill=blue!20!white] (1.0, 2.99) circle (3pt);
    \draw[black] (-0.50, 0.25) rectangle (1.50, 1.75);
    \draw[black] (-0.60, 1.25) rectangle (1.60, 2.75);
    \draw[black] (-0.50, 2.25) rectangle (1.50, 3.75);
    \draw[black] (3, 0) rectangle node {M3} (4, 1);
    \draw[black] (3, 1.5) rectangle node {M2} (4, 2.5);
    \draw[black] (3, 3) rectangle node {M1} (4, 4);
    \draw[->] (1.50, 1.00) -- (3, 0.5);
    \draw[->] (1.60, 2.00) -- (3, 2.0);
    \draw[->] (1.50, 3.00) -- (3, 3.5);
    \node[draw, circle] (A) at (6, 2) {$\frac{1}{N} \sum$};
    \draw[->] (4, 0.5) -- (A);
    \draw[->] (4, 2.0) -- (A);
    \draw[->] (4, 3.5) -- (A);
    \draw[->] (A) -- ($(A)+(1,0)$);
    \draw[->] (A) -- ($(A)+(1,0)$) node[yshift=6pt] {y};
    \end{tikzpicture}
    }~\scalebox{0.60}{\hspace{0.4cm}\begin{tikzpicture}
    \draw[black, fill=black!20!white] (0.0, 1.00) circle (3pt);
    \draw[black, fill=red!20!white] (0.0, 2.00) circle (3pt);
    \draw[black, fill=green!20!white] (0.0, 3.00) circle (3pt);
    \draw[black, fill=blue!20!white] (0.5, 0.50) circle (3pt);
    \draw[black, fill=yellow!20!white] (0.5, 1.50) circle (3pt);
    \draw[black, fill=black!20!white] (0.5, 2.50) circle (3pt);
    \draw[black, fill=purple!20!white] (0.5, 3.50) circle (3pt);
    \draw[black, fill=orange!20!white] (1.0, 0.99) circle (3pt);
    \draw[black, fill=green!20!white] (1.0, 1.99) circle (3pt);
    \draw[black, fill=blue!20!white] (1.0, 2.99) circle (3pt);
    \draw[black] (-0.50, 0.25) rectangle (1.50, 3.75);
    \draw[black] (3, 2.5) rectangle node {M1} (4, 3.5);
    \draw[black] (6, 2.5) rectangle node {M2} (7, 3.5);
    \draw[black] (9, 2.5) rectangle node {M3} (10, 3.5);
    \draw[->] (1.5, 2.00) -- (3, 3) node[midway,sloped,yshift=4pt] {\footnotesize uniform};
    \draw[->] (4, 3) -- (6, 3) node[midway,sloped,yshift=10pt] {\footnotesize \begin{tabular}{c} misclassified \\ by M1 \end{tabular}};
    \draw[->] (7, 3) -- (9, 3) node[midway,sloped,yshift=10pt] {\footnotesize \begin{tabular}{c} misclassified \\ by M1, M2 \end{tabular}};
    \node[draw, circle] (A) at (5, 1) {$\frac{1}{N} \sum$};
    \draw[->] (3.5, 2.5) -- (A);
    \draw[->] (6.5, 2.5) -- (A);
    \draw[->] (9.5, 2.5) -- (A);
    \draw[->] (A) -- ($(A)+(1, 0)$) node[xshift=3pt] {y};
    \end{tikzpicture}}
    \caption{Comparison between bagging (left) and `vanilla' boosting (right) techniques. The bagging ensemble trains the models in parallel over a subset of the dataset drawn uniformly; each prediction is then merged via an average function. The boosting ensemble trains the models sequentially, the first predictor draws the samples uniformly, and the subsequent models draw the elements from a probability distribution biased toward previously misclassified items.}
    \label{fig:bagging_boosting_stacking}
\end{figure}
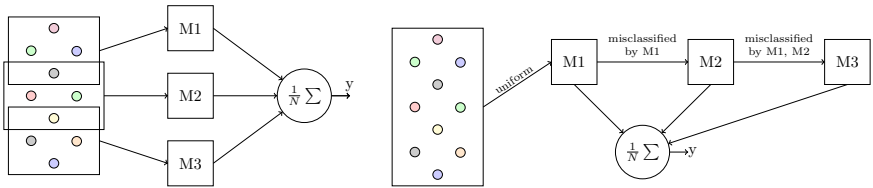

There are numerous schemes for combining predictors, with \emph{bagging} being the most straightforward and commonly used. Bagging, short for bootstrap aggregation, involves the creation of multiple homogeneous model instances trained on bootstrapped datasets. An instance of a bagging scheme is the random forest, which involves bagging decision trees trained on differing sample subsets (in some cases, random forests may favor a random patch data selection strategy over bagging). Another predictor combination scheme is \emph{boosting}, which involves training a sequence of predictors via subsampling data according to the following strategy: an initial predictor is trained on a uniformly drawn subset of samples, while the $i$-th instance of the predictor is trained on a subset of elements that the previous ensemble classifier incorrectly predicted. The ensemble is itself the convex cumulative sum over predictors. Numerous variations of boosting exist, one of the most notable being AdaBoost \cite{freund1997decision}. Contrary to vanilla boosting, AdaBoost employs an exponential loss such that the ensemble error function allows for the fact that it is only the sign of outcome that is significant. These two scheme are illustrated in Figure \ref{fig:bagging_boosting_stacking}. The other major ensemble scheme is {\em stacking} in which a collection of heterogeneous classifiers trained on the same dataset are combined via an optimised meta-classifier. 

The combination rule merges the output of individual models $h_1, ..., h_m$. In classification tasks i.e. where the label output is discrete $y \in C = \{c_1, ..., c_k\}$, the most commonly used rule is majority voting. This is calculated as $y_\text{ens} = \arg \max_{c \in C} \sum_{i=1}^m \llbracket h_i(x) = c \rrbracket$. Where there exists prior knowledge regarding the performance of individual predictors, positive weights $w_i$ can be assigned, such that the output is a weighted majority vote. The ensemble prediction in this case will be $y_\text{ens} = \arg \max_{c \in C} \sum_{i=1}^m w_i \llbracket h_i(x) = c \rrbracket$. Alternatively, the \emph{borda count} method sorts labels in descending order by likelihood, with the ensemble prediction being the highest ranking sum. Nevertheless, averaging functions can also be utilised for ensemble classifiers for small ensemble sizes \cite{hastie2009elements}. For regression tasks where $y \in \Real$, common combination rules are (possibly weighted) mean, minimum, and maximum.

\section{Discussion}

Ensemble techniques, while well-established in the classical realm, have been largely overlooked in the quantum literature, leaving a number of open questions in this setting, such as whether bagging techniques, which reduce variance, can be deployed as effectively as boosting techniques, which reduce bias (both of which are also data-manifold and base-model dependent). It is also unclear as to the relative resource saving in terms of circuit size (number of qubits) and depth (number of gates), and also samples required for training, that can be obtained by using an ensemble of quantum neural networks instead of a single, large quantum network. Furthermore, it is not currently well understood the extent to which an ensemble system can mitigate hardware noise. Our experiments are designed to explore these questions.

To investigate the first two aspects, we conduct a suite of experiments within a simulation environment, employing seven distinct ensemble schemes with varying strategies for data selection, model training and decision combination applied to four synthetic and real-world datasets, encompassing both regression and classification tasks. Specifically, we analyze: a synthetic linear regression dataset, the Concrete Compressive Strength regression dataset, the Diabetes regression dataset, and the Wine classification dataset, which are widely used benchmarks for evaluating machine learning models. 

Six of the proposed techniques are classified as bagging methods, employing bootstrapped data to generate the ensemble, while the seventh is a sequential boosting technique, namely AdaBoost. In particular, we implemented the AdaBoost.R2 version \cite{drucker1997improving} for the regression tasks and the AdaBoost SAMME.R version \cite{hastie2009multi} for the classification problem. The bagging ensembles are characterized by two parameters: the sample ratio $r_n \in [0,1]$, which determines the percentage of training samples used for each base predictor (with replacement), and the feature ratio $r_f \in [0,1]$, which indicates the percentage of features used for each predictor (without replacement). We test six bagging schemes by varying $(r_n, r_f) \in \{0.2, 1.0\} \times \{0.3, 0.5, 0.8\}$. For both the classification and regression tasks, the outputs of the base predictors are combined via averaging, as suggested in \cite{hastie2009elements}. In the case of the AdaBoost ensemble, the training set for each base predictor has the same size and dimensionality as the original training set. However, the samples are not uniformly drawn but are selected and weighted based on the probability of misclassification by previous classifiers composing the cumulative ensemble; single predictors are hence combined using a weighted average. Each ensemble system comprises 10 base predictors. The characteristics of these ensemble schemes are summarized in Table \ref{tab:simulated_ensemble}, where FM identifies the baseline quantum neural network model, whereas Bag\_$r_f$\_$r_n$ represents a bagging model with $r_f$ percentage of the features and $r_n$ percentage of the samples. Our experiments aim to evaluate the performance of each of the ensemble frameworks in comparison to the baseline model, as well as to assess the overall resource saving, including the number of qubits and overall parametric requirements.

\begin{table}[htbp]
    \centering
    \begin{tabular}{llllll}
        \toprule
        \multirow{2}{*}{Model} & \multicolumn{2}{c}{Data Loading} & \multirow{2}{*}{Ensemble} & \multirow{2}{*}{\#BP} & \multirow{2}{*}{Rule} \\\cline{2-3}
                        & RSBS ($r_f$) & BST ($r_n$) &         &    &       \\ \midrule
        FM              & -     & -     & -        & -  & -     \\
        Bag\_0.3\_0.2   & 0.3   & 0.2   & Bagging  & 10 & Avg   \\
        Bag\_0.3\_1.0   & 0.3   & 1.0   & Bagging  & 10 & Avg   \\
        Bag\_0.5\_0.2   & 0.5   & 0.2   & Bagging  & 10 & Avg   \\
        Bag\_0.5\_1.0   & 0.5   & 1.0   & Bagging  & 10 & Avg   \\
        Bag\_0.8\_0.2   & 0.8   & 0.2   & Bagging  & 10 & Avg   \\
        Bag\_0.8\_1.0   & 0.8   & 1.0   & Bagging  & 10 & Avg   \\
        AdaBoost        & 1.0   & 1.0   & AdaBoost & 10 & W.Avg \\
        \bottomrule
    \end{tabular}
    \caption{Characteristics of the baseline benchmark  model (0) and ensemble systems (I to VII). The ensemble system is identified by its broad data loading method (BST for Boosting and RSBS for Random Subspace), predictor composition \& training type (Ensemble), number of base predictors (\#BP), composition rule (Rule, with Avg representing the average function and W.Avg representing weighted average).}
    \label{tab:simulated_ensemble}
\end{table}

\begin{figure}[htbp]
    \centering
    \subfloat[]{\label{fig:ibm_lima}\includegraphics[width=.3\linewidth]{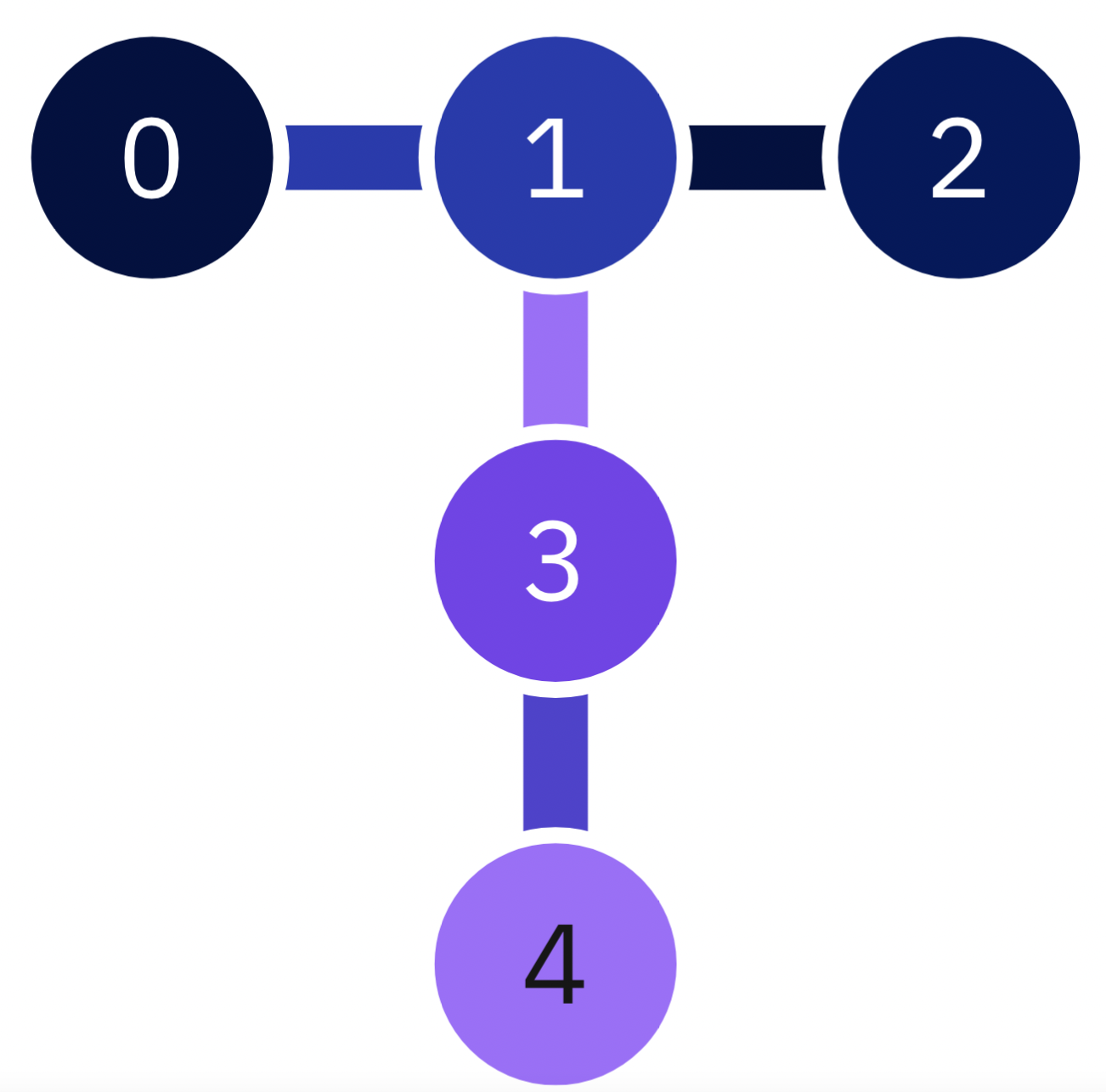}}
    \qquad \qquad
    \subfloat[]{\label{fig:ibm_lagos}\includegraphics[width=.3\linewidth]{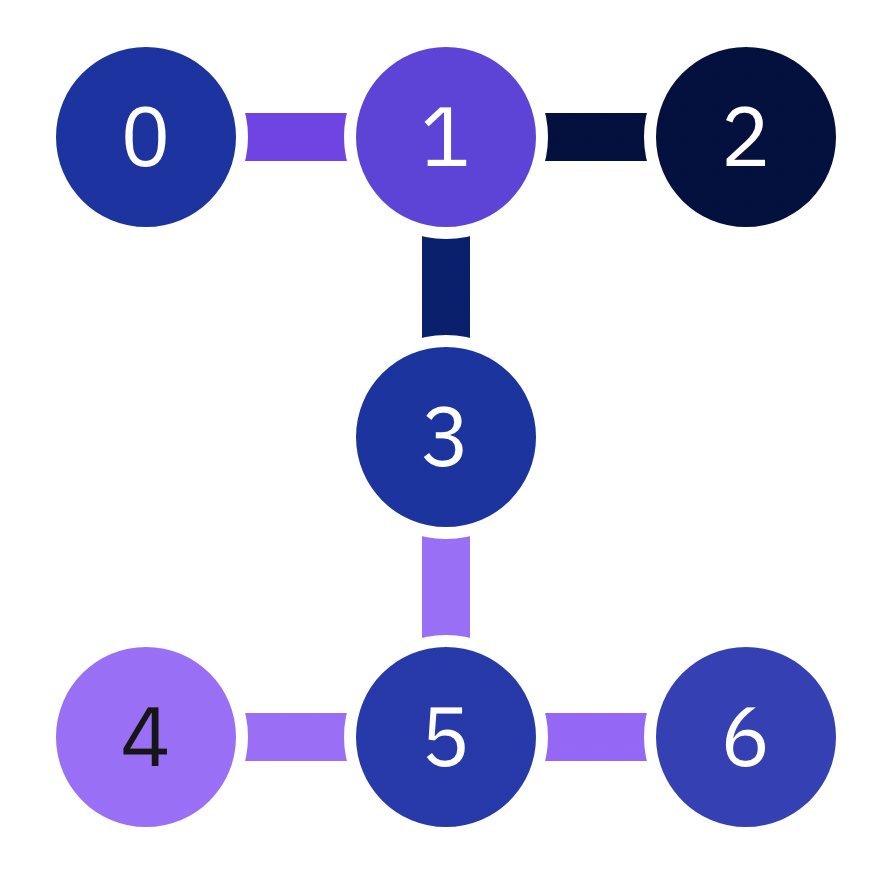}}
    \caption{Topology of (a) IBM Lima QPU, (b) IBM Lagos QPU}
    \label{fig:QPUs}
\end{figure}

To investigate the impact of quantum hardware noise, we conduct two additional experiments. The first one is performed in a noisy simulated setting with PennyLane-Qiskit plugin, which allows to simulate of a noisy device by selecting a fake backend from IBM Quantum's suite. It mimics the behaviors of IBM Quantum systems using system snapshots. The latter contains information about the quantum system such as coupling map, basis gates, qubit properties (T1, T2, error rate, etc.). These elements are useful for incorporating a realistic noise model into our experiment and thus performing noisy simulations of the system. We selected IBM Quantum Lima as the quantum backend to mimic. It is a 5-qubit superconducting-based quantum computer, and its topology is depicted in Figure \ref{fig:QPUs}a. In this experiment, the baseline FM is compared to the Bagging models, with the aim to assess their robustness in the presence of quantum hardware noise and qubit coupling. The second noisy experiment was performed directly on the IBM Lagos QPU real hardware. Such a device is a 7-qubit superconducting-based quantum computer. The topology of Lagos is depicted in Figure \ref{fig:QPUs}b. Specifically, here we compare the performance of the baseline model FM with that of the Bag\_0.8\_0.2 configuration on the linear regression dataset. Our goal with these two experiments is to determine whether ensemble techniques can effectively mitigate quantum noise, and whether the difference in performance between single predictors and ensemble systems is more pronounced within a simulated, noise-free environment in comparison with noisy executions on quantum hardware.

\subsection{Experimental setup}\label{sec:methods}

This section outlines experimental protocols used to evaluate the performance of the various ensemble approaches in terms of both the experimental structure and specific parameters/settings used to configure the algorithm and hardware. 

\paragraph{Choice of quantum neural networks}  We utilize a quantum neural network of the form $f(x; \theta) = \Trace[U^\dagger(\theta)V^\dagger(x) \rho_0 V(x) U(\theta) O]$, which operates on $n$ qubits, with $n$ corresponding to the number of features in the classification/regression problem. For the feature map, we opted for the simple parametric transformation $V(x) = \bigotimes_{i=1}^n R_y^{(i)}(x_i)$. This choice was motivated by the findings in \cite{kubler2021inductive}, suggesting that more complex feature maps can lead to unfavorable generalization properties, the incorporation of which may thus unnecessarily bias our findings. (In \cite{lloyd2020quantum}, various feature maps are compared). 

\begin{figure}[tbp]
    \centering
    \includegraphics[width=0.9\textwidth]{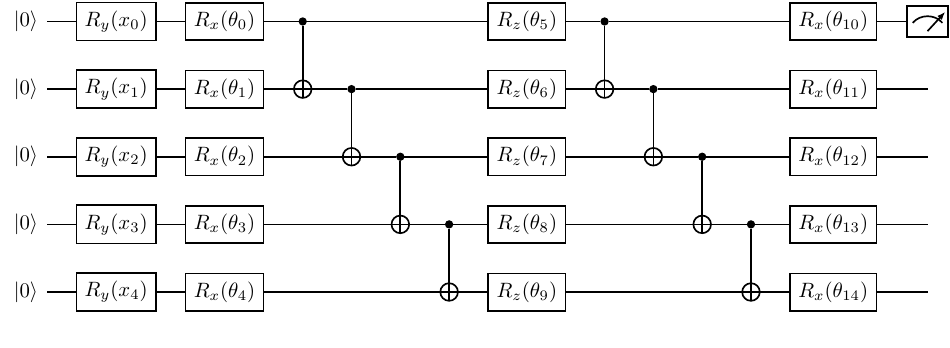}
    \caption{Quantum Neural Network used to classify the linear regression dataset, having $5$ qubits and $\ell=1$ layers. The rotational gates parameterized by the feature $x_i$ form the feature map, while those parameterized via the $\theta$s form the ansatz.}
    \label{fig:qnn}
\end{figure}

The ansatz is implemented with the parametric transformations structured layer-wise with, for  $\ell$ the number of layers, a total of  $3\ell n$ parameters. It is thus defined as:
\begin{align}
    \nonumber
    U_\ell(\theta) = & 
    \prod_{k=1}^\ell 
    \Bigg[
    \left(\bigotimes_{i=1}^n R_x^{(i)}(\theta_{3kn+2n+i})\right)
    \left(\prod_{i=1}^{n-1} \mathrm{CX}^{(i, i+1)}\right)
    \left(\bigotimes_{i=1}^n R_z^{(i)}(\theta_{3kn+n+i})\right) \\
    & \qquad\qquad\qquad\qquad\qquad
    \left(\prod_{i=1}^{n-1} \mathrm{CX}^{(i, i+1)}\right)
    \left(\bigotimes_{i=1}^n R_x^{(i)}(\theta_{3kn+i})\right)
    \Bigg]
\end{align}
The role of CNOT gates is the introduction of entanglement in the system, which would otherwise be efficiently classical simulable. 
We select as the observable $O = \sigma_z^{(0)}$, which operates on a single qubit. Local observables like this one are less susceptible to the barren plateau problem than global ones, for example, $O = \otimes_{i=1}^n \sigma_z^{(i)}$ (as noted in \cite{cerezo2021cost}). Note that the measurement on the first qubit depends nontrivially on the overall state due to the entanglement introduced by the chosen unitary. The quantum neural network described in our investigation is pictured in Figure \ref{fig:qnn}. 

\paragraph{Training of the model} To train models, we utilize a standard state-of-the-art  gradient descent-based algorithm, ADAM. The MSE was selected as the loss function and error metric to evaluate the performances of the models in the regression tasks, as it is a standard error metric in supervised learning. MSE was selected as the loss function to train the networks because it is more sensitive to larger errors. Categorical Cross Entropy (CCE) was used as the loss function for the classification task instead, while Accuracy score was employed as the error metric to assess the goodness of the classification. Given the output $f$ of the model, the computation of its gradient $\nabla f$, which is required to calculate the gradient of the loss function, is accomplished using the parameter-shift rule \cite{schuld2019evaluating}, since the commonly-used finite difference method $\nabla f(x; \theta) \approx (f(x;\theta)-f(x;\theta+\epsilon))/\epsilon$ is highly susceptible to hardware noise. The optimization hyper-parameters used are the learning rate, set to $0.1$, and the number of training epochs, which were selected through empirical investigation. Specifically, we carry out 150 training epochs to obtain the simulated noise-free results, while for noisy simulations and real QPU-based results, we perform just 100 and 10 epochs respectively, due to technological constraints on current hardware.

\paragraph{Datasets} We assess the performance of our approach using both synthetic and real-world datasets, across both regression and classification problems. The linear regression dataset is artificially generated with parametric control over the number of samples $n$, the dimensionality $d$, and the noise variance $\sigma$. It is procedurally generated by randomly sampling a weight vector $w$ uniformly over $[-1,1]^d$ such that the training set $\{(x^{(i)}, y^{(i)})\}_{i=1}^n$ is constructed with $x^{(i)}$ uniformly sampled from $[-1,1]^d$, $y^{(i)} = w\cdot x^{(i)}+\epsilon^{(i)}$, and $\epsilon^{(i)}$ sampled from a normal distribution with zero mean and variance $\sigma$. In our case we have $n = 250$ (jointly the training and testing datasets), $d = 5$ and $\sigma=0.1$. The other datasets involved in the experiments are the \emph{Concrete Compressive Strength} dataset, the \emph{Diabetes} dataset, and the \emph{Wine} dataset.
The first of these is a multivariate regression problem calculating the strength of the material based on its age and ingredients. The second is a multivariate regression problem correlating the biological and lifestyle characteristic of patients to their insulin levels. The third one is a multivariate, three-class classification problem investigating the geographic origin of wine samples from their chemical characteristics. All are freely available and open source. Table \ref{tab:datasets} summarizes the characteristics of these datasets; more details can be found in \cite{shaikhina2017handling,gupta2018selection,nelli2018machine}. Every dataset is divided into 80\% train samples and 20\% test samples. Moreover, in a data preprocessing phase, raw data were scaled in the range $[-1,1]$ to best suit the output of the quantum neural networks; the scaler was fitted using training data only. No other preprocessing technique, i.e. PCA, has been applied. 

\begin{table}[htbp]
    \centering
    \begin{tabular}{lllrrl}
        \toprule
        Dataset & Source & Nature & \# Features & \# Samples & Task \\\midrule
        Linear & - & Synthetic & 5 & 250 & Regression \\
        Concrete & UCI & Real-world & 8 & 1030 & Regression \\
        Diabetes & Scikit-Learn & Real-world & 10 & 442 & Regression \\
        Wine & UCI & Real-world & 13 & 178 & Classification \\\bottomrule
    \end{tabular}
    \caption{Characteristics of the datasets analyzed. UCI stands for the open source \emph{UCI Repository} available at \protect\url{archive.ics.uci.edu}. \emph{Scikit-Learn} is an open-source software library for Python3. The number of features does not include the target.}
    \label{tab:datasets}
\end{table}

\paragraph{Implementation details} Our implementation\footnote{The source code used to obtain our results can be freely accessed at \url{https://github.com/incud/Classical-ensemble-of-Quantum-Neural-Networks}.} is written in Python3, and utilizes Pennylane as a framework to define and simulate quantum circuits, with the Pennylane-Qiskit plugin used to execute circuits on IBM Quantum devices via the Qiskit software stack. To improve simulation times, we employed the JAX linear algebra framework as the simulation backend. By using JAX, the quantum circuit can be just-in-time compiled to an intermediate representation called XLA, which can significantly speed up simulation times (by up to a factor of 10). Our simulations were run on a commercial computer with an AMD Ryzen 7 5800X (8-core CPU with a frequency of 3.80 GHz) and 64 GB of RAM.   
The first experiment on the noise canceling properties of ensemble systems was simulated with Qiskit's \texttt{FakeLimaV2} backend, which mimics the behaviour of IBM Quantum's Lima QPU. It consists of 5 qubits arranged in the topology $\{(0,1);(1,2);(1,3);(3,4)\}$. The single-gate and CNOT fidelities of this QPU were $4.79e^{-4}$ and $1.07e^{-2}$, respectively. The second experiment on the noise canceling properties was conducted on the \texttt{ibm\_lagos} quantum processing unit, which consists of 7 qubits arranged in the topology $\{(0,1);(1,2);(1,3); (3,5); (4,5); (5,6)\}$. The single-gate fidelity and CNOT fidelity of this QPU did not exceed $2.89e^{-4}$ and $8.63e^{-3}$, respectively (according to the latest calibration available).

\subsection{Simulated Noiseless Experiments}

Initially, we evaluate our method in a simulated environment, one free of noise, such that  the output estimation is infinitely precise. This differs significantly from execution on a NISQ quantum processing unit, which introduces various types of hardware error (such as decoherence and infidelity of operations) as well as sampling error caused via the measurement operation. 
We examine the performance of both the baseline models and ensemble systems in a scenario where the number of layers (i.e. quantum neural network depth) is gradually increased.
To establish  robustness to random initialization of parameters (that is, susceptibility to local minima effects), each simulation is repeated ten times. 

\subsubsection{Experiment I}
The first experiment seeks to perform linear regression on a synthetic noisy 5-dimensional dataset. The function generating the targets is as follows: $y = w \cdot x + \epsilon$, where $x \in (-1,1)^5 \subseteq \Real{}^5$,  $w \in \Real{}^5$ is randomly generated from a uniform distribution having as support the range $-1$ to $1$, and $\epsilon$ is a Gaussian noise of mean zero and standard deviation $0.1$. The total number of samples composing this synthetic dataset is 250. Each experimental data point instantiates a layer number, a number of bagged features, and a percentage of training data points available to the ensemble.  

\begin{figure}[htbp]
    \centering
    \includegraphics[width=1\textwidth]{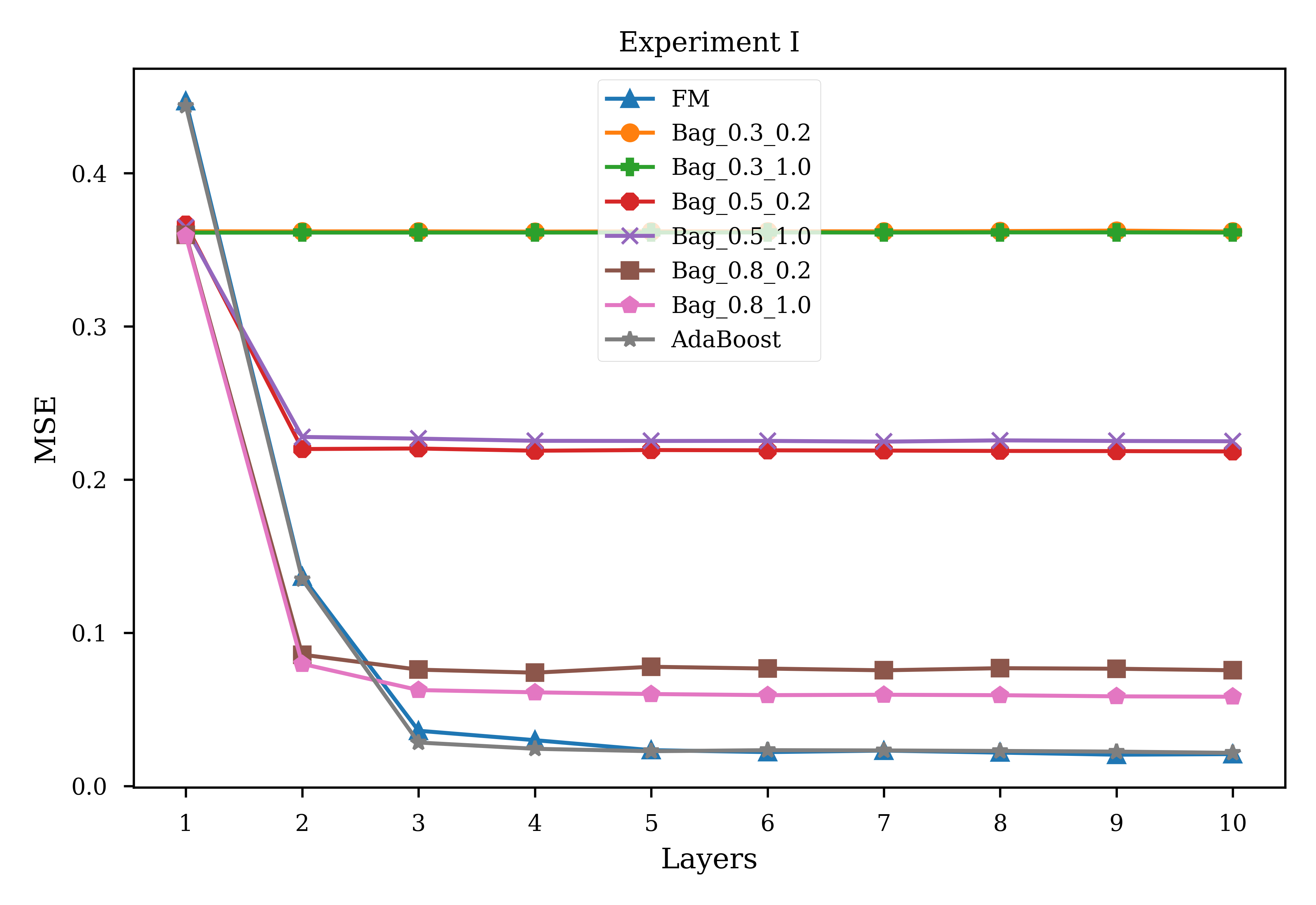}
    \caption{Evolution of MSE with respect to the number of quantum neural network layers in Experiment I. Each experimental data point instantiates a layer number, a number of bagged features and a percentage of training data points available to the ensemble.}
    \label{fig:linear_all}
\end{figure}

The results of the first experiment are indicated in Figure~\ref{fig:linear_all}. Both FM and AdaBoost achieve the lowest MSE on test of about 0.021 at 10 layers, reaching a performance plateau at 5 layers. The bagging models utilising 80\% of the features are able to reach satisfactory results with 10 layers, which are only 0.03 - 0.05 points higher than the error obtained by the best performing models. In general, it appears that quantum bagging models with a high number of features are able to generalize well on unseen data in this setting, even with only 20\% of the training samples (unsurprisingly, the performance of bagging models with only 20\% of training samples are worse than those of the counterparts using 100\% of the training samples). Nevertheless, they still achieve remarkable results and show impressive generalization capabilities, confirming the effectiveness of bagged quantum models in generalizing well with relatively little training data \cite{caro2022generalization}.

It is also notable that all of the bagging models have a lower MSE test error as compared to FM and AdaBoost when the number of layers is low. In particular, with just 1 layer, all of the bagging models outperform FM and AdaBoost. However, as the number of layers increases, the performances of bagging models begin to plateau more rapidly than FM and Adaboost which, in contrast, continue their trend of decreasing error with increasing circuit depth. This is due to bagging models reaching their performance saturation point more quickly as the number of layers increases, so that additional layers do not significantly improve their predictive power. The primary cause for this phenomenon is the impossibility of bagging models in utilizing all the available information from the features of the samples; an increment in the number of bagging estimators may alleviate such phenomenon. On the other hand, FM and AdaBoost have the advantage of fully leveraging all the features of the samples for the prediction. As a result, they gain benefits from higher circuit depth, enabling them to progressively diminish their error rates.

Finally, the decreasing error trend seen in the more complex bagging models as well as the FM and AdaBoost models is not visible in relation bagging with 30\% of the features. We conjecture that since this bagging configuration utilises only 1 qubit, it cannot appropriately model the evolution of the quantum state with respect to the input. Hence, despite leveraging 10 different submodels of 1 qubit (i.e., one feature) each, the performance of bagging models with 30\% of the features cannot improve as the number of layers increases (adding more layers in this case translates in performing rotations on the single qubit only, without the possibility of further CNOTs or other entangling gate operations). This result hence highlights the importance of entanglement in quantum neural network models as a means of improving performance.
 
In order to provide a comprehensive evaluation of the models' performances, we hereafter report the results obtained by varying not only the characteristics of the single base learners (i.e. the number of layers, the number of samples $r_n$, and the number of features $r_f$), but also the ensemble size (i.e., the number of base learners that are averaged). As illustrated in Table~\ref{tab:ensemble_size_I}, the ensemble works as expected, since the test error decreases while increasing the ensemble size.

\begin{table}[h!]
    \centering
    \renewcommand{\arraystretch}{1.7}
    \begin{tabular}{ccccccc}
        \toprule
        & & \multicolumn{5}{c}{Layers}\\ 
        \# Estimators & Model & 2 & 4 & 6 & 8 & 10\\\midrule
        \multirow{6}{*}{4} & Bag\_0.3\_0.2 & 0.390 & 0.390 & 0.390 & 0.390 & 0.390 \\
        & Bag\_0.3\_1.0 & 0.378 & 0.378 & 0.378 & 0.378 & 0.378 \\
        & Bag\_0.5\_0.2 & 0.243 & 0.241 & 0.240 & 0.240 & 0.240 \\
        & Bag\_0.5\_1.0 & 0.241 & 0.237 & 0.237 & 0.238 & 0.236 \\
        & Bag\_0.8\_0.2 & 0.090 & 0.078 & 0.079 & 0.082 & 0.080 \\
        & Bag\_0.8\_1.0 & 0.084 & 0.069 & 0.068 & 0.068 & 0.065 \\
        \hline
        \multirow{6}{*}{6} & Bag\_0.3\_0.2 & 0.375 & 0.375 & 0.375 & 0.375 & 0.376 \\
        & Bag\_0.3\_1.0 & 0.373 & 0.373 & 0.373 & 0.373 & 0.373 \\
        & Bag\_0.5\_0.2 & 0.237 & 0.235 & 0.236 & 0.236 & 0.235 \\
        & Bag\_0.5\_1.0 & 0.240 & 0.237 & 0.237 & 0.238 & 0.237 \\
        & Bag\_0.8\_0.2 & 0.102 & 0.090 & 0.093 & 0.094 & 0.092 \\
        & Bag\_0.8\_1.0 & 0.090 & 0.073 & 0.071 & 0.071 & 0.069 \\
        \hline
        \multirow{6}{*}{8} & Bag\_0.3\_0.2 & 0.365 & 0.365 & 0.365 & 0.365 & 0.365 \\
        & Bag\_0.3\_1.0 & 0.365 & 0.365 & 0.365 & 0.365 & 0.365 \\
        & Bag\_0.5\_0.2 & 0.230 & 0.228 & 0.229 & 0.229 & 0.228 \\
        & Bag\_0.5\_1.0 & 0.234 & 0.231 & 0.231 & 0.232 & 0.231 \\
        & Bag\_0.8\_0.2 & 0.091 & 0.080 & 0.082 & 0.083 & 0.082 \\
        & Bag\_0.8\_1.0 & 0.084 & 0.066 & 0.064 & 0.064 & 0.063 \\
        \hline
        \multirow{6}{*}{10} & Bag\_0.3\_0.2 & 0.362 & 0.362 & 0.362 & 0.362 & 0.362 \\
        & Bag\_0.3\_1.0 & 0.361 & 0.361 & 0.361 & 0.361 & 0.361 \\
        & Bag\_0.5\_0.2 & 0.220 & 0.219 & 0.219 & 0.219 & 0.218 \\
        & Bag\_0.5\_1.0 & 0.228 & 0.225 & 0.225 & 0.226 & 0.225 \\
        & Bag\_0.8\_0.2 & 0.086 & 0.074 & 0.077 & 0.077 & 0.075 \\
        & Bag\_0.8\_1.0 & 0.080 & 0.061 & 0.059 & 0.059 & 0.058 \\
        \bottomrule
    \end{tabular}
    \caption{MSE of bagging ensembles by varying the number of bagging estimators in Experiment I. Only even number of layers are reported for brevity.}
    \label{tab:ensemble_size_I}
\end{table}

\subsubsection{Experiment II}
The second experiment seeks to assess the performance of the respective ensemble techniques on the Concrete Compressive Strength dataset, which consists in 1030 samples of 8 features. The target value to predict in this regression case is hence the concrete compressive strength, measured in Megapascal (MPa), a highly nonlinear function of age and composition of the material.

\begin{figure}[htbp]
    \centering
    \includegraphics[width=1\textwidth]{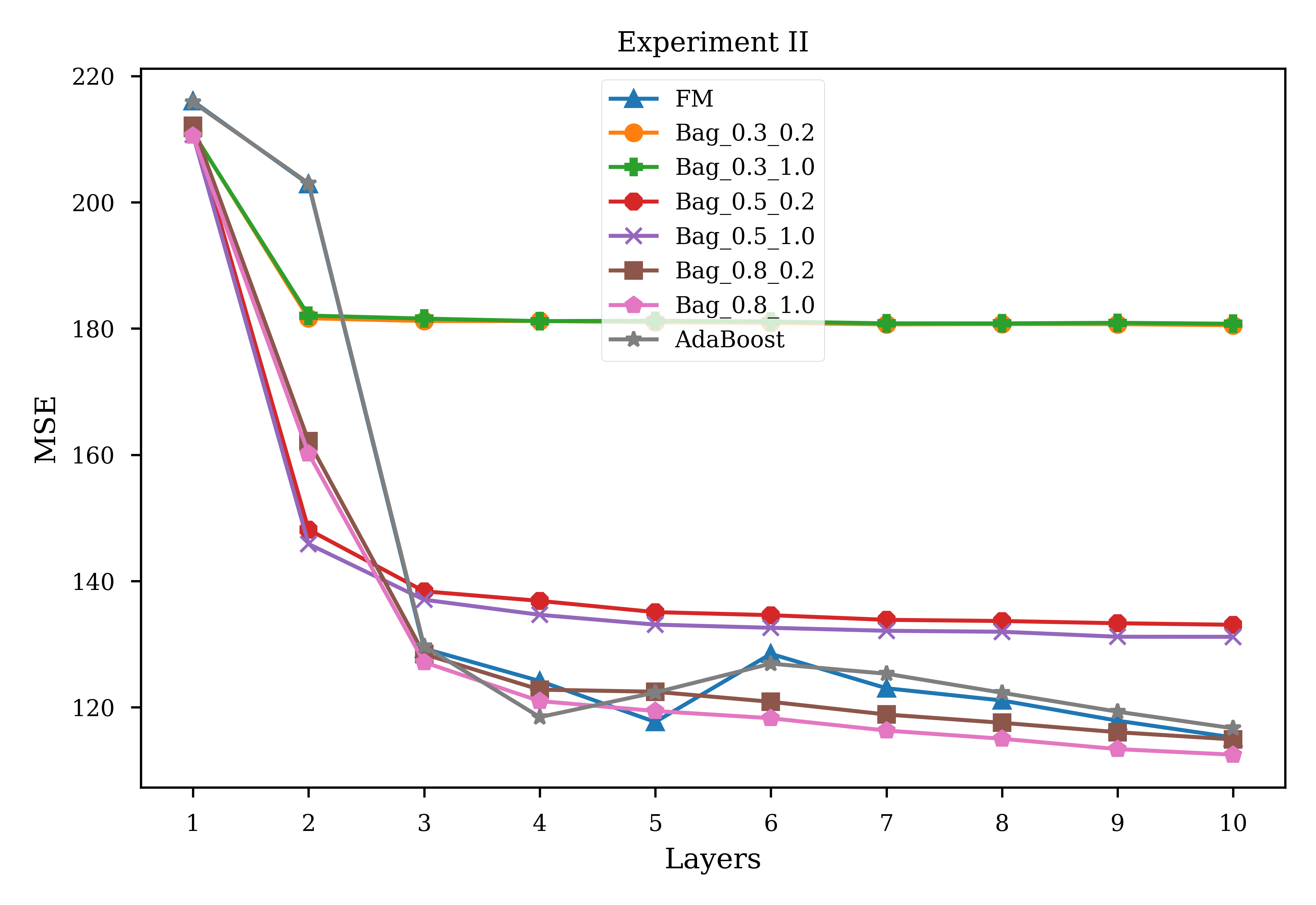}
    \caption{Evolution of MSE with respect to the number of quantum neural network layers in Experiment II.}
    \label{fig:concrete_all}
\end{figure}

The results of the regression experiment are in line with the findings of Experiment I, and are reported in Figure~\ref{fig:concrete_all}. FM, AdaBoost and the two bagging models applied in relation to 80\% of features achieve comparable results at 10 layers, with the Bag.\_0.8\_1.0 configuration obtaining the lowest MSE, followed by Bag.\_0.8\_0.2, FM and finally by AdaBoost. Also in this case, the differential between bagging models with 20\% of samples and with 100\% of samples is marginal, confirming the effectiveness of bagging quantum models in relation to reduced training dataset size. In contrast with Experiment I, bagging models having 30\% of available features now have 2 qubits, and therefore demonstrate a relative improvement in test error when $l=2$. However, their expressive power soon saturates and their error curves plateau. 

In general, the generalization capability of bagging models decreases monotonically with the number of layers, in contrast to FM and AdaBoost. In fact, they exhibit episodes of increased test errors when utilising 5 (and up to 7) layers, while bagging appears to be able to evade this outcome. Such an increase in the test error of FM and AdaBoost may be due to the complex landscape of the cost function, which in some cases could be challenging to optimize and lead to bad generalization.

All of the bagging models analyzed still outperform FM and AdaBoost at a low number of layers, suggesting that they may be the right choice for implementation on NISQ devices, or else when there is any necessity of implementing low-depth quantum circuits. As in the first experiment, it is also of interest to note that all the bagging models with $l=1$ here have very similar MSE values, while their performances vary as the number of layers increases. This may indicate that the MSE value reached at $l=1$ is the optimal for that family of bagging models, given their expressibility. Moreover, a sharp decrease in MSE beyond the first layers would appear to be a common pattern, both with respect to the ensembles and the FM model. For example, at $l \geq 3$, the MSE of FM and AdaBoost dramatically decrease, while bagging models with 50\% of the features exhibit this trend between $l=1$ and $l=2$. (A future analysis of this topic might seek to exploit this characteristic in order to predict {\em a priori} how many layers one would need to attain an error level within a given bound).

As for Experiment I, we report here the MSE of the bagging models with respect to the variation of the ensemble size (Table~\ref{tab:ensemble_size_II}). It confirms the trend for which as the ensemble size increases, the MSE on test decreases.

\begin{table}[h!]
    \centering
    \renewcommand{\arraystretch}{1.7}
    \begin{tabular}{ccccccc}
        \toprule
        & & \multicolumn{5}{c}{Layers}\\ 
        \# Estimators & Model & 2 & 4 & 6 & 8 & 10\\\midrule
        \multirow{6}{*}{4} & Bag\_0.3\_0.2 & 188.1 & 188.0 & 188.0 & 187.9 & 187.5 \\ 
        & Bag\_0.3\_1.0 & 188.0 & 187.3 & 187.8 & 187.4 & 187.4 \\ 
        & Bag\_0.5\_0.2 & 152.1 & 140.7 & 138.2 & 137.0 & 136.2 \\ 
        & Bag\_0.5\_1.0 & 149.0 & 137.1 & 135.2 & 134.6 & 134.0 \\ 
        & Bag\_0.8\_0.2 & 167.1 & 125.7 & 122.2 & 118.0 & 115.7 \\ 
        & Bag\_0.8\_1.0 & 163.8 & 122.0 & 118.8 & 115.9 & 112.8 \\
        \hline
        \multirow{6}{*}{6} & Bag\_0.3\_0.2 & 184.7 & 184.2 & 184.3 & 184.0 & 183.7 \\
        & Bag\_0.3\_1.0 & 183.7 & 182.7 & 182.9 & 182.6 & 182.6 \\
        & Bag\_0.5\_0.2 & 150.5 & 139.1 & 136.6 & 135.5 & 134.6 \\
        & Bag\_0.5\_1.0 & 147.1 & 135.9 & 133.7 & 133.1 & 132.4 \\
        & Bag\_0.8\_0.2 & 164.3 & 125.8 & 122.8 & 119.3 & 116.9 \\
        & Bag\_0.8\_1.0 & 161.2 & 122.1 & 119.2 & 116.3 & 113.7 \\
        \hline
        \multirow{6}{*}{8} & Bag\_0.3\_0.2 & 183.0 & 182.7 & 182.5 & 182.1 & 182.0 \\
        & Bag\_0.3\_1.0 & 183.0 & 182.0 & 181.9 & 181.6 & 181.5 \\
        & Bag\_0.5\_0.2 & 148.6 & 137.2 & 134.0 & 134.0 & 133.3 \\
        & Bag\_0.5\_1.0 & 145.4 & 134.1 & 132.0 & 131.3 & 130.6 \\
        & Bag\_0.8\_0.2 & 163.4 & 123.6 & 121.0 & 117.6 & 115.0 \\
        & Bag\_0.8\_1.0 & 161.0 & 120.9 & 117.8 & 114.6 & 112.0 \\
        \hline
        \multirow{6}{*}{10} & Bag\_0.3\_0.2 & 181.6 & 181.2 & 180.9 & 180.7 & 180.5 \\
        & Bag\_0.3\_1.0 & 182.1 & 181.2 & 181.1 & 180.8 & 180.8 \\
        & Bag\_0.5\_0.2 & 148.2 & 136.9 & 134.6 & 133.7 & 133.0 \\
        & Bag\_0.5\_1.0 & 145.9 & 134.7 & 132.6 & 131.9 & 131.1 \\
        & Bag\_0.8\_0.2 & 162.2 & 122.8 & 120.9 & 117.6 & 114.9 \\
        & Bag\_0.8\_1.0 & 160.2 & 121.0 & 118.3 & 115.0 & 112.5 \\
        \bottomrule
    \end{tabular}
    \caption{MSE of bagging ensembles by varying the number of bagging estimators in Experiment II. Only even number of layers are reported for brevity.}
    \label{tab:ensemble_size_II}
\end{table}

\subsubsection{Experiment III}
The dataset used in Experiment III is the reference Diabetes dataset from Scikit-learn, consisting of 10 numerical features, including age, sex, body mass index, blood serum measurements, and also a target variable, a quantitative measure of disease progression one year after baseline. The dataset is composed of 442 instances and is often used for non-trivial regression analysis in ML.

\begin{figure}[htbp]
    \centering
    \includegraphics[width=1\textwidth]{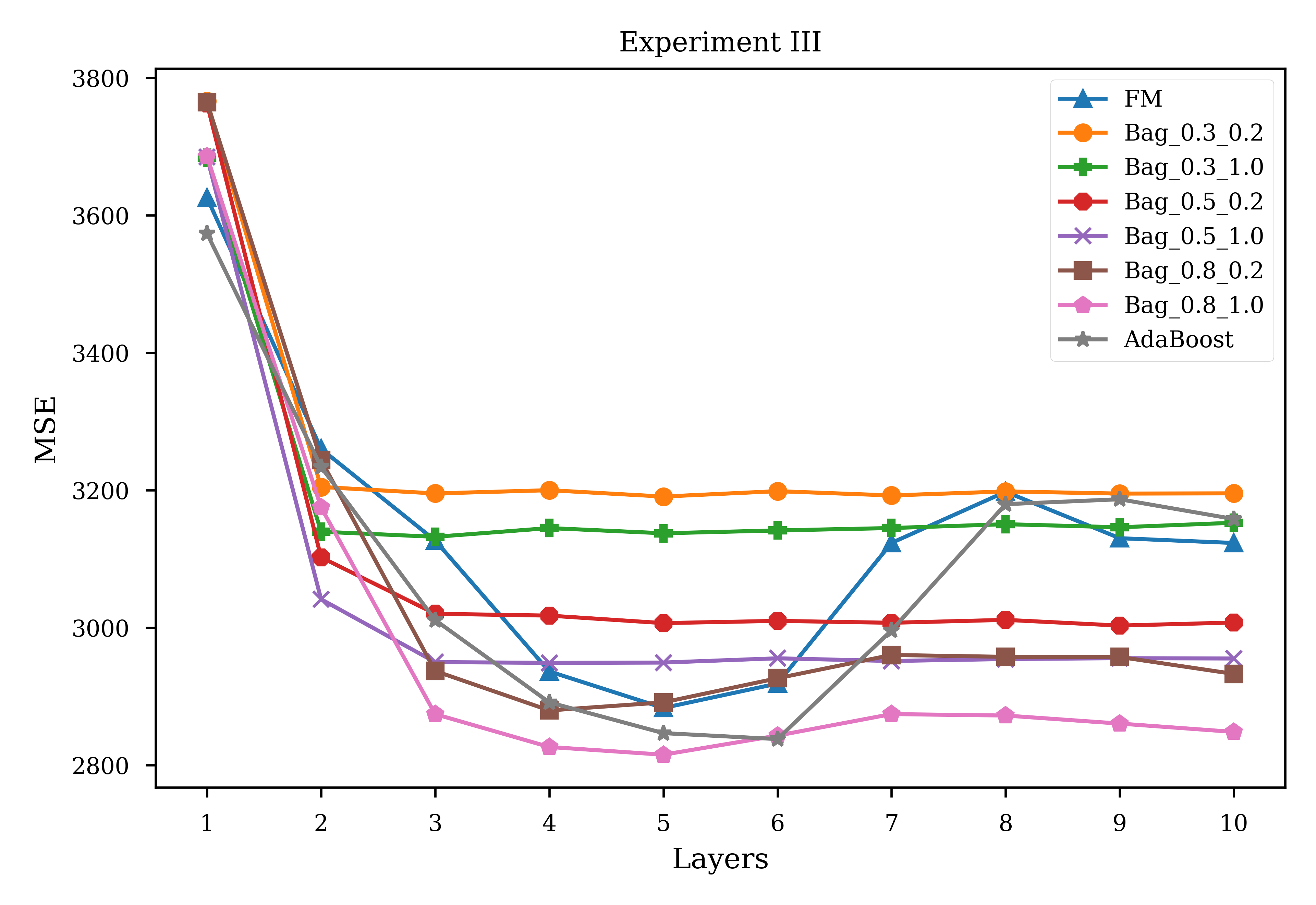}
    \caption{Evolution of MSE with respect to the number of quantum neural network layers in Experiment III.}
    \label{fig:diabete_all}
\end{figure}

Figure~\ref{fig:diabete_all} illustrates the results of this experiment. The performance of the quantum models is notably different from those of the previous two experiments. It may be seen that the best-performing models are the bagging models containing 80\% of the features for almost any number of layers, while FM and AdaBoost achieve satisfactory results up to 6 layers, at which point their MSE begins to increase. At $l=10$, every model has stabilized, however. Bag.\_0.8\_1.0 and Bag.\_0.8\_0.2 have an MSE of respectively 8.8\% and 6.1\% lower than that of FM. AdaBoost has an MSE comparable to the error of Bag.\_0.3\_1.0, being only 0.9\% higher than FM. Bagging models with 50\% of the features have surprisingly good results, better than those of FM and very close to bagging models with 80\% of the features.

As in Experiment I and II, a very sharp MSE reduction between $l=1$ and $l=3$ is evident for all of the models. Less complex models like bagging with 30\% and 50\% of the features immediately reach a plateau, while the error curves for bagging with 80\% of the features, FM and AdaBoost evolves as the number of parameters increases. Considering layer numbers between $l=6$ and $l=8$, it is possible that the capacity of FM and AdaBoost models saturates as the number of model parameters increases, and thus they perform poorly on both training and test data; they struggle in navigating the cost function and escaping from local minima during training. In particular, their learning is constrained to such an extent that they almost reach the same performance level of the simplest bagging models with 30\% of the features. The latter show no indication of bad generalization however, in common with bagging models having 50\% of the features. Bagging with 80\% of the features shows light signs of bad generalization and learning saturation when $l>6$, but still achieve the best results from among all of the tested algorithms.
The robustness of bagging models to learning saturation with respect to AdaBoost and FM arises from their ability to reduce variance via averaging of decorrelated error across the predictions of each submodel. By contrast, when the number of layers is high, AdaBoost and FM utilise a model that is too complex and expressive for the underlying task, leading to learning saturation, bad generalization performances and difficulties in approximating the underlying function. In concordance with Experiment II, this result confirm the effectiveness of bagging models in improving the predictive performance of QNN models, especially in cases where the optimization of the cost function becomes challenging, in line with the classical counterpart.

In addition, this experiment also highlights more markedly the discrepancy between the error level of bagging models with the same number of features but a distinct number of training samples. The difference between the MSE of the bagging model with 30\% and 20\% of samples and that with 100\% of samples is now far more apparent, suggesting that when the variance of the dataset is very high, even bagging models require a sufficient threshold of training samples to perform well in the NISQ setting. 

Regarding the performances of the bagging models with different ensemble sizes, the results are displayed in Table~\ref{tab:ensemble_size_III}. The findings further support the previously observed trend that as the size of the ensemble increases, the MSE decreases on the test dataset.

\begin{table}[ht!]
    \centering
    \renewcommand{\arraystretch}{1.7}
    \begin{tabular}{ccccccc}
        \toprule
        & & \multicolumn{5}{c}{Layers}\\ 
        \# Estimators & Model & 2 & 4 & 6 & 8 & 10\\\midrule
        \multirow{6}{*}{4} & Bag\_0.3\_0.2 & 3438.6 & 3428.7 & 3421.7 & 3426.0 & 3418.7 \\ 
        & Bag\_0.3\_1.0 & 3321.8 & 3323.6 & 3319.5 & 3333.3 & 3328.4 \\ 
        & Bag\_0.5\_0.2 & 3236.1 & 3169.1 & 3173.7 & 3171.9 & 3165.2 \\  
        & Bag\_0.5\_1.0 & 3121.1 & 3041.4 & 3053.2 & 3058.5 & 3051.6 \\ 
        & Bag\_0.8\_0.2 & 3340.9 & 2959.9 & 2986.8 & 2998.4 & 2975.1 \\ 
        & Bag\_0.8\_1.0 & 3272.6 & 2867.8 & 2868.8 & 2889.0 & 2874.7 \\ 
        \hline
        \multirow{6}{*}{6} & Bag\_0.3\_0.2 & 3326.2 & 3324.8 & 3325.6 & 3324.0 & 3319.7 \\ 
        & Bag\_0.3\_1.0 & 3218.5 & 3225.2 & 3222.0 & 3231.5 & 3227.7 \\ 
        & Bag\_0.5\_0.2 & 3152.9 & 3077.6 & 3084.4 & 3082.7 & 3080.2 \\ 
        & Bag\_0.5\_1.0 & 3049.6 & 2983.8 & 2993.1 & 2995.3 & 2990.1 \\ 
        & Bag\_0.8\_0.2 & 3275.4 & 2928.5 & 2956.2 & 2973.3 & 2956.4 \\ 
        & Bag\_0.8\_1.0 & 3198.8 & 2845.5 & 2852.5 & 2870.0 & 2847.6 \\ 
        \hline
        \multirow{6}{*}{8} & Bag\_0.3\_0.2 & 3259.5 & 3249.9 & 3250.3 & 3250.1 & 3246.5 \\ 
        & Bag\_0.3\_1.0 & 3174.6 & 3180.9 & 3176.2 & 3186.1 & 3188.6 \\ 
        & Bag\_0.5\_0.2 & 3130.5 & 3053.3 & 3057.5 & 3056.5 & 3052.5 \\ 
        & Bag\_0.5\_1.0 & 3061.6 & 2978.0 & 2986.1 & 2987.7 & 2985.3 \\ 
        & Bag\_0.8\_0.2 & 3265.1 & 2900.8 & 2933.5 & 2958.3 & 2934.7 \\ 
        & Bag\_0.8\_1.0 & 3209.1 & 2838.3 & 2860.0 & 2884.8 & 2861.5 \\ 
        \hline
        \multirow{6}{*}{10} & Bag\_0.3\_0.2 & 3204.9 & 3200.2 & 3198.8 & 3198.4 & 3195.6 \\ 
        & Bag\_0.3\_1.0 & 3139.8 & 3145.1 & 3141.6 & 3150.7 & 3152.8 \\ 
        & Bag\_0.5\_0.2 & 3102.5 & 3017.7 & 3010.1 & 3011.5 & 3007.6 \\ 
        & Bag\_0.5\_1.0 & 3041.5 & 2949.0 & 2955.6 & 2954.5 & 2955.3 \\ 
        & Bag\_0.8\_0.2 & 3243.9 & 2879.8 & 2926.6 & 2957.7 & 2932.9 \\ 
        & Bag\_0.8\_1.0 & 3175.1 & 2826.6 & 2843.1 & 2872.3 & 2848.5 \\ 
        \bottomrule
    \end{tabular}
    \caption{MSE of bagging ensembles by varying the number of bagging estimators in Experiment III. Only even number of layers are reported for brevity.}
    \label{tab:ensemble_size_III}
\end{table}

\subsubsection{Experiment IV}

For the classification task in Experiment IV, we used the reference UCI Wine dataset. It is a multi-class classification dataset corresponding to the results of a chemical analysis of wines grown within a specific region of Italy. It consists of 13 numerical features representing various chemical properties, such as alcohol, malic acid, and ash content, and a target variable indicating the class of the wine. The dataset has 178 samples and is a common baseline ML benchmark for low-parametric complexity classifiers.

\begin{figure}[htbp]
    \centering
    \includegraphics[width=1\textwidth]{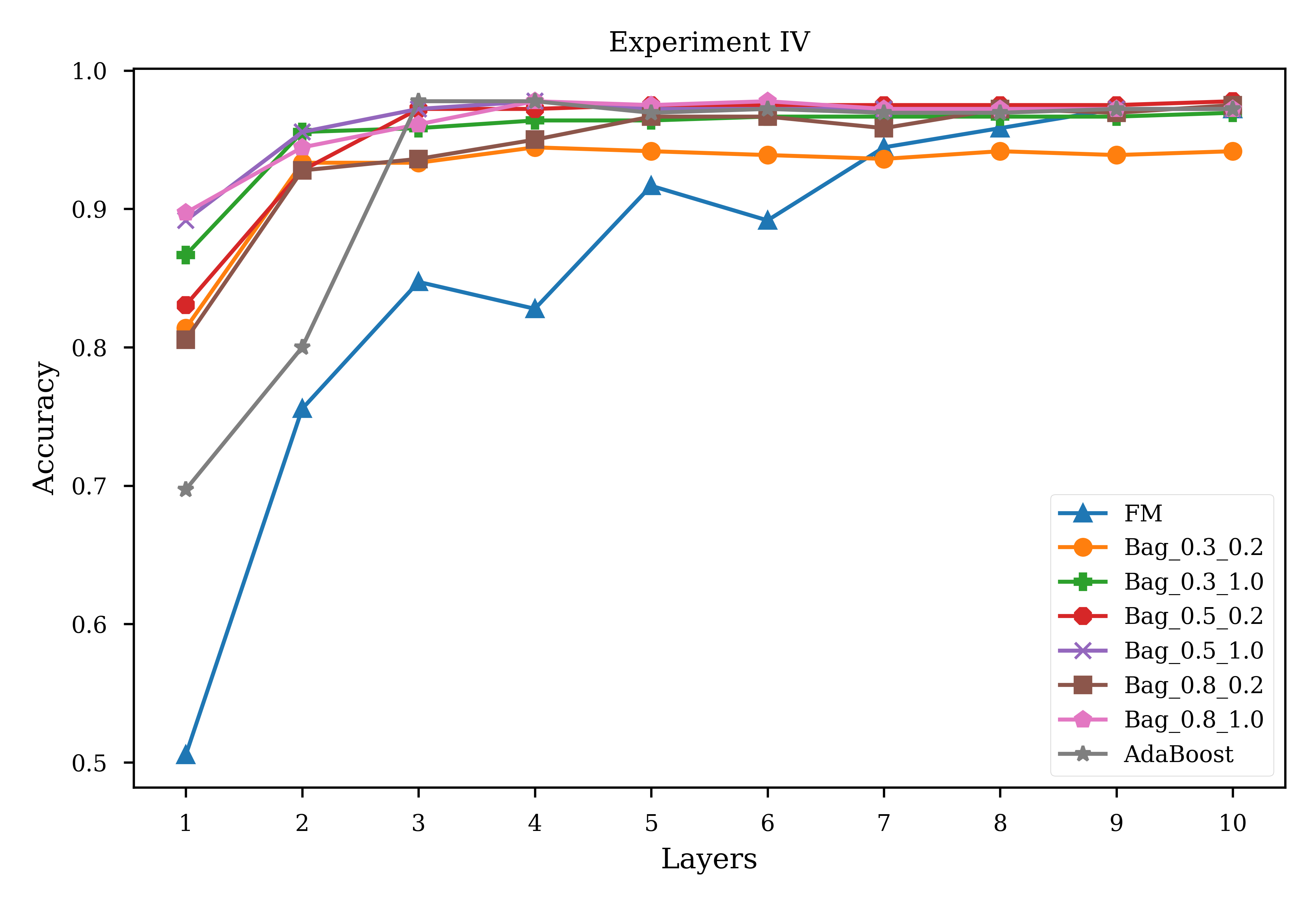}
    \caption{Evolution of Accuracy score with respect to  quantum neural network depth in Experiment IV.}
    \label{fig:wine_all}
\end{figure}

Results from Experiment IV are reported in Figure~\ref{fig:wine_all}. Although they cannot be directly compared to the previous results due to the intrinsically different nature of the problem, there are few comparative insights that can be gained from the respective plot of Accuracy curves. First, all the models except bagging with 30\% of the features achieve the same accuracy score of 97.2\% using 10 layers. The performances of Bag.\_0.3\_0.2 and Bag.\_0.3\_1.0 are still relatively strong, however, having an accuracy score of 94.2\% and 96.9\% respectively. Given the very low complexity of these two models, this is a striking result.

A further notable aspect of the Accuracy curves is that all ensemble models converge with far fewer layers than FM. In particular, they require 3 layers in order to reach a performance plateau on average, after which they saturate and the accuracy score reaches saturation as well. By contrast, FM struggles to achieve a comparable accuracy score, only achieving accuracy greater than 90\% when $l \geq 7$. This means that the ensemble models are able to learn and capture the complex relationships between the input features far more efficiently than FM, which requires a much deeper architecture to attain comparable results. This observation is particularly relevant when considering the implementation of these models on NISQ devices, where the number of qubits and the coherence time are severely limited.

As expected, bagging models with 100\% of the samples obtain almost everywhere (especially with few layers), a higher accuracy score than their counterparts with 20\% of the features given the same number of layers. This suggests that using more training samples can improve the performance of ensemble models provided that the number of layers is low, as it allows them to better capture the underlying patterns of class discriminability in the data.

Finally, Table~\ref{tab:ensemble_size_IV} shows the performance of the bagging models with varying ensemble sizes. The outcomes confirm the previously observed pattern that increasing the ensemble size leads to an increment in the Accuracy calculated on the test set. This indicates that the use of larger ensembles can contribute to improving the accuracy and robustness of the quantum predictive models.

\begin{table}[ht!]
    \centering
    \renewcommand{\arraystretch}{1.7}
    \begin{tabular}{ccccccc}
        \toprule
        & & \multicolumn{5}{c}{Layers}\\ 
        \# Estimators & Model & 2 & 4 & 6 & 8 & 10\\\midrule
        \multirow{6}{*}{4} & Bag\_0.3\_0.2 & 0.875 & 0.897 & 0.892 & 0.900 & 0.900 \\ 
        & Bag\_0.3\_1.0 & 0.908 & 0.911 & 0.911 & 0.917 & 0.911 \\ 
        & Bag\_0.5\_0.2 & 0.897 & 0.947 & 0.953 & 0.953 & 0.947 \\  
        & Bag\_0.5\_1.0 & 0.928 & 0.967 & 0.969 & 0.967 & 0.969 \\ 
        & Bag\_0.8\_0.2 & 0.889 & 0.936 & 0.939 & 0.964 & 0.964 \\ 
        & Bag\_0.8\_1.0 & 0.947 & 0.969 & 0.969 & 0.972 & 0.975 \\ 
        \hline
        \multirow{6}{*}{6} & Bag\_0.3\_0.2 & 0.908 & 0.933 & 0.933 & 0.931 & 0.931 \\ 
        & Bag\_0.3\_1.0 & 0.922 & 0.936 & 0.936 & 0.933 & 0.939 \\ 
        & Bag\_0.5\_0.2 & 0.919 & 0.964 & 0.964 & 0.967 & 0.972 \\ 
        & Bag\_0.5\_1.0 & 0.947 & 0.969 & 0.969 & 0.969 & 0.969 \\ 
        & Bag\_0.8\_0.2 & 0.911 & 0.944 & 0.947 & 0.969 & 0.956 \\ 
        & Bag\_0.8\_1.0 & 0.953 & 0.969 & 0.983 & 0.978 & 0.978 \\ 
        \hline
        \multirow{6}{*}{8} & Bag\_0.3\_0.2 & 0.925 & 0.933 & 0.933 & 0.933 & 0.933 \\ 
        & Bag\_0.3\_1.0 & 0.942 & 0.956 & 0.958 & 0.958 & 0.961 \\ 
        & Bag\_0.5\_0.2 & 0.931 & 0.969 & 0.964 & 0.972 & 0.969 \\ 
        & Bag\_0.5\_1.0 & 0.958 & 0.972 & 0.972 & 0.972 & 0.972 \\ 
        & Bag\_0.8\_0.2 & 0.908 & 0.947 & 0.961 & 0.967 & 0.969 \\ 
        & Bag\_0.8\_1.0 & 0.942 & 0.967 & 0.986 & 0.975 & 0.975 \\ 
        \hline
        \multirow{6}{*}{10} & Bag\_0.3\_0.2 & 0.933 & 0.944 & 0.939 & 0.942 & 0.942 \\ 
        & Bag\_0.3\_1.0 & 0.956 & 0.964 & 0.967 & 0.967 & 0.969 \\ 
        & Bag\_0.5\_0.2 & 0.928 & 0.972 & 0.975 & 0.975 & 0.978 \\ 
        & Bag\_0.5\_1.0 & 0.956 & 0.978 & 0.972 & 0.972 & 0.972 \\ 
        & Bag\_0.8\_0.2 & 0.928 & 0.950 & 0.967 & 0.972 & 0.975 \\ 
        & Bag\_0.8\_1.0 & 0.944 & 0.978 & 0.978 & 0.972 & 0.972 \\ 
        \bottomrule
    \end{tabular}
    \caption{Accuracy of bagging ensembles by varying the number of bagging estimators in Experiment IV. Only even number of layers are reported for brevity.}
    \label{tab:ensemble_size_IV}
\end{table}

\subsection{Resource efficiency of quantum neural network ensembles}

Besides performance, resource efficiency is a key argument for the  utilization of quantum neural network ensembles. Efficiency can be measured by various metrics: for example, number of qubits, gates, parameters, and training samples required to achieve comparable performance.

\begin{figure}[htbp]
    \centering
    \includegraphics[width=0.49\textwidth]{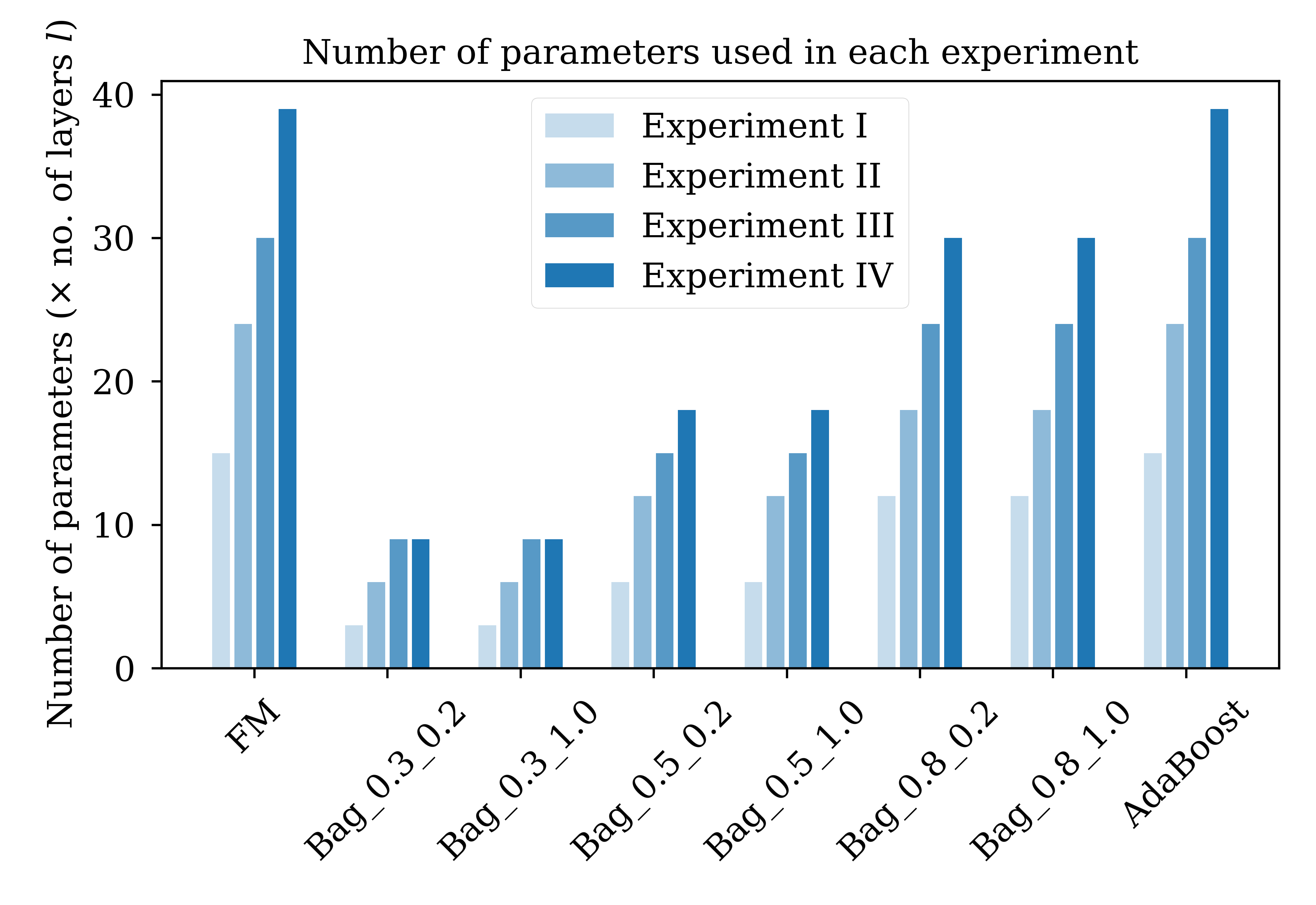}
    \includegraphics[width=0.49\textwidth]{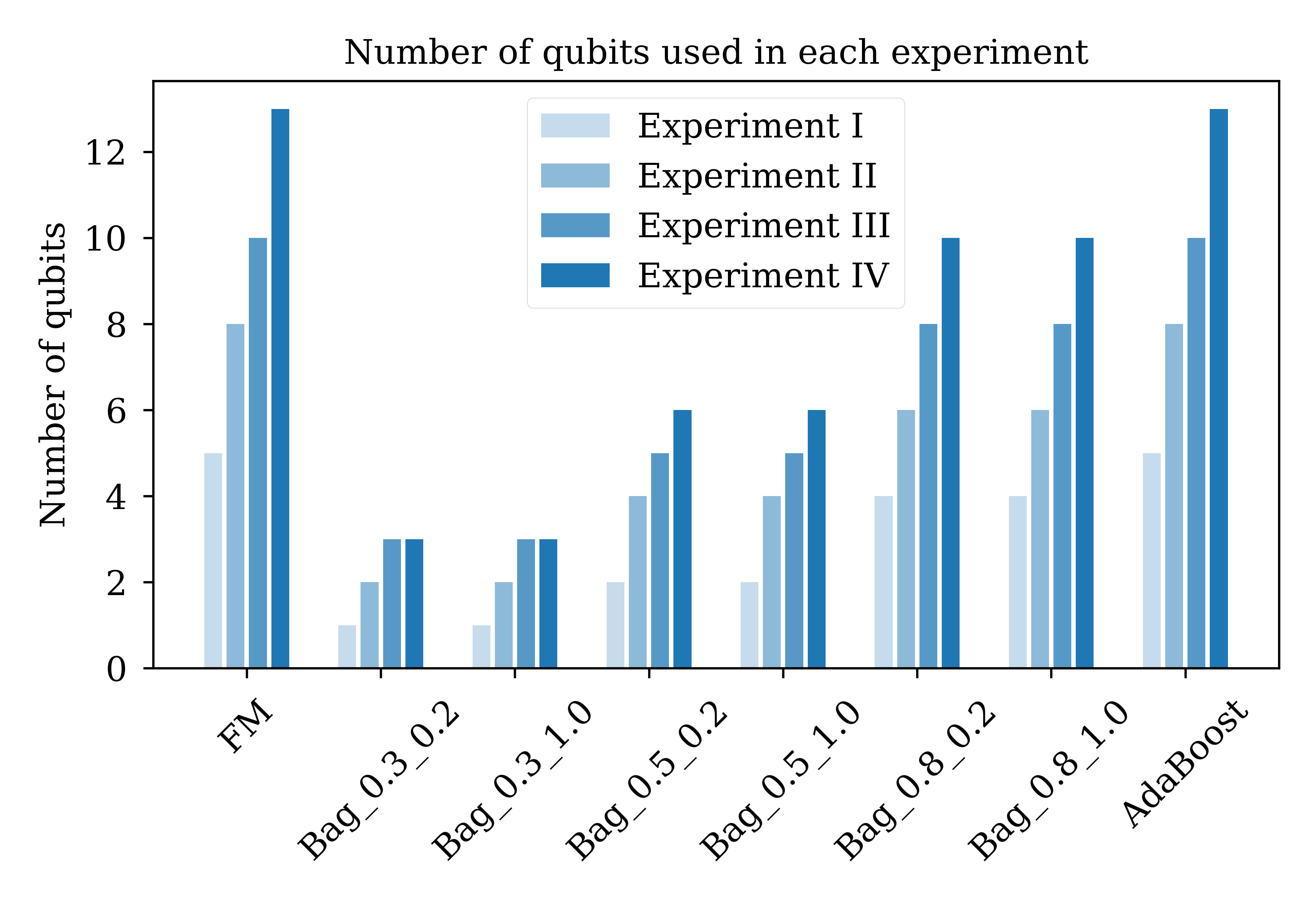}
    \caption{Number of qubits \&  parameters employed in individual experiments.}
    \label{fig:net_struct}
\end{figure}

To determine the potential savings in the number of qubits we here deploy the random subspace technique (also known as {\em attribute bagging} or {\em attribute bootstrap aggregation}). Our experiments (cf Figure \ref{fig:net_struct}) suggest a potential saving of 20\% to 80\% of the total qubit budget via this approach. However, such a saving is made at the cost of the ensemble was a whole having the potential for less rich class-discrimination behaviour, dependent on both the sampling required to achieve full feature coverage and the nature of the underlying data manifold. A positive consequence of reducing the number of qubits, though,  is that each quantum circuit will have fewer gates and parameters, resulting in improved noise robustness on real hardware (i.e less decoherence, higher overall fidelity), as well as faster gradient calculation (individual gradient calculations require $P+1$ quantum circuit evaluations for $P$ parameters). This allows for a saving of the parameter budget of up to 75\% in the indicated experimental regime, while the saving on gates corresponds proportionately (cf Figure \ref{fig:qnn}). Savings for each dataset and ensemble technique are as depicted in Figure \ref{fig:net_struct}. 

\subsection{Simulated Noisy Experiments}

For the noisy simulated experiment, we compare the performance of the FM baseline with all the Bagging techniques on the same synthetic linear regression dataset used in Experiment I, in order to assess the ensemble's noise mitigation properties. Unfortunately, further experiments on the other datasets are currently unfeasible due to technological constraints, since noisy simulations with Qiskit's backend on large datasets are very time-consuming above 5-6 qubits. Moreover, such constraints allow to simulate up to 3 layers only; simulation of more complex models would demand a significant amount of runtime and computational resources that are currently unfeasible. Nevertheless, they are enough to demonstrate the behavior and goodness of the ensemble techniques in the presence of quantum hardware noise, as well as the evolution of the models' performance as the number of layers increases. Each experiment is repeated 5 times to ensure statistical validity. 

\begin{figure}[htbp]
    \centering
    \subfloat[]{\label{fig:noisy_simulation:a}\includegraphics[width=.49\linewidth]{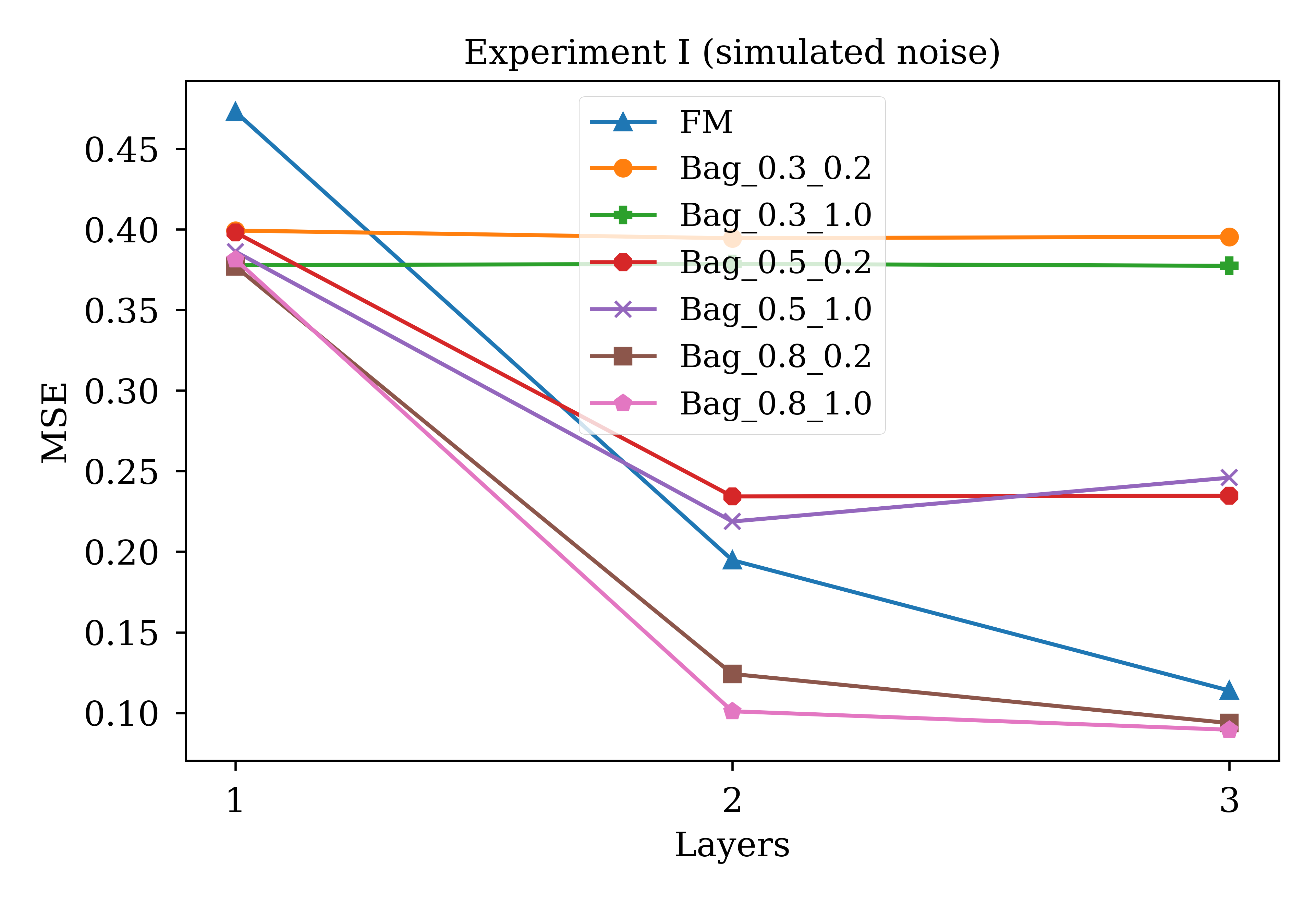}}
    \subfloat[]{\label{fig:noisy_simulation:b}\includegraphics[width=.49\linewidth]{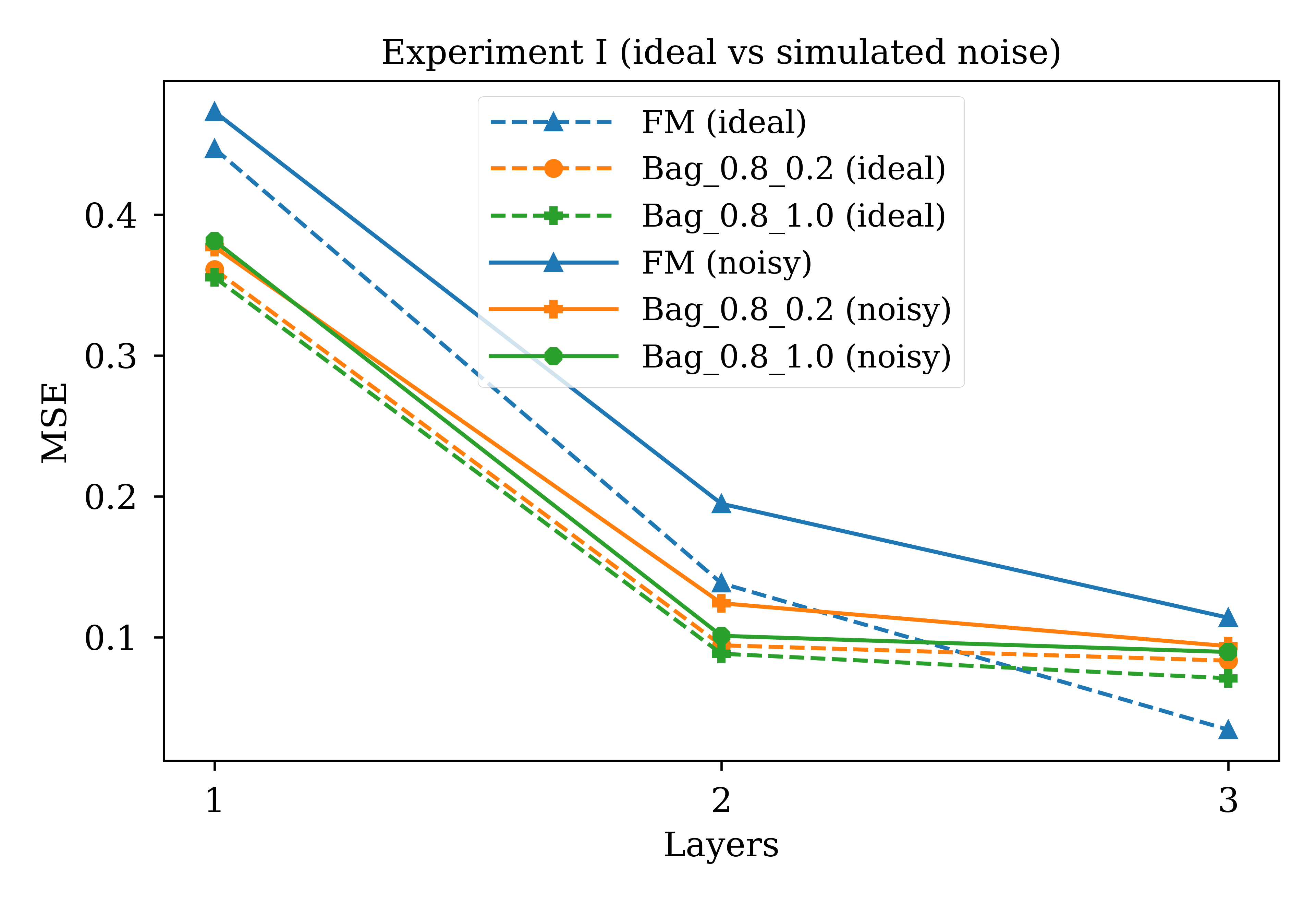}}
    \caption{Comparison of average performance of the baseline model and the Bagging ensemble models on simulated IBM Lima quantum hardware. (\ref{fig:noisy_simulation:a}) shows the difference in terms of MSE over 5 executions with all the bagging techniques. (\ref{fig:noisy_simulation:b}) shows the performance of FM and bagging models with 80\% of the features both in an ideal and simulated noisy setting.}
    \label{fig:noisy_simulation}
\end{figure}

From Figure \ref{fig:noisy_simulation:a}, it is evident that FM is highly affected by noise, while sufficiently complex Bagging techniques are more resilient to such errors and manage to achieve better results, even outperforming FM in the case of Bagging with 80\% of the features. This discrepancy in performance is expected to accentuate as the number of layers (i.e., the depth) of the circuit or the number of ensemble estimators increase. These results confirm the goodness of the Bagging approach in mitigating the effect of noise for QNNs if compared to the outcomes of the same experiment in an ideal setting, as illustrated in Figure \ref{fig:noisy_simulation:b} for the best-performing techniques (FM and Bagging with 80\% of the features).

\subsection{Experiments executed on superconducting-based QPU}

For the real-hardware evaluation, we compare the performance of the baseline quantum neural network with the Bag\_0.8\_0.2 ensemble on the same synthetic linear regression dataset used in Experiment I. We selected the Bag\_0.8\_0.2 model as representative ensemble technique for its outstanding performance in the simulated experiments despite the low number of training samples. To ensure statistical validity, we repeat each experiment 10 times. However, due to technological constraints on real quantum hardware, we analyze only the linear dataset with a quantum neural network having a single layer.

\begin{figure}[htbp]
    \centering
    \subfloat[]{\label{fig:ibm_execution:a}\includegraphics[width=.49\linewidth]{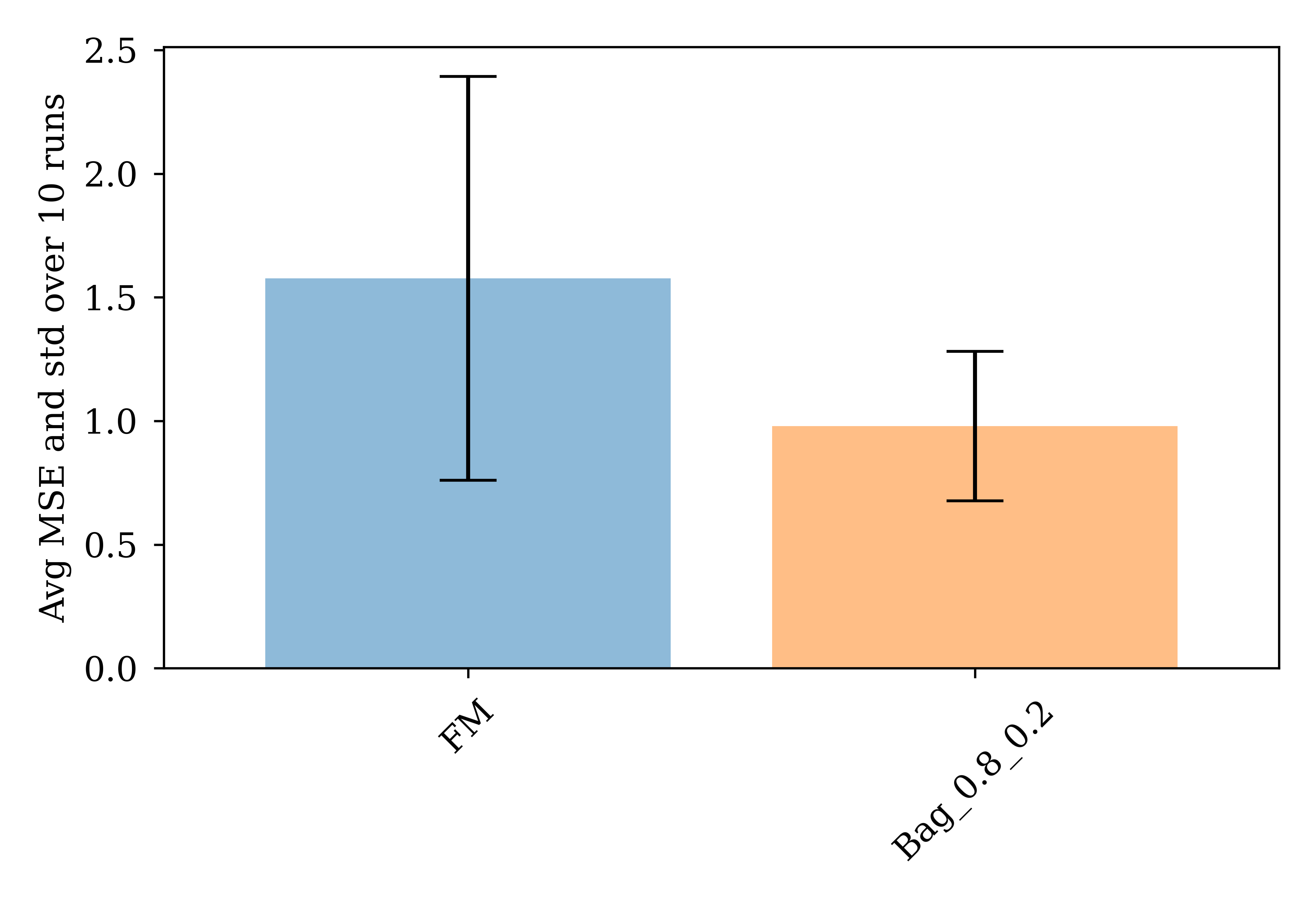}}
    \subfloat[]{\label{fig:ibm_execution:b}\includegraphics[width=.49\linewidth]{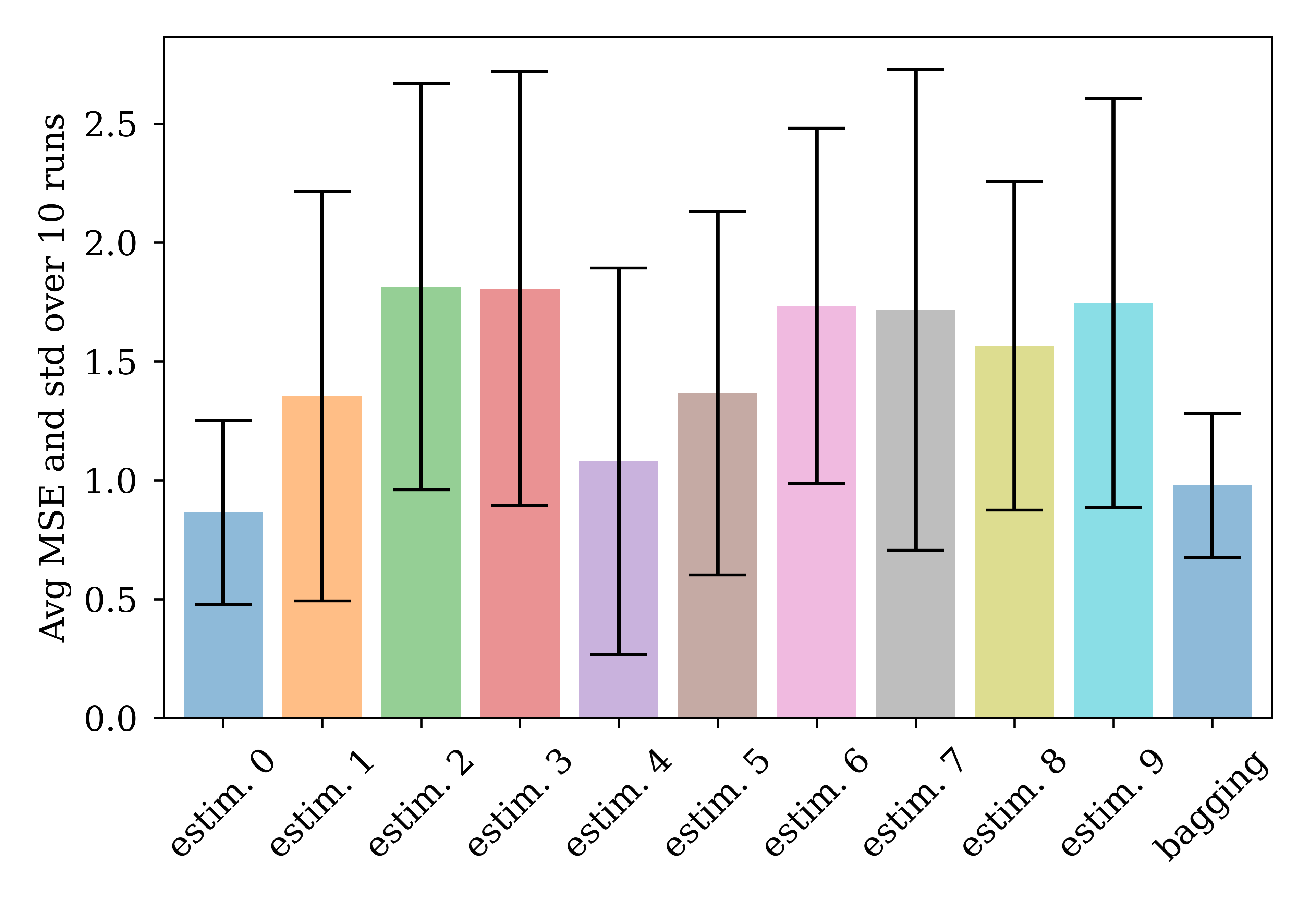}}
    \caption{Comparison of average performance of the baseline model and the Bag\_0.8\_0.2 ensemble technique on IBM quantum hardware. (\ref{fig:ibm_execution:a}) shows the difference in terms of MSE over 10 executions. (\ref{fig:ibm_execution:b}) shows the performance of the bagging model with respect to its estimators.}
    \label{fig:ibm_execution}
\end{figure}

Figure \ref{fig:ibm_execution} presents the real-world experimental findings, which indicate that the bagging ensemble reduces the expected MSE by one-third and the expected variance by half when executed on quantum hardware, compared to the baseline model. Such results demonstrate that the noise-canceling capabilities of ensemble technique can be effectively exploited to work on NISQ devices in realistic settings. Additionally, the performance of the ten bagging models varied significantly, underlining the need to reinitialise the ensemble multiple times and validate it against a suitable validation dataset to ensure that the best model is selected.

\section{Conclusion}

We propose the use of ensemble techniques for practical implementation of quantum machine learning models on NISQ hardware. In particular, we justify the application of these techniques based on their capacity for significant reduction in resource usage, including in respect to the overall qubit, parameter, and gate budget, which is achieved via the random subspace (attribute bagging) technique. This resource-saving is especially crucial for noisy hardware, which is typically limited to a small number of qubits, being vulnerable to decoherence, noise, and operational errors. Consequently, the contribution of ensemble techniques may be seen as a form of quantum noise reduction. 
 
To establish this, we evaluated and compared various configurations of bagging and boosting ensemble techniques on synthetic and real-world datasets, tested in a simulated noise-free environment, in a simulated noisy setting and on a superconducting-based QPU by IBM, and subtending a range of layer depths. 

Our experimental findings showed that bagging ensembles can effectively train quantum neural network instances using fewer features and qubits, which leads to ensemble models with superior performance compared to the baseline model. Reducing the number of features in bagging models of quantum neural networks directly translates into a reduction in the number of qubits, that is a desirable characteristics for practical quantum applications. Ensembles of quantum neural network can also help addressing some of the toughest challenges associated with noise and decoherence in NISQ devices, as well as to mitigate barren plateau effects. These can be key considerations in the development of quantum machine learning models, particularly when working with limited resources on modern quantum systems. 

Moreover, bagging models were found to be extremely robust to learning saturation, being able to effectively capture the underlying patterns in the data with high generalization ability. This makes them better suited for tasks where generalization is important, such as in real-world applications. As in the classical case, the increase of the ensemble size in case of bagging also helps to achieve better performances for QNNs. However, it is important to notice that the effectiveness of bagging quantum models diminishes with a decrement in the number of features, which suggests that complex bagging models are still needed to obtain satisfactory results. Using only a subset of the features can reduce the computational complexity of the model and help in the optimization process, but it may also result in a loss of information and a decrease in performance. On the contrary, the number of training samples do not seem to have a deep impact on bagging quantum models, hence this bagging strategy may be used when executing quantum neural network instances on real hardware in order to deal with long waiting queues and job scheduling issues. In this regard, having a low number of training data leads to faster training procedures and quantum resource savings. The training of ensembles can also be done in parallel on multiple QPUs in a distributed learning fashion. Therefore, it is important to strike a balance between model complexity and performance to achieve the best possible outcomes.

Additionally, the fact that the bagging models outperform FM and AdaBoost at low number of layers suggests that the former models are better suited for low-depth quantum circuits, which have limited capacity and are prone to noise and errors. For quantum machine learning tasks with NISQ devices, using bagging models with a low number of layers may be a good strategy to achieve good generalization performance while minimizing the impact of noise and errors in the circuit.

Overall, our results suggest that ensembles of quantum neural network models can be a promising avenue for the development of practical quantum machine learning applications on NISQ devices, both from a performance and resource usage perspective. A careful evaluation of the trade-offs between model complexity, performance, quantum resources available and explainability may be necessary to make an informed decision. 

In a future work, we plan to further investigate the relationship between ensembles and quantum noise, which is a key consideration when developing quantum neural network models. 
In addition, it would be relevant to discuss how our proposed approach compares to classical ensembles for real-world applications; it would contribute to a more comprehensive and insightful understanding of the potential advantages and limitations of our approach.
To sum up, our findings could potentially contribute to the development of more efficient and accurate quantum machine learning algorithms, which could have significant implications for real-world applications.

\section*{Acknowledgements}

The contribution of M. Panella in this work was supported by the ``NATIONAL CENTRE FOR HPC, BIG DATA AND QUANTUM COMPUTING'' (CN1, Spoke 10) within the Italian ``Piano Nazionale di Ripresa e Resilienza (PNRR)'', Mission 4 Component 2 Investment 1.4 funded by the European Union - {NextGenerationEU} - CN00000013 - CUP B83C22002940006. MG and SV are supported by CERN through CERN Quantum Technology Initiative. Access to the IBM Quantum Services was obtained through the IBM Quantum Hub at CERN. The views expressed are those of the authors and do not reflect the official policy or position of IBM and the IBM~Q team. MI is part of the Gruppo Nazionale Calcolo Scientifico of ``Istituto Nazionale di Alta Matematica Francesco Severi''. AM is supported by Foundation for Polish Science (FNP), IRAP project ICTQT, contract no. 2018/MAB/5, co-financed by
EU Smart Growth Operational Programme. 

\section*{Declaration}

\subsection*{Authors' contributions}

MI, MG, and AC had the initial idea, implemented the interface for executing experiments on the IBM QPUs, performed the experiments, and analyzed the data. MG, SV, DW, AM, and MP supervised the project. All authors contributed to the manuscript.

\subsection*{Availability of data and materials} 

The source code used to obtain our results can be freely accessed at \url{https://github.com/incud/Classical-ensemble-of-Quantum-Neural-Networks}.

\subsection*{Conflict of interest}

The authors declare no conflict of interest. 



\begin{appendices}

\section{Detailed plots}\label{apx:detailed_plots}

\begin{figure}[htbp]
\begin{tabular}[b]{@{}c@{}}
\includegraphics[width=.48\linewidth]{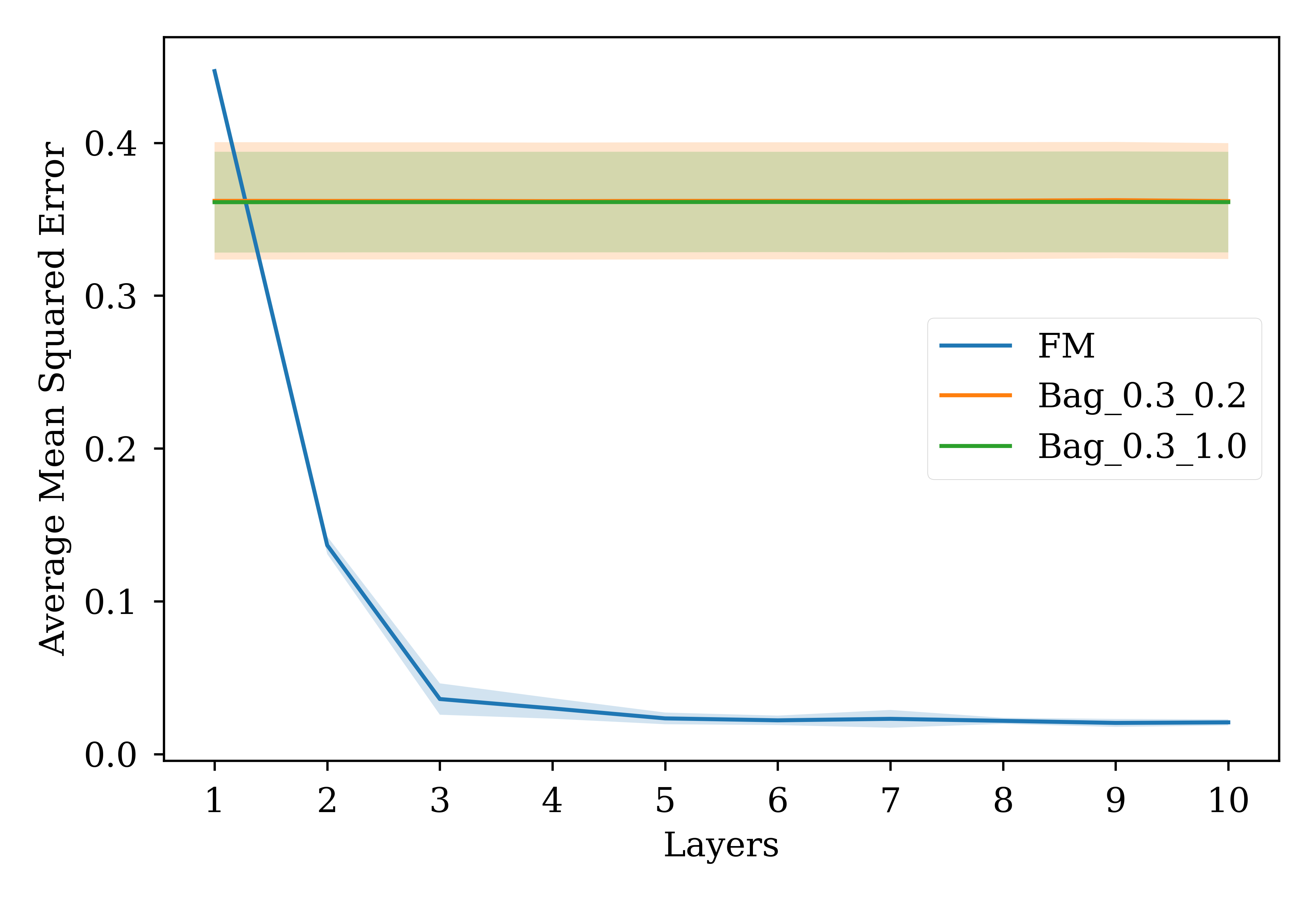}%
\includegraphics[width=.48\linewidth]{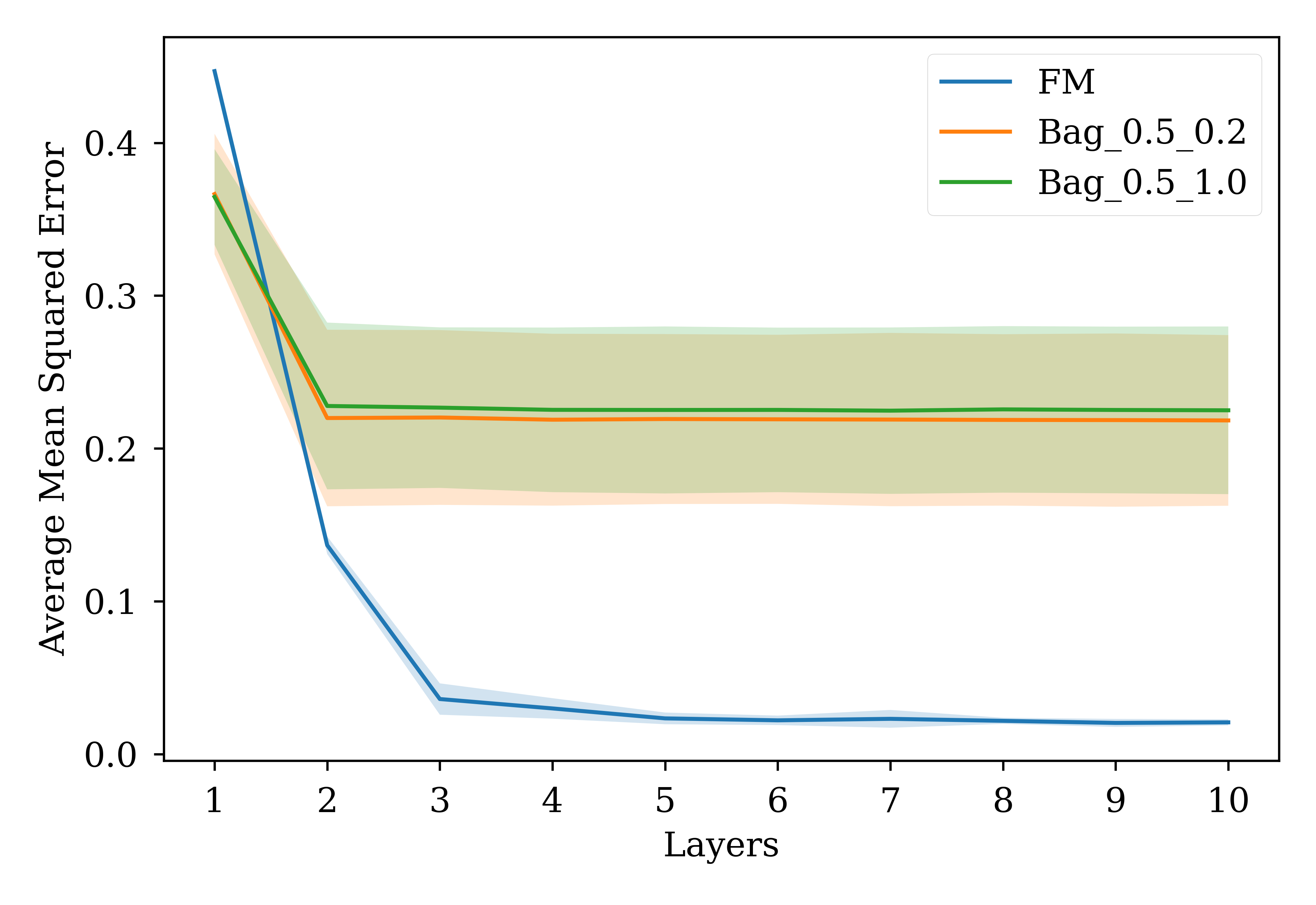}\\[-3pt]
\includegraphics[width=.48\linewidth]{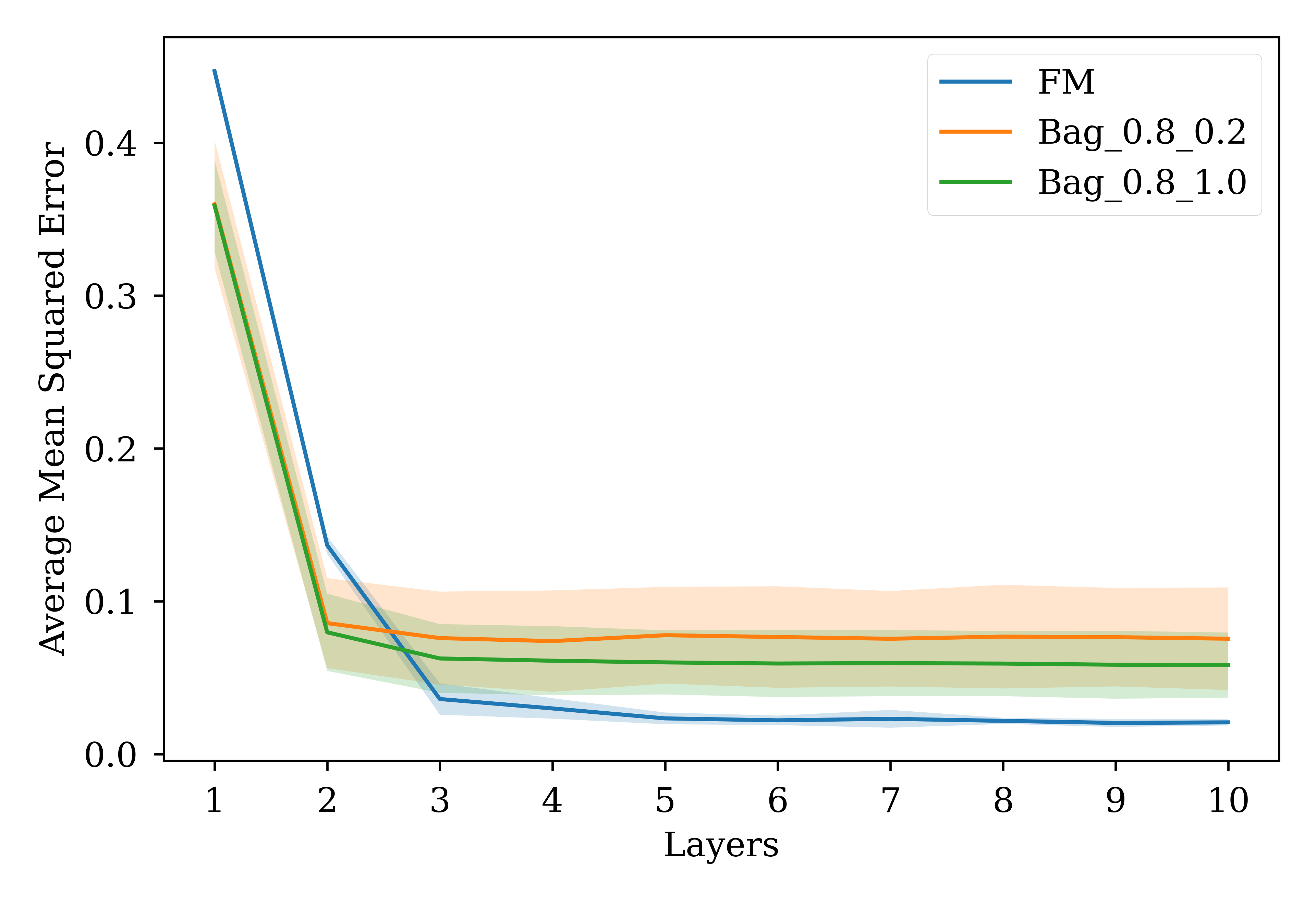}%
\includegraphics[width=.48\linewidth]{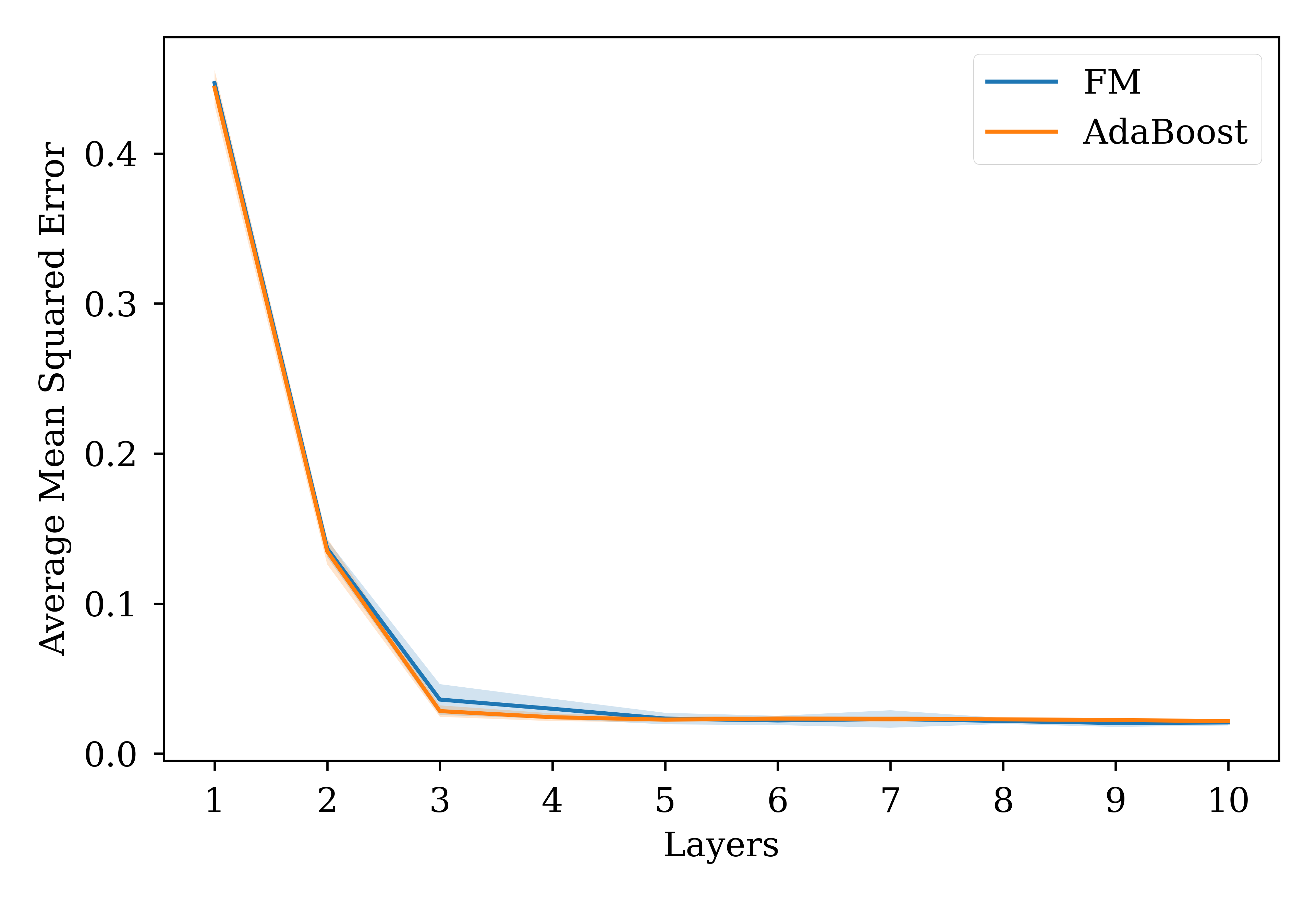}
\end{tabular}%
\caption{Comparison of the performance of the baseline model and ensemble systems on the Linear Regression dataset. It exhibits the MSE and standard deviation, with a semi-transparent area, of the ensemble schemes in comparison to the baseline models. The top-left image shows ensembles with Random Subspace at 30\% of the features, top-right shows ensembles with Random Subspace at 50\%, bottom-left displays ensembles with Random Subspace at 80\%, and bottom-right illustrates AdaBoost.}%
\label{fig:linear_erb}
\end{figure}

\begin{figure}[htbp]
\begin{tabular}[b]{@{}c@{}}
\includegraphics[width=.48\linewidth]{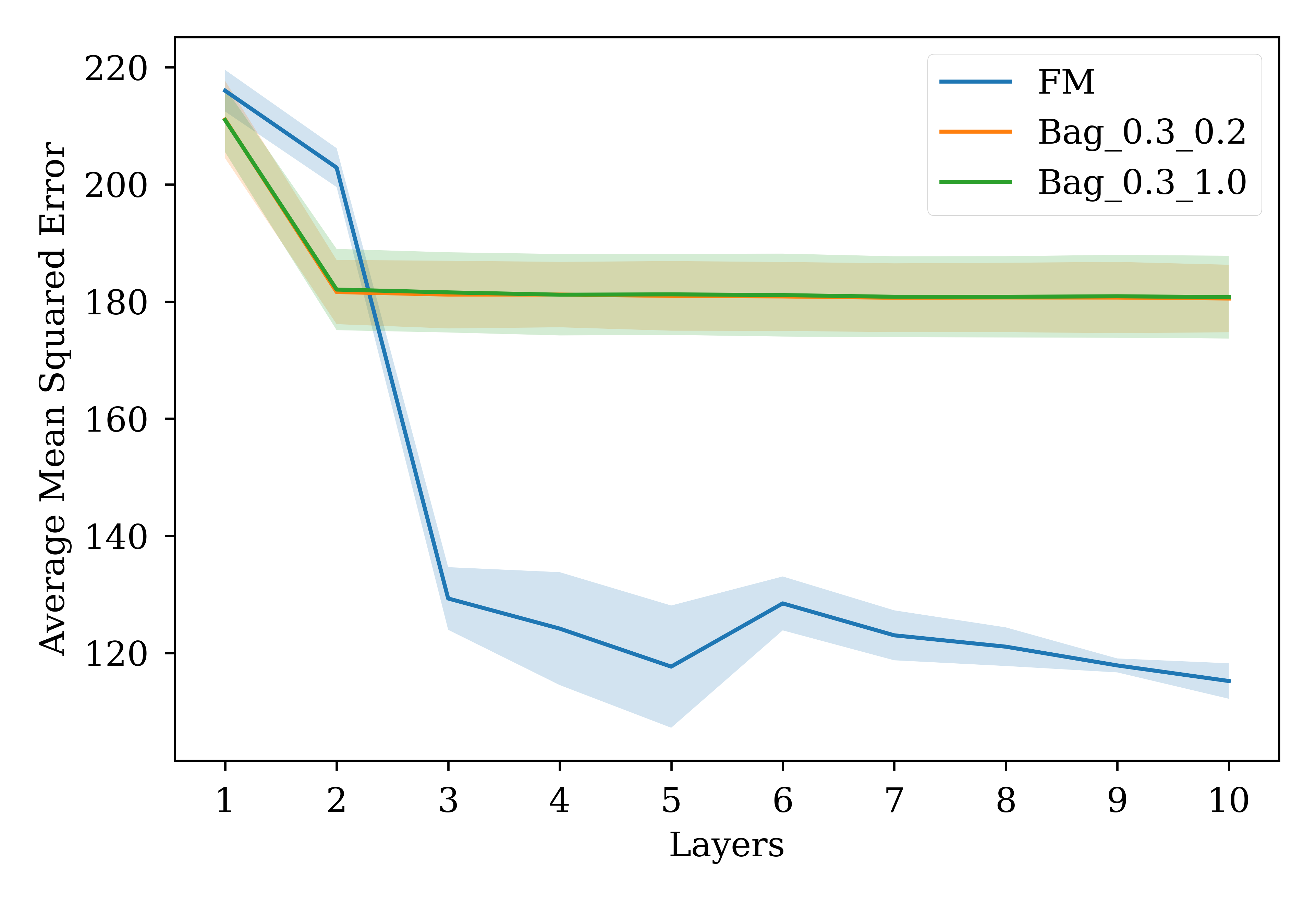}%
\includegraphics[width=.48\linewidth]{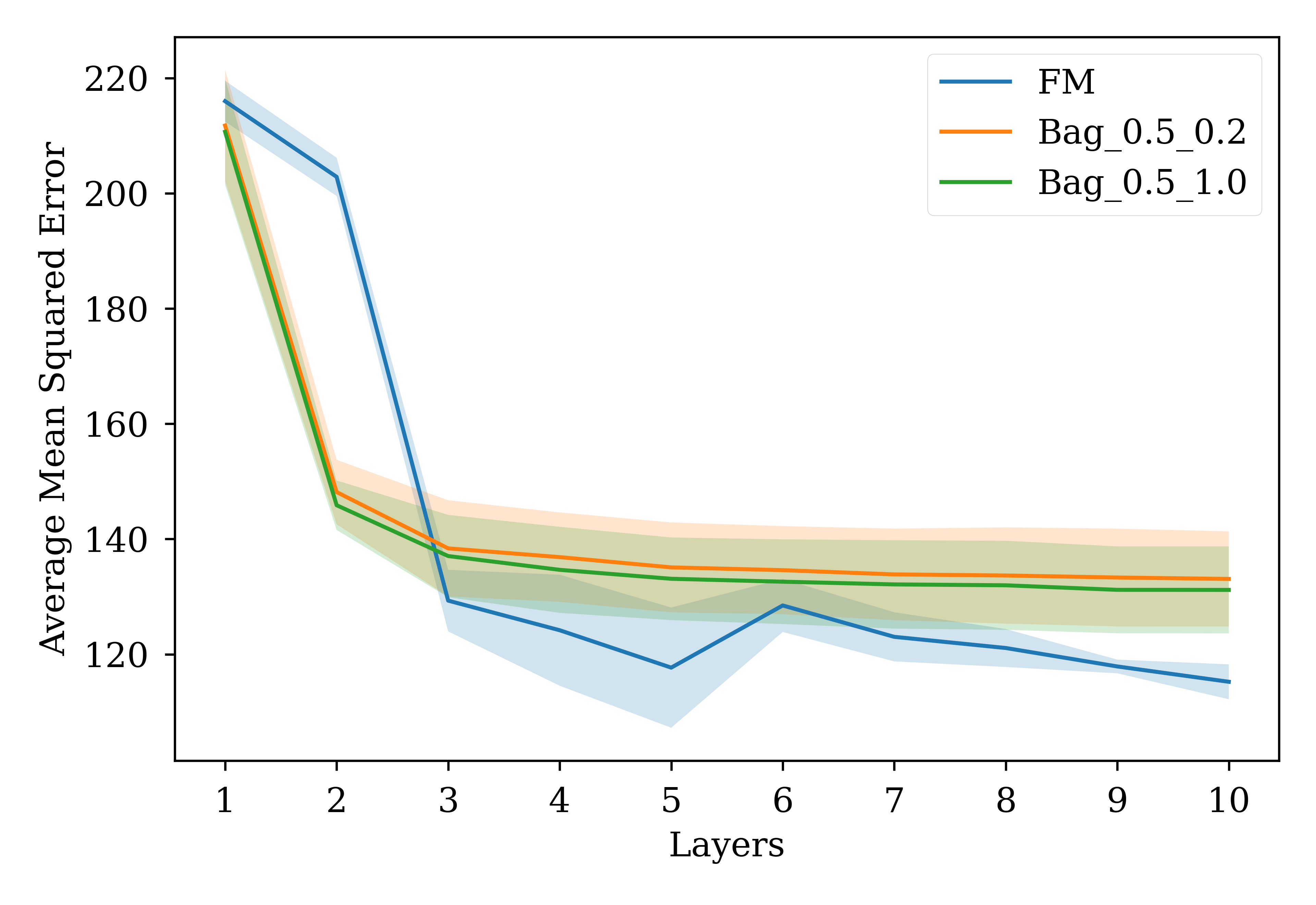}\\[-3pt]
\includegraphics[width=.48\linewidth]{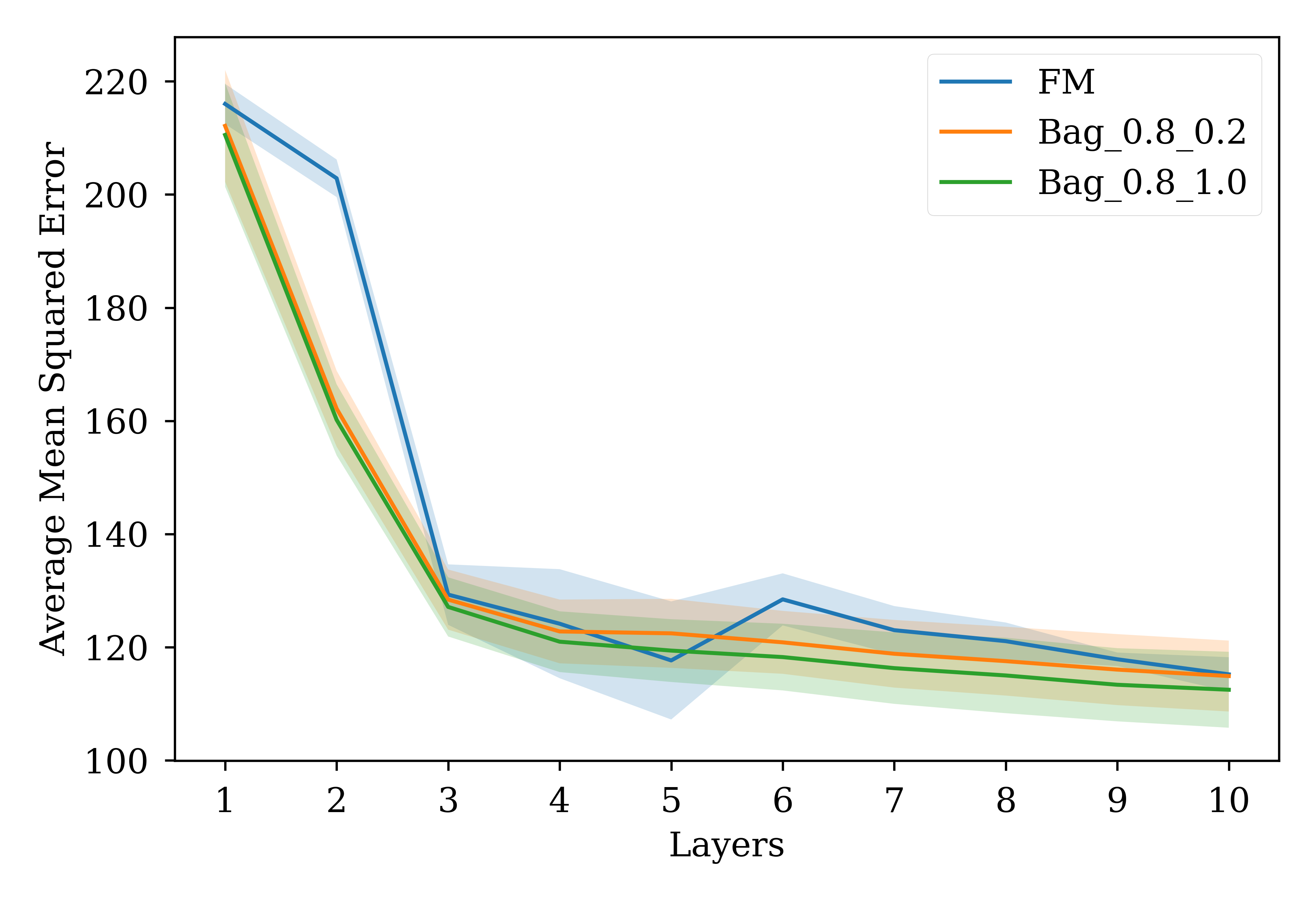}%
\includegraphics[width=.48\linewidth]{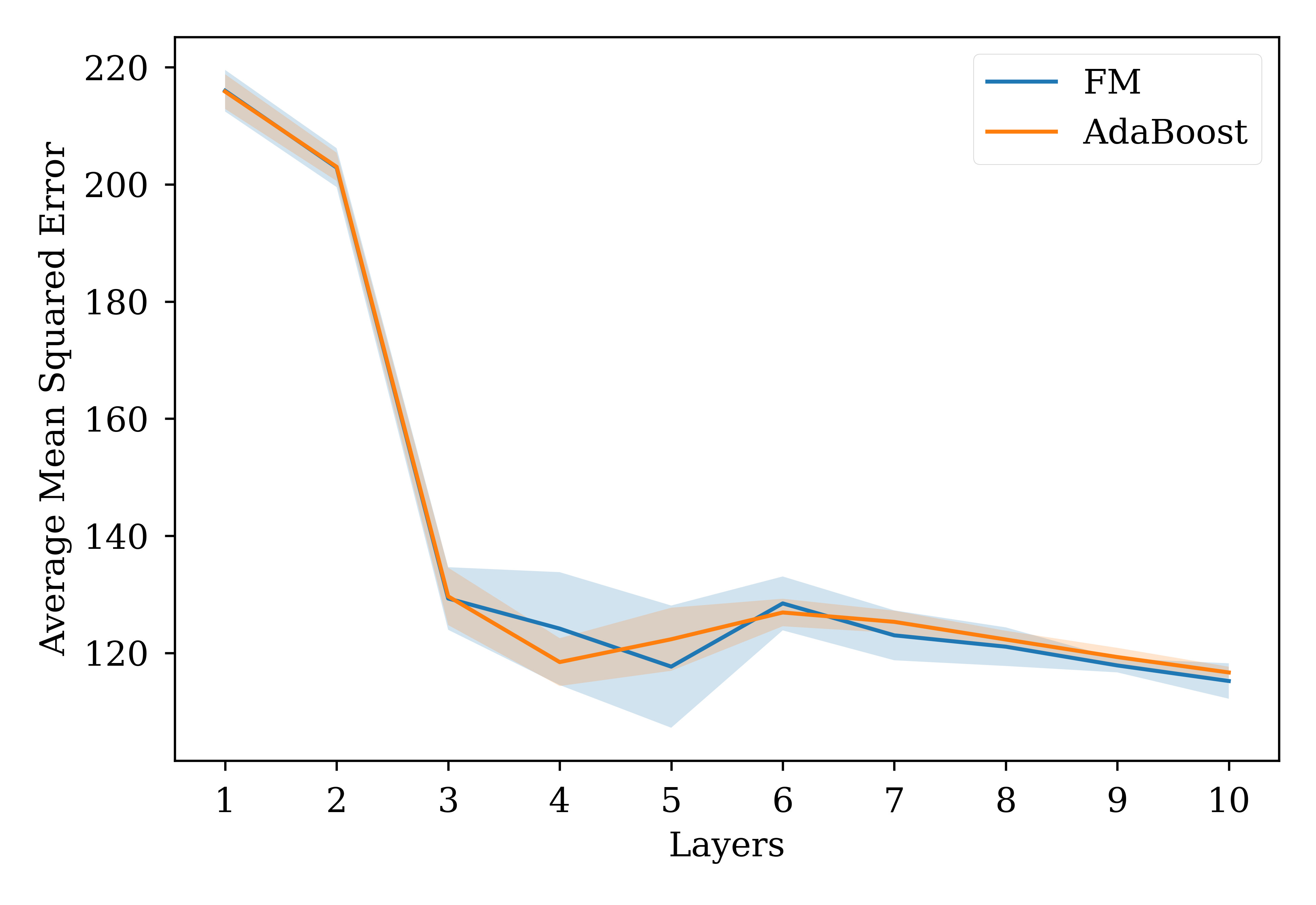}
\end{tabular}%
\caption{Comparison of the performance of the baseline model and ensemble systems on the Concrete Compressive Strength dataset. It exhibits the MSE and standard deviation, with a semi-transparent area, of the ensemble schemes in comparison to the baseline models. The top-left image shows ensembles with Random Subspace at 30\% of the features, top-right shows ensembles with Random Subspace at 50\%, bottom-left displays ensembles with Random Subspace at 80\%, and bottom-right illustrates AdaBoost.}%
\label{fig:concrete_erb}
\end{figure}

\begin{figure}[htbp]
\centering
\begin{tabular}[b]{@{}c@{}}
\includegraphics[width=.48\linewidth]{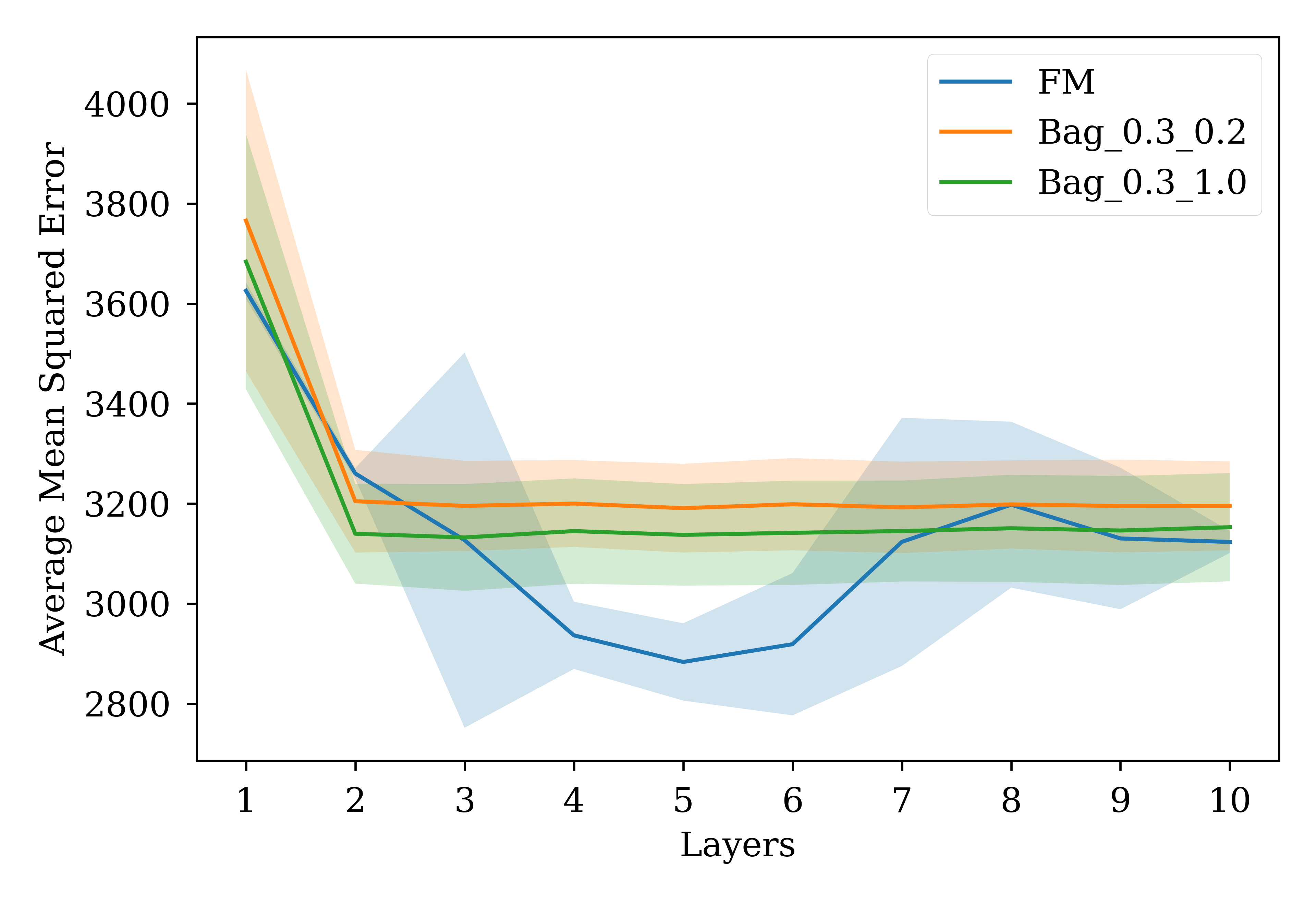}%
\includegraphics[width=.48\linewidth]{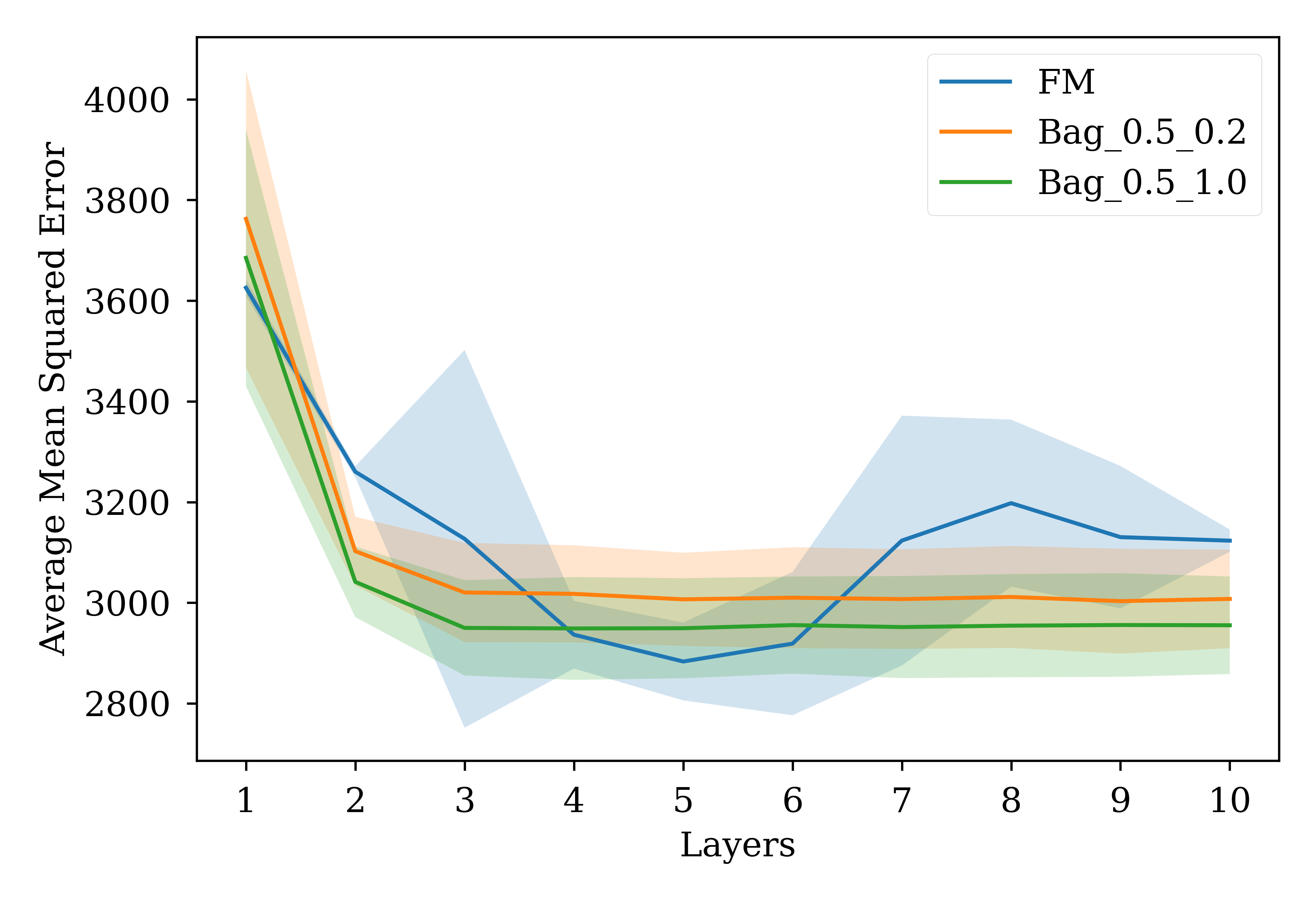}\\[-3pt]
\includegraphics[width=.48\linewidth]{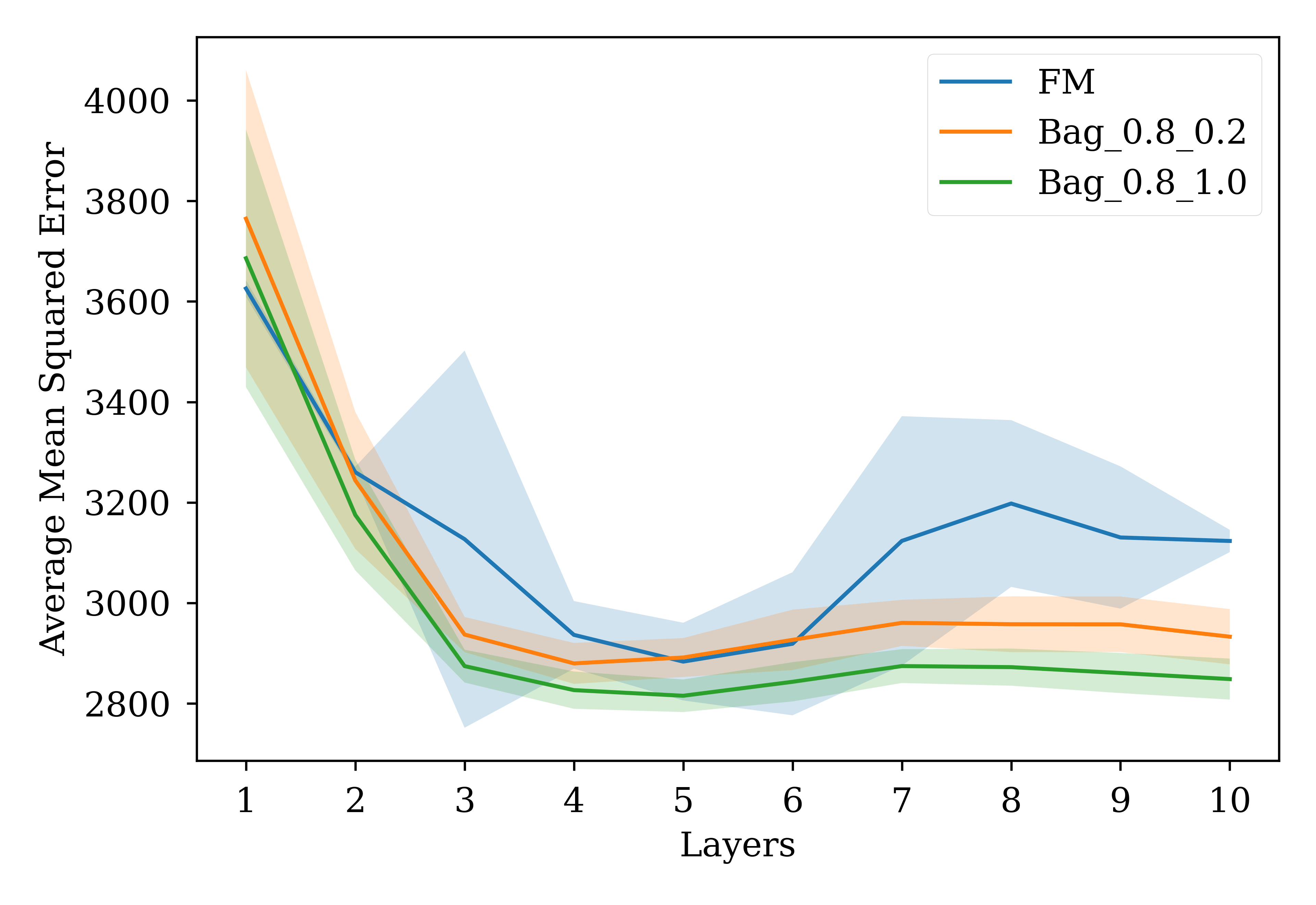}%
\includegraphics[width=.48\linewidth]{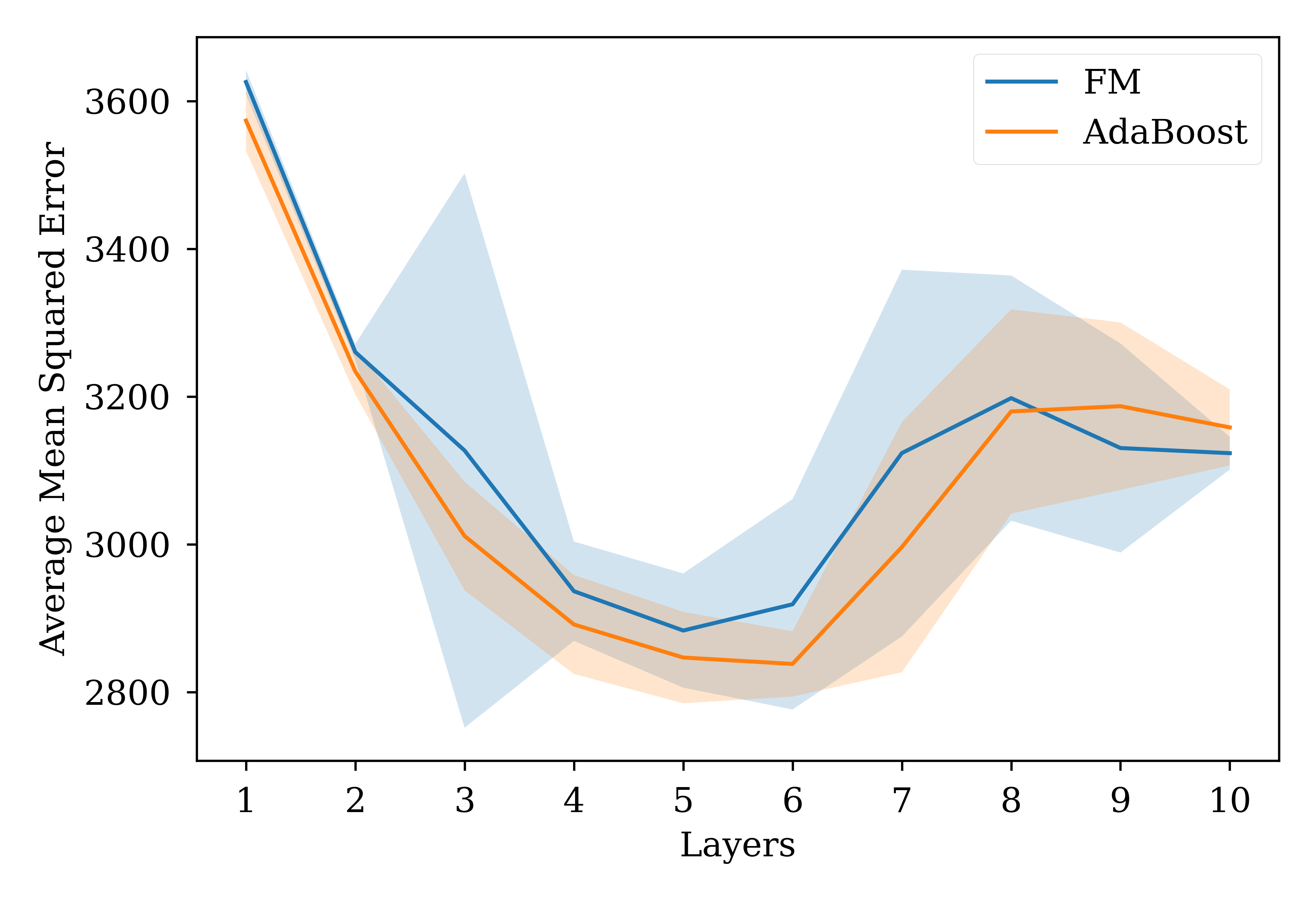}
\end{tabular}%
\caption{Comparison of the performance of the baseline model and ensemble systems on the Diabetes dataset. It exhibits the MSE and standard deviation, with a semi-transparent area, of the ensemble schemes in comparison to the baseline models. The top-left image shows ensembles with Random Subspace at 30\% of the features, top-right shows ensembles with Random Subspace at 50\%, bottom-left displays ensembles with Random Subspace at 80\%, and bottom-right illustrates AdaBoost.}%
\label{fig:diabete_erb}
\end{figure}

\begin{figure}[tbp]
\centering
\begin{tabular}[b]{@{}c@{}}
\includegraphics[width=.48\linewidth]{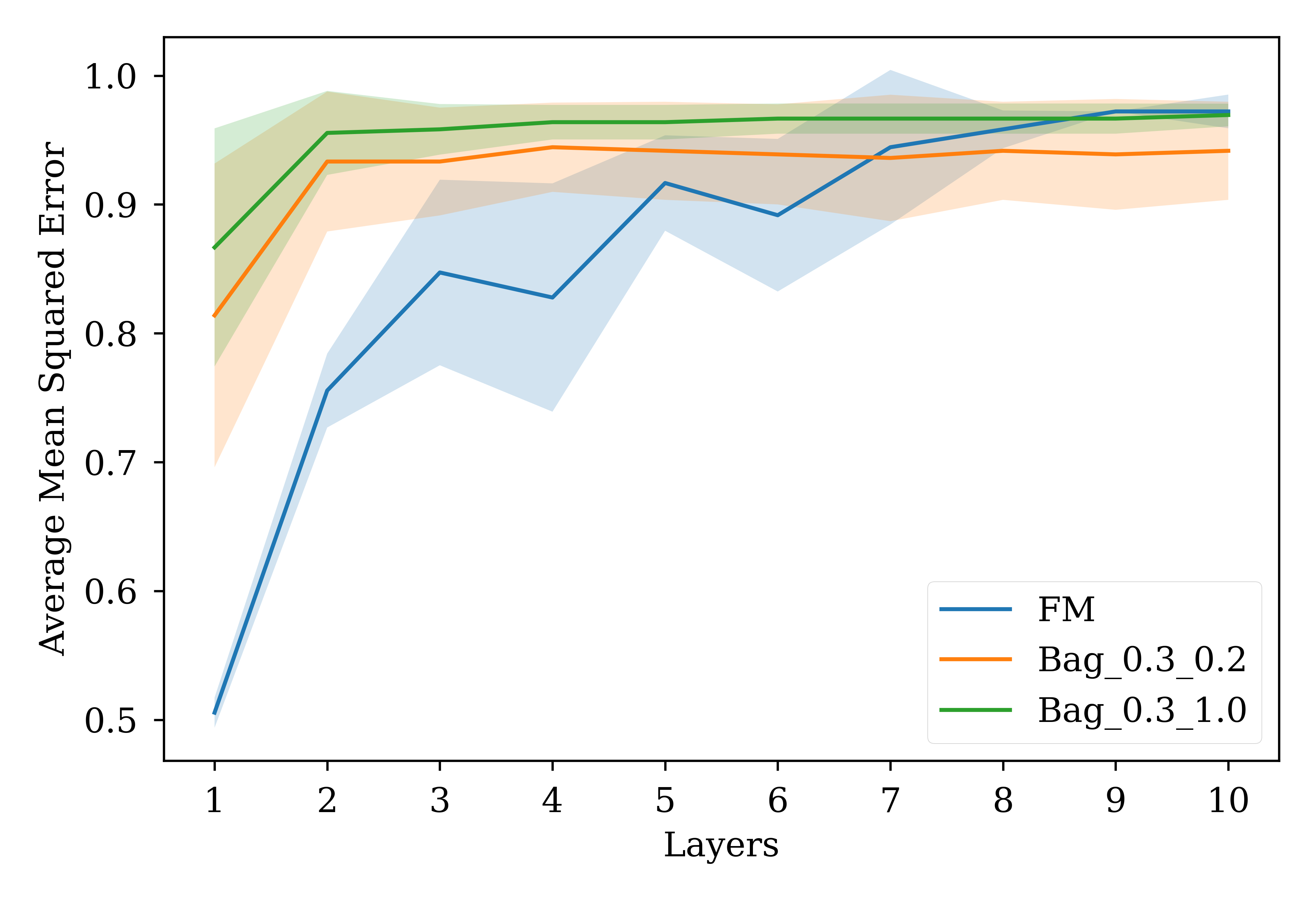}
\includegraphics[width=.48\linewidth]{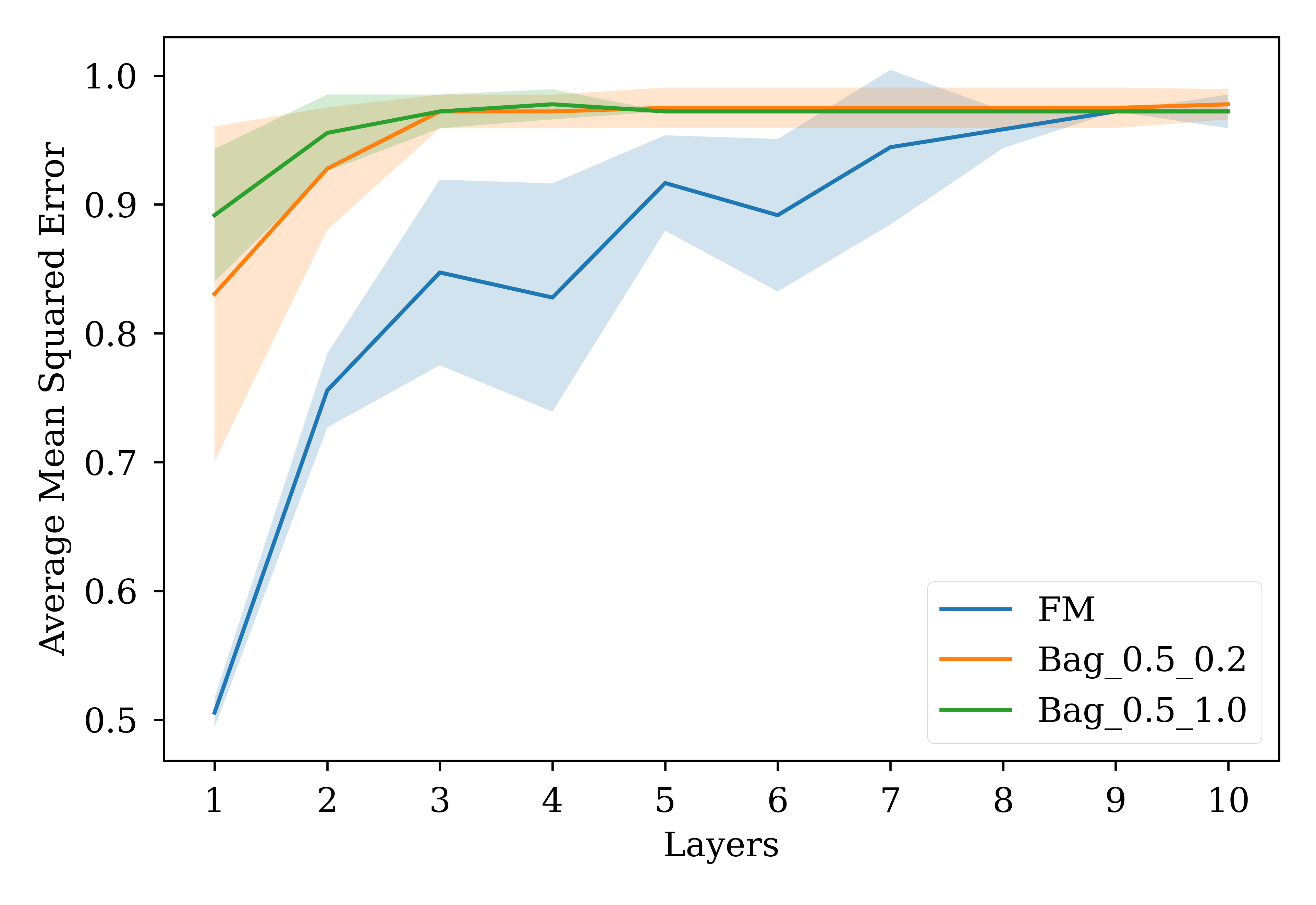}\\[-3pt]
\includegraphics[width=.48\linewidth]{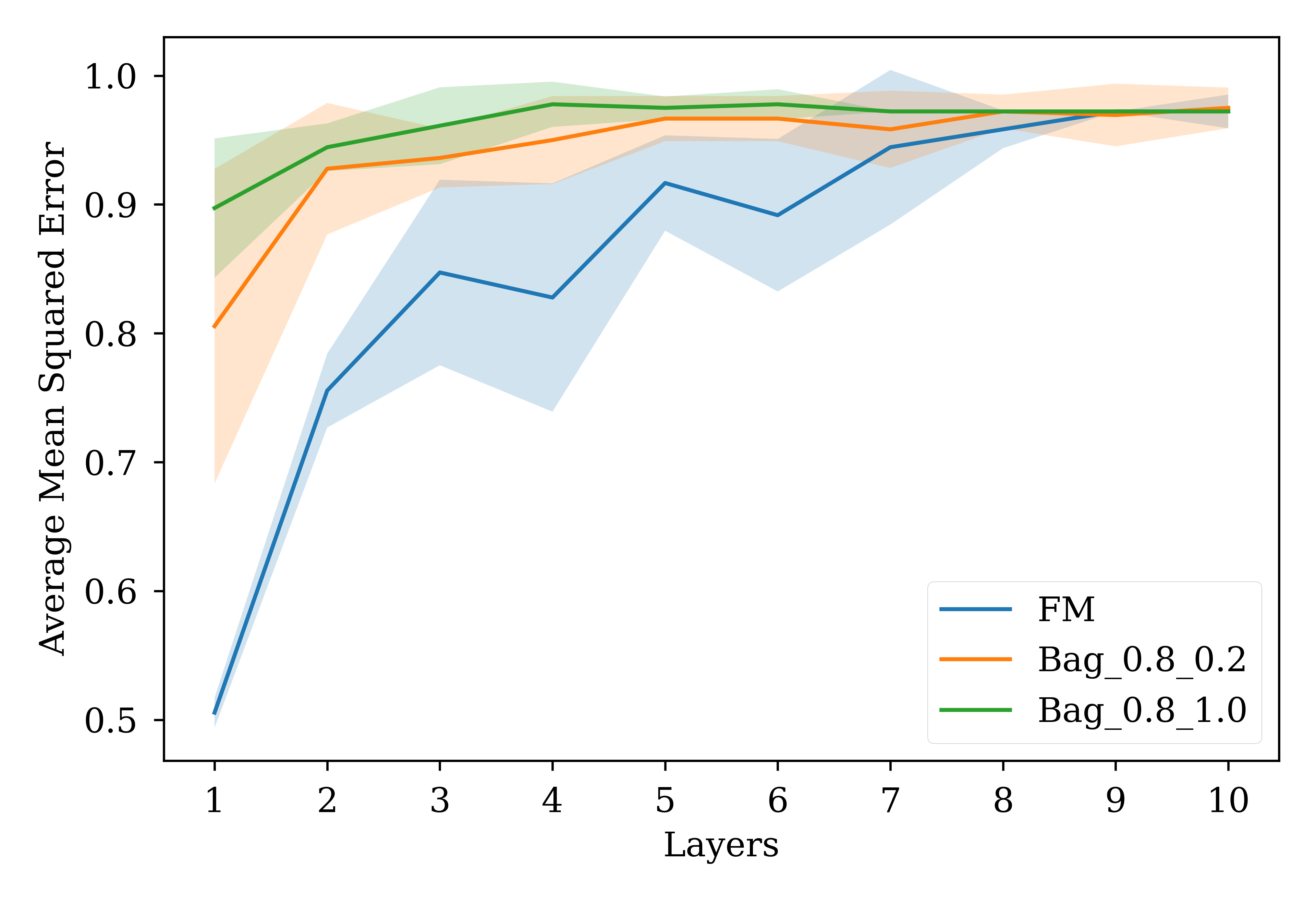}%
\includegraphics[width=.48\linewidth]{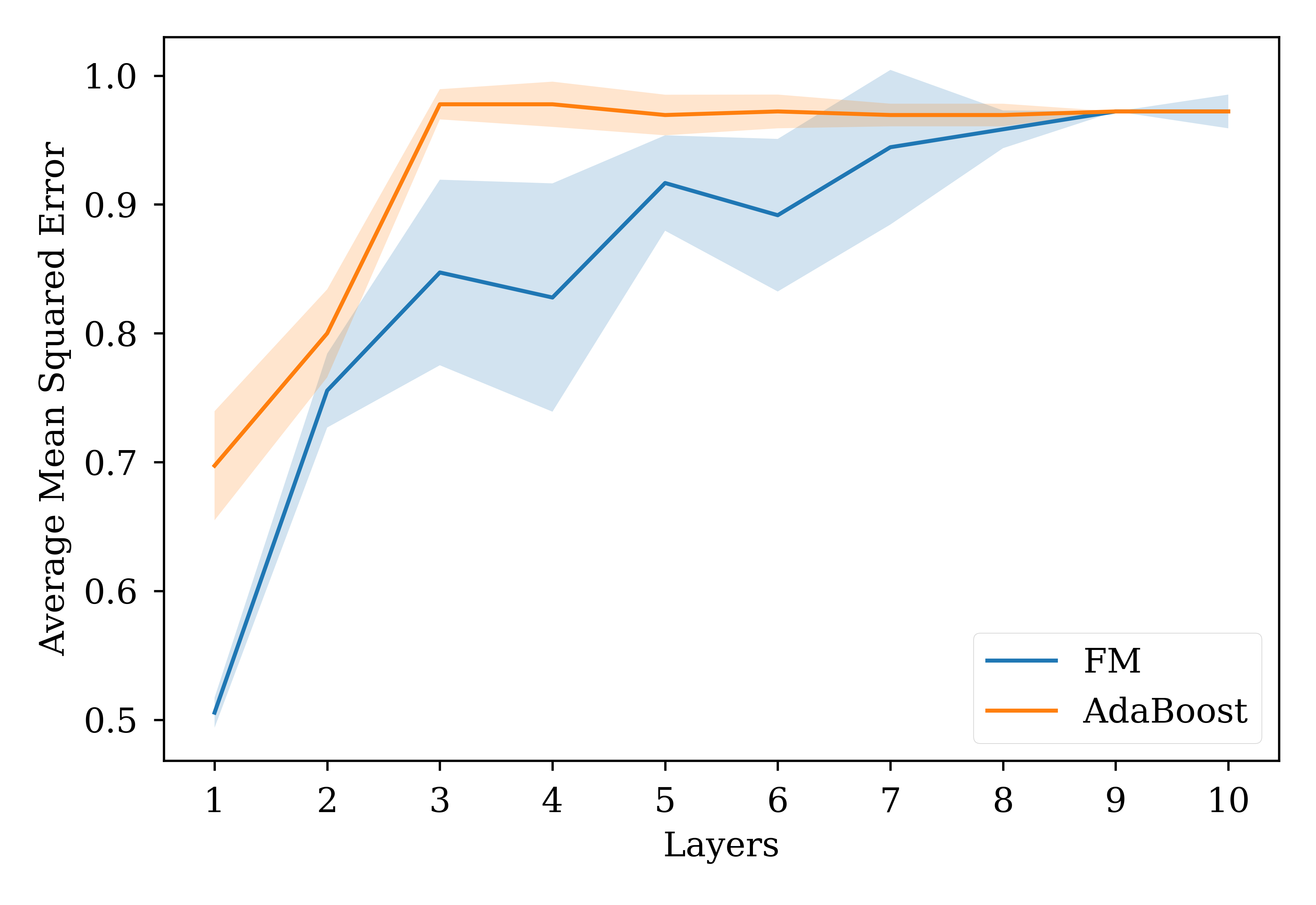}
\end{tabular}%
\caption{Comparison of the performance of the baseline model and ensemble systems on the Wine dataset. It exhibits the average accuracy and standard deviation, with a semi-transparent area, of the ensemble schemes in comparison to the baseline models. The top-left image shows ensembles with Random Subspace at 30\% of the features, top-right shows ensembles with Random Subspace at 50\%, bottom-left displays ensembles with Random Subspace at 80\%, and bottom-right illustrates AdaBoost.}%
\label{fig:wine_erb}
\end{figure}

We provide some additional plots of the simulated experiments. In particular, we compare the different configurations of bagging and boosting techniques and their variance. Figure \ref{fig:linear_erb}, \ref{fig:concrete_erb}, \ref{fig:diabete_erb}, \ref{fig:wine_erb} shows the results for the Linear, Concrete, Diabetes, and Wine datasets, respectively. 

Here we also report the training MSE and Accuracy for Experiments I-III and IV, respectively. The error trends are similar to the ones on the test sets and confirm the good convergence of all the models.

\begin{figure}[tbp]
\centering
\begin{tabular}[b]{@{}c@{}}
\includegraphics[width=.48\linewidth]{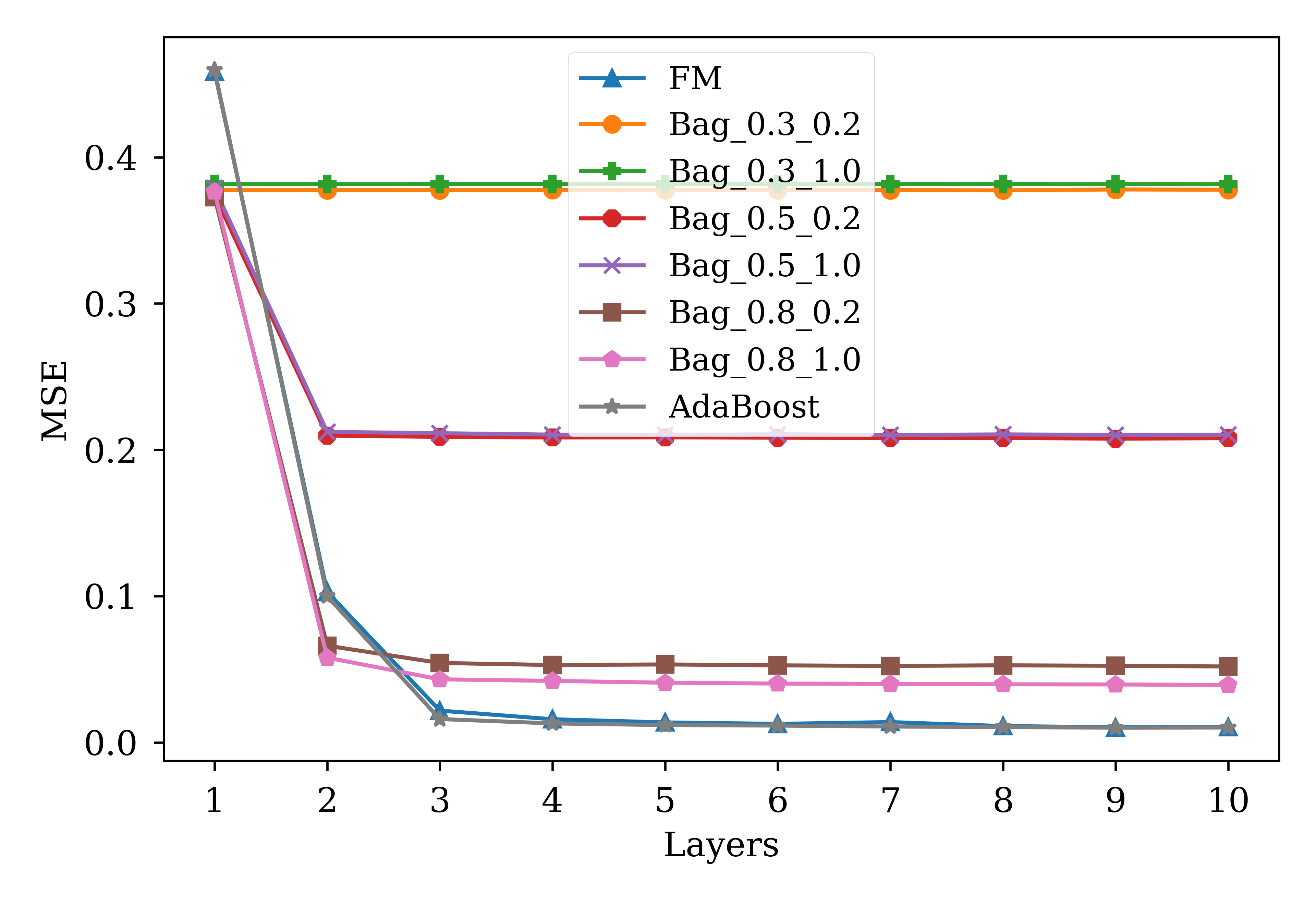}
\includegraphics[width=.48\linewidth]{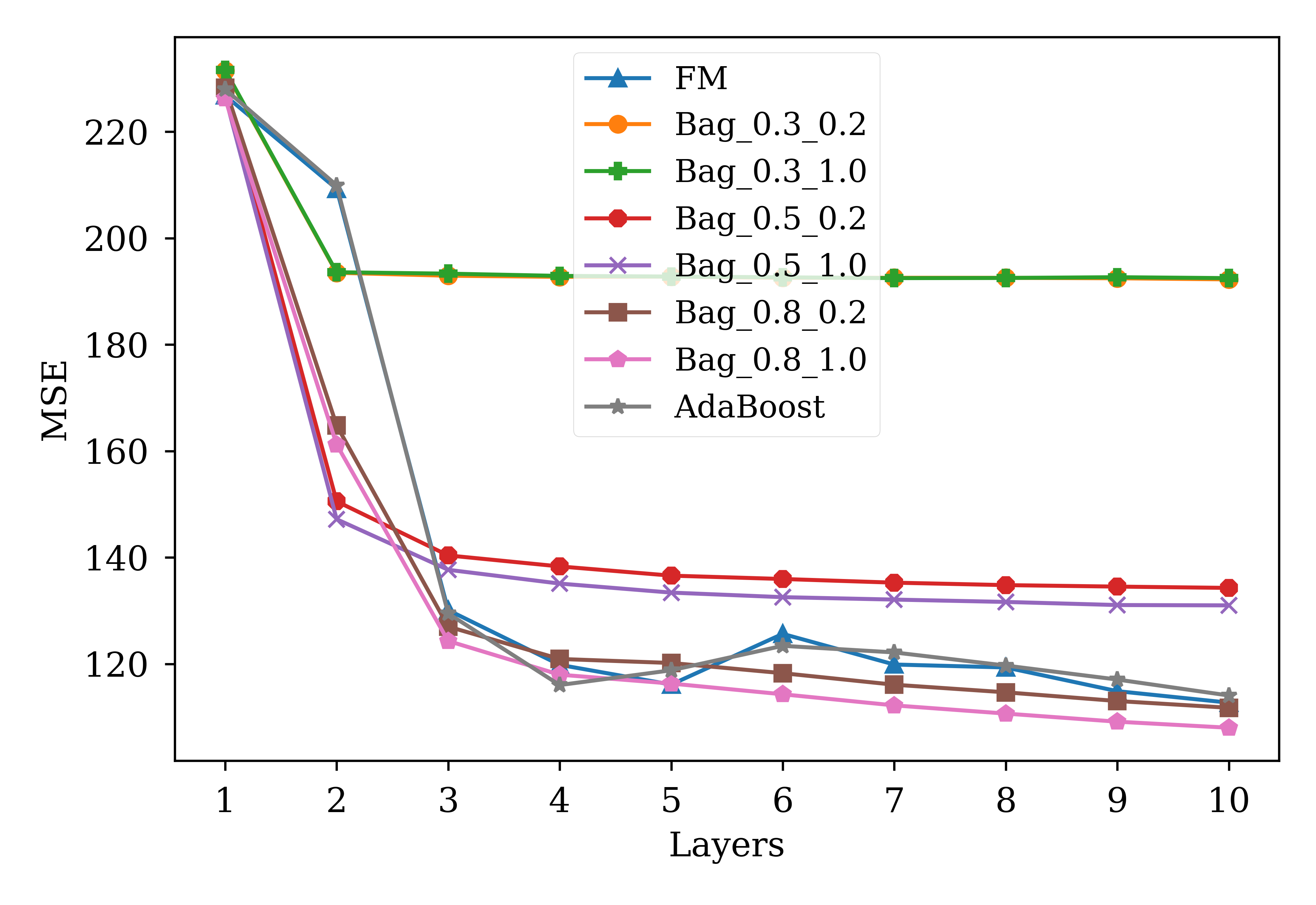}\\[-3pt]
\includegraphics[width=.48\linewidth]{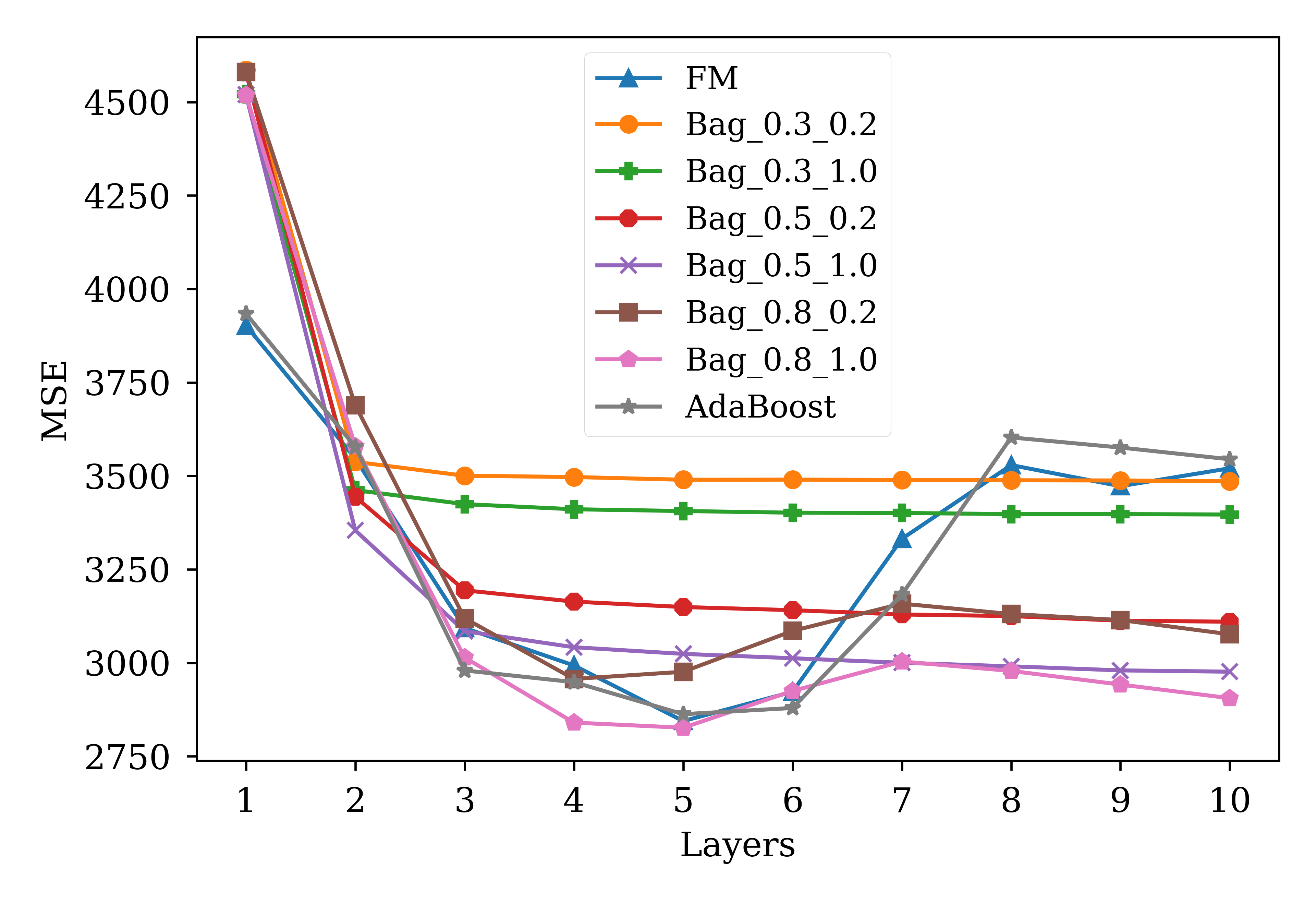}%
\includegraphics[width=.48\linewidth]{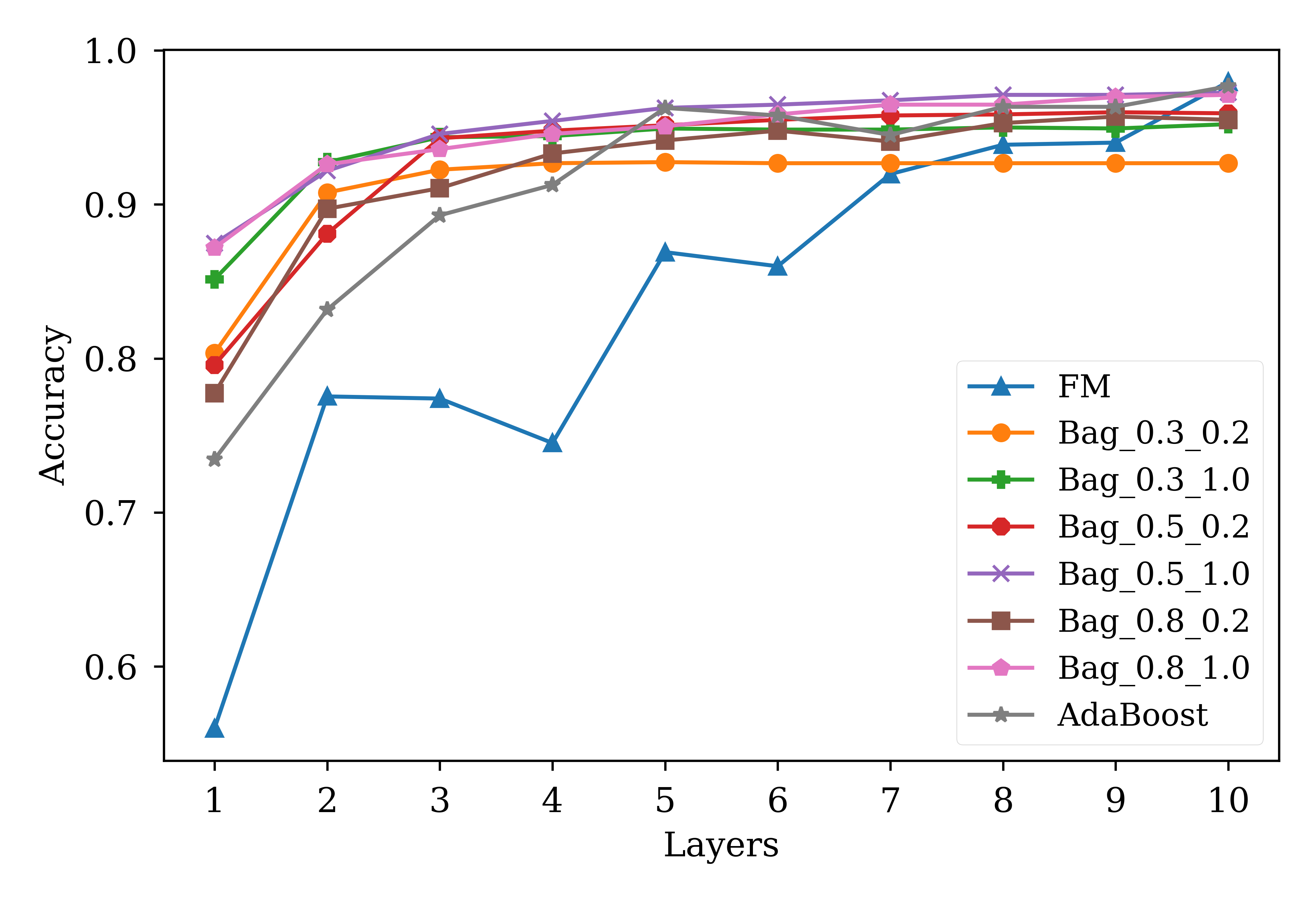}
\end{tabular}%
\caption{Training MSE of Experiment I (top-left), Experiment II (top-right), Experiment III (bottom-left) and Accuracy of Experiment IV (bottom-right).}%
\label{fig:training_mse_all}
\end{figure}

\section{Correlation analysis}

In order to assess the relationship between MSE or Accuracy and ensemble size, we calculated the Spearman correlation coeFfficient. It is a non-parametric measure of the strength and direction of monotonic association between two variables. With this analysis, we aimed to quantitatively determine whether changes in the ensemble size were associated with consistent trends in either the MSE or Accuracy values. In particular, the Spearman correlation provides a value between -1 and 1, where -1 indicates a perfect negative monotonic relationship, 1 indicates a perfect positive monotonic relationship, and 0 indicates no monotonic relationship.
Tables \ref{tab:spearman_I}, \ref{tab:spearman_II}, \ref{tab:spearman_III}, and \ref{tab:spearman_IV} illustrate the results of this analysis for the Linear, Concrete, Diabetes, and Wine datasets, respectively. The correlation observed in the tables indicates that larger ensemble sizes tend to result in lower MSE or higher Accuracy, although more tests should be conducted to further study such a phenomenon and obtain statistically significant results with more samples.

\begin{table}[h!]
    \centering
    \renewcommand{\arraystretch}{1.7}
    \begin{tabular}{cccccc}
        \toprule
        & \multicolumn{5}{c}{Layers}\\ 
        Model & 2 & 4 & 6 & 8 & 10\\\midrule
        Bag\_0.3\_0.2 & -1.00 & -1.00 & -1.00 & -1.00 & -1.00 \\ 
        Bag\_0.3\_1.0 & -1.00 & -1.00 & -1.00 & -1.00 & -1.00 \\ 
        Bag\_0.5\_0.2 & -1.00 & -1.00 & -1.00 & -1.00 & -1.00 \\ 
        Bag\_0.5\_1.0 & -1.0 & -0.95 & -0.95 & -0.95 & -0.79 \\ 
        Bag\_0.8\_0.2 & -0.40 & -0.40 & -0.40 & -0.40 & -0.40 \\ 
        Bag\_0.8\_1.0 & -0.63 & -0.80 & -0.80 & -0.80 & -0.80 \\
        \bottomrule
    \end{tabular}
    \caption{Spearman correlation coefficient between MSE and ensemble size in Experiment I (p-value $\leq$ 0.05). Only even number of layers are reported for brevity.}
    \label{tab:spearman_I}
\end{table}

\begin{table}[h!]
    \centering
    \renewcommand{\arraystretch}{1.7}
    \begin{tabular}{cccccc}
        \toprule
        & \multicolumn{5}{c}{Layers}\\ 
        Model & 2 & 4 & 6 & 8 & 10\\\midrule
        Bag\_0.3\_0.2 & -1.00 & -1.00 & -1.00 & -1.00 & -1.00 \\ 
        Bag\_0.3\_1.0 & -1.00 & -1.00 & -1.00 & -1.00 & -1.00 \\ 
        Bag\_0.5\_0.2 & -1.00 & -1.00 & -0.79 & -1.00 & -1.00 \\ 
        Bag\_0.5\_1.0 & -0.80 & -0.80 & -0.80 & -0.80 & -0.80 \\ 
        Bag\_0.8\_0.2 & -1.00 & -0.80 & -0.80 & -0.74 & -0.80 \\ 
        Bag\_0.8\_1.0 & -1.00 & -0.60 & -0.60 & -0.60 & -0.60 \\
        \bottomrule
    \end{tabular}
    \caption{Spearman correlation coefficient between MSE and ensemble size in Experiment II (p-value $\leq$ 0.05). Only even number of layers are reported for brevity.}
    \label{tab:spearman_II}
\end{table}

\begin{table}[h!]
    \centering
    \renewcommand{\arraystretch}{1.7}
    \begin{tabular}{cccccc}
        \toprule
        & \multicolumn{5}{c}{Layers}\\ 
        Model & 2 & 4 & 6 & 8 & 10\\\midrule
        Bag\_0.3\_0.2 & -1.00 & -1.00 & -1.00 & -1.00 & -1.00 \\ 
        Bag\_0.3\_1.0 & -1.00 & -1.00 & -1.00 & -1.00 & -1.00 \\ 
        Bag\_0.5\_0.2 & -1.00 & -1.00 & -1.00 & -1.00 & -1.00 \\ 
        Bag\_0.5\_1.0 & -0.80 & -1.00 & -1.00 & -1.00 & -1.00 \\ 
        Bag\_0.8\_0.2 & -1.00 & -1.00 & -1.00 & -1.00 & -1.00 \\ 
        Bag\_0.8\_1.0 & -0.80 & -1.00 & -0.80 & -0.40 & -0.40 \\
        \bottomrule
    \end{tabular}
    \caption{Spearman correlation coefficient between MSE and ensemble size in Experiment III (p-value $\leq$ 0.05). Only even number of layers are reported for brevity.}
    \label{tab:spearman_III}
\end{table}

\begin{table}[h!]
    \centering
    \renewcommand{\arraystretch}{1.7}
    \begin{tabular}{cccccc}
        \toprule
        & \multicolumn{5}{c}{Layers}\\ 
        Model & 2 & 4 & 6 & 8 & 10\\\midrule
        Bag\_0.3\_0.2 & 1.00 & 0.95 & 0.95 & 1.00 & 1.00 \\ 
        Bag\_0.3\_1.0 & 1.00 & 1.00 & 1.00 & 1.00 & 1.00 \\ 
        Bag\_0.5\_0.2 & 0.80 & 1.00 & 0.94 & 1.00 & 0.80 \\ 
        Bag\_0.5\_1.0 & 0.80 & 1.00 & 0.89 & 0.95 & 0.89 \\ 
        Bag\_0.8\_0.2 & 0.80 & 1.00 & 1.00 & 0.80 & 0.80 \\ 
        Bag\_0.8\_1.0 & 0 & 0.32 & 0.80 & 0 & 0 \\
        \bottomrule
    \end{tabular}
    \caption{Spearman correlation coefficient between Accuracy and ensemble size in Experiment IV (p-value $\leq$ 0.05). Zero values indicate that the alternative hypothesis is rejected, i.e. the coefficient is 0 and hence there is no correlation. This outcome was observed exclusively with the more complex ensemble, and can be attributed to its ability to attain favorable Accuracy values even when employing a relatively small number of estimators. Only even number of layers are reported for brevity.}
    \label{tab:spearman_IV}
\end{table}

\end{appendices}

\end{document}